\documentclass[aps, prd, reprint,10pt, notitlepage, a4paper,floats, floatfix,amsmath, amssymb, amsfonts,superscriptaddress,showpacs, showkeys,nofootinbib,longbibliography]{revtex4-2}

\pdfoutput=1

\usepackage{epsfig}
\usepackage{graphics}
\usepackage{graphicx}
\usepackage{amsmath,amssymb,mathrsfs}
\usepackage{amsfonts}
\usepackage[usenames,dvipsnames]{xcolor}
\usepackage{wasysym}
\usepackage{times}
\usepackage{mathptmx}
\usepackage{gensymb}
\usepackage{appendix}
\usepackage{listings}
\usepackage{url}
\usepackage[normalem]{ulem}
\usepackage{alltt}
\usepackage[colorlinks]{hyperref}
\usepackage{cleveref}
\usepackage{longtable}
\usepackage{enumitem}
\setlist{nosep}
\usepackage{color}
\usepackage{calc}
\usepackage{tensor}
\usepackage{bm}
\usepackage{times}
\usepackage{multirow}
\usepackage[varg]{txfonts}
\usepackage{float}
\usepackage{dcolumn}
\usepackage[nolist,nohyperlinks]{acronym}
\usepackage{xspace}
\usepackage[english]{babel}
\usepackage[abs]{overpic}
\usepackage{pict2e}
\usepackage[caption=false]{subfig}
\allowdisplaybreaks[1]
\usepackage[utf8]{inputenc}
\usepackage{gensymb}
\usepackage{bm}
\usepackage{stackengine}
\usepackage{boldline,multirow}
\usepackage{braket}
\usepackage{longtable}
\usepackage{tabularx}
\usepackage{rotating}

\DeclareMathAlphabet{\mathcalstd}{OMS}{cmsy}{m}{n}
\DeclareMathAlphabet{\mathpzc}{OT1}{pzc}{m}{it}

\newcommand{\AEI}{Max Planck Institute for Gravitational Physics (Albert Einstein Institute), Am M{\"u}hlenberg 1, Potsdam, 14476, Germany}
\newcommand{\Maryland}{Department of Physics, University of Maryland, College Park, MD 20742, USA}

\definecolor{RED}{HTML}{F5054F}
\definecolor{LIGHT_BLUE}{HTML}{448EE4}
\definecolor{DARK_BLUE}{HTML}{0343df}

\hypersetup{citecolor=LIGHT_BLUE, urlcolor=DARK_BLUE, linkcolor=RED}

\begin{document}

\title{Testing the spin-induced multipole moments of compact binary coalescences using the flexible theory-independent framework}

\author{Elise M. S{\"a}nger}
\email{elise.saenger@aei.mpg.de}
\affiliation{\AEI}

\author{Alessandra Buonanno}
\affiliation{\AEI}
\affiliation{\Maryland}

\author{Ajit Kumar Mehta}
\affiliation{Chennai Mathematical Institute, Siruseri 603013, Chennai, India}
\affiliation{International Centre for Theoretical Sciences, Tata Institute of Fundamental Research, Bangalore 560089, India}

\author{Jan Steinhoff}
\affiliation{\AEI}

\date{\today}

\begin{abstract}

According to the no-hair theorem, the multipole moments of an electrically neutral black hole in general relativity are entirely determined by its mass and spin. However, this is not in general true for compact objects: The multipole moments of neutron stars or exotic compact objects can depend on their formation history and internal processes, which are encoded in an equation of state. Furthermore, their spin-induced multipole moments differ from those of black holes of the same mass and spin, leaving an imprint on the dynamics and emitted gravitational waves of the binary. Gravitational waves can thus be used to test the nature of compact binary coalescences. Here, we present a test of the spin-induced quadrupole and octupole moments of compact objects based on the flexible theory-independent (FTI) framework. FTI is a parameterized inspiral test of general relativity which enables the addition of post-Newtonian coefficient deviations to the gravitational-wave phase of a generic aligned-spin frequency-domain waveform model. We use this test on synthetic signals to study the measurability of spin-induced quadrupole and octupole moments. Next, we apply the test to a subset of signals observed by the LIGO-Virgo-KAGRA Collaboration. Lastly, we present forecasts for next-generation ground-based detectors, such as Einstein Telescope and Cosmic Explorer. Our estimates suggest that these detectors will be capable of placing stringent constraints on the spin-induced quadrupole and octupole moments of $\mathcal{O}(10^{-2})$ and $\mathcal{O}(10^{-1})$ respectively, which is two orders of magnitude tighter than current constraints, and thus on the nature of black holes in a compact binary coalescence.
\end{abstract}

\maketitle

\section{Introduction} \label{sec:intro}

One characteristic property of black holes (BHs) in general relativity (GR) is the no-hair theorem~\cite{Carter:1971zc, Hansen:1974zz}, which manifests itself in the fact that all multipole moments of a Kerr BH are determined by its mass $\mathcal{M}$ and dimensionless spin $\chi = S/\mathcal{M}^2$. The multipole moments of a Kerr BH can be written as
\begin{equation}
    \mathcal{M}_l^\mathrm{BH} + i\mathcal{S}_l^\mathrm{BH} = \mathcal{M}^{l+1}(i\chi)^l,
\end{equation}
where $\mathcal{M}_l^\mathrm{BH}$ and $\mathcal{S}_l^\mathrm{BH}$ are the mass and current multipole moments. Since all multipole moments depend only on the mass and spin, measuring three of them---for instance, the mass $\mathcal{M}_0 = \mathcal{M}$, the spin $\mathcal{S}_1 = S$, and the spin-induced quadrupole moment (SIQM) $\mathcal{M}_2$---can serve as a null test to confirm the nature of BHs in GR~\cite{Krishnendu:2017shb}. Furthermore, neutron stars and exotic compact objects can have spin-induced moments that differ from those of Kerr BHs. 
Testing for the SIQM could therefore be used to classify the objects by type in the lower mass gap~\cite{Lyu:2023zxv} or to identify exotic compact objects.

Deviations from the Kerr multipole moments result in differences in the emitted gravitational waves (GWs). During the inspiral phase of compact binary coalescence, these differences can be modeled using the post-Newtonian (PN) approximation. These deviations then lead to modifications in the PN coefficients compared to binary BH (BBH) waveforms~\cite{Arun:2008kb, Mishra:2016whh}. A SIQM test that implements these modifications has already been proposed~\cite{Krishnendu:2017shb} and has been used to analyze events observed by the collaboration of LIGO~\cite{LIGOScientific:2014pky}, Virgo~\cite{VIRGO:2014yos}, and \hyphenation{KAGRA}KAGRA~\cite{KAGRA:2020tym} (LVK) detectors~\cite{Krishnendu:2019tjp, LIGOScientific:2020zkf, LIGOScientific:2020tif, LIGOScientific:2021sio, LIGOScientific:2025brd}. One highlight is the stringent constraint on the dimensionless SIQM of order unity for the primary, highly-spinning, compact object in GW241011\_233834~\cite{LIGOScientific:2025brd}. It has also been used for population inference to constrain the fraction of non-black holes in the observed compact binary population~\cite{Saleem:2021vph}. This test was later extended to include higher harmonics and spin precession~\cite{Divyajyoti:2023izl}, and to test for the spin-induced octupole moment (SIOM)~\cite{Krishnendu:2018nqa, Saini:2023gaw}. SIQM corrections have been incorporated into the spin-precession dynamics and have been shown to increase the test's sensitivity~\cite{Lyu:2023zxv}. A version of the test was recently developed for eccentric binaries~\cite{Naqvi:2025gly, Krishnendu:2025awt}. Versions of this test have also been used to make projections for the future measurability of the spin-induced multipole moments for next-generation ground-based (XG)~\cite{Krishnendu:2018nqa, Branchesi:2023mws, ET:2025xjr} and space-based~\cite{Krishnendu:2019ebd, Kong:2024ssa, Piarulli:2025rvr} instruments.

The Flexible Theory-Independent (FTI) method~\cite{Mehta:2022pcn} is a theory-agnostic, parameterized inspiral test. It modifies the frequency-domain phase during inspiral by adding generic corrections to the PN expansion coefficients. This allows the test to be sensitive to a variety of deviations from GR. It has already been used on observations by the LVK Collaboration to put constraints on possible deviations from the PN predictions~\cite{LIGOScientific:2018dkp, LIGOScientific:2019fpa, LIGOScientific:2020aai,  LIGOScientific:2020zkf, LIGOScientific:2020tif, LIGOScientific:2021sio, Sanger:2024axs, LIGOScientific:2025cmm, LIGOScientific:2025wao, LIGOScientific:2026fcf, LIGOScientific:2026qqv}. These bounds can then be used to constrain on some specific modified theories of gravity~\cite{LIGOScientific:2021sio, Sanger:2024axs, LIGOScientific:2026fcf}. The versatility of the FTI framework allows it to be easily extended to test for the multipole moments of compact objects. In this study, we implement a test for the SIQM and SIOM using the FTI framework.

While the existing SIQM and SIOM tests and the test using the FTI method presented here are not identical, they are both based on deviations in the PN coefficients. The former is specific to the \textsc{IMRPhenom} waveform family, while the latter can be used with any frequency-domain spin-aligned waveform model and usually employs the \textsc{SEOBNRv5} waveform family. The implementations also differ slightly, primarily during the transition from inspiral to merger-ringdown. Seemingly minor variations in the underlying GR waveform model and the exact implementation of the test can lead to different systematic errors when applying the tests. Therefore, it is important to develop independent tests to distinguish between true violations of GR and erroneous deviations from GR. Additionally, the construction of the FTI framework allows for greater flexibility in the baseline waveform model and the tapering between inspiral and merger-ringdown, which can be leveraged to identify systematic errors. Within the LVK Collaboration, the FTI SIQM test using \textsc{SEOBNRv5} has already been applied to the analysis of GW241011\_233834~\cite{LIGOScientific:2025brd} and the fourth Gravitational-Wave Transient Catalog (GWTC-4.0)~\cite{LIGOScientific:2026fcf}, giving results consistent with those of the IMRPhenomX-based SIQM test.

The rest of this paper is structured as follows. In Sec.~\ref{sec:FTI}, we briefly discuss the FTI method and its application to measuring the SIQM and SIOM of compact objects. Section~\ref{sec:injections} studies the measurability of the SIQM and SIOM using simulated signals (injections). We also examine systematic errors and the effect of flexibility in the choice of tapering on these errors. Then, in Sec.~\ref{sec:real-events}, we apply the SIQM and SIOM tests to selected signals observed by the LVK Collaboration. Lastly, in Sec.~\ref{sec:3G}, we use Bayesian inference to forecast how well XG detectors will constrain the SIQM and SIOM.

\section{The FTI model} \label{sec:FTI}

In GR, GWs from quasi-circular BBHs depend on eight intrinsic parameters, namely the masses $m_i$ and spin vectors $\vec{S}_i$ of the compact objects $i=1,2$. For aligned-spin systems, this can be reduced to four parameters $\vec{\theta} = \{ m_1, m_2, \chi_1, \chi_2 \}$, where $\chi_i = \vec{S}_i \cdot \vec{L} / (|\vec{L}|m_i^2)$ are the dimensionless spins aligned with the orbital angular momentum $\vec{L}$. It can be useful to express the masses using the following quantities: the mass ratio $q = m_2/m_1 \leq 1$, the symmetric mass ratio $\eta = q/(1+q)^2$, the binary's total mass $M = m_1 + m_2$, and the chirp mass $\mathcal{M}_c = \eta^{3/5}M$. Another useful quantity is the effective spin parameter $\chi_\mathrm{eff} = (m_1\chi_1 + m_2\chi_2)/M$. In addition to these intrinsic parameters, the GW signal depends on seven extrinsic parameters: $\vec{\xi} = \{ \imath, \varphi_c, \alpha, \delta, \beta, d_L, t_c \}$, where these are the angular orientation of the line of sight measured in the source frame ($\imath, \varphi_c$), the sky location of the source in the detector frame ($\alpha, \delta$), the polarization angle $\beta$, 
the luminosity distance of the source $d_L$, and the time of arrival $t_c$. Thus, for BBHs with aligned spins in a quasi-circular orbit, the total number of parameters required to describe the system is 11.

During quasi-circular, adiabatic inspiral, the waveform can be calculated using the PN formalism~\cite{Blanchet:2013haa}. The frequency-domain phase can then be obtained from the time domain PN waveform via the stationary-phase approximation~\cite{Sathyaprakash:1991mt, Cutler:1994ys, Buonanno:2009zt}. In GR, the frequency-domain phase is given by
\begin{multline} \label{eq:GR-phase}
    \psi_{\ell m}^\mathrm{GR}(f, \vec{\theta}) = 2\pi f t_c - \varphi_c - \frac{\pi}{4} \\ + \frac{3}{128\eta v^5}\frac{m}{2} \left[ \sum_{n=0}^7 \psi_n^\mathrm{GR}(\vec{\theta}) v^n + \sum_{n=5}^6 \psi_{n(l)}^\mathrm{GR}(\vec{\theta}) v^n \log v \right],
\end{multline}
where $v \equiv (2\pi fM/m)^{1/3}$ with $f$ the GW frequency. The subscript $\ell m$ denotes the $(\ell, m)$-mode from the mode decomposition of the GW signal into spin-weighted spherical harmonics. The functions $\psi_n^\mathrm{GR}$ and $\psi_{n(l)}^\mathrm{GR}$ are the $(n/2)$-PN phase coefficients in GR. The complete expressions for these coefficients can be found in Ref.~\cite{Mehta:2022pcn}.

\subsection{Generic modifications}

The FTI method~\cite{Mehta:2022pcn} generalizes the frequency-domain phase during inspiral. We adapt the GR waveform model from Eq.~\eqref{eq:GR-phase} by generalizing the PN coefficients. This gives corrections to the phase of the form 
\begin{multline}
    \delta \psi_{\ell m}(f, \vec{\theta}; \delta\psi_n, \delta\psi_{n(l)}) = \\
    \frac{3}{128\eta v^5}\frac{m}{2} \left[ \sum_{n=-2}^7 \delta\psi_n(\vec{\theta}) v^n + \sum_{n=5}^6 \delta\psi_{n(l)}(\vec{\theta}) v^n \log v \right], \label{eq:generic-phase-correction}
\end{multline}
where $\delta\psi_n$ and $\delta\psi_{n(l)}$ are deviations from the $(n/2)$-PN phase coefficients.

In FTI, we assume that each deviation coefficient $\delta\psi_n(\vec{\theta})$, $\delta\psi_{n(l)}(\vec{\theta})$ can be represented by a deviation parameter $\delta\varphi_n$, $\delta\varphi_{n(l)}$ that is the fractional deviation from the corresponding PN coefficient in GR, so we have that
\begin{align}
    \delta\psi_n(\vec{\theta}; \delta\varphi_n) &\equiv \delta\varphi_n \psi_n^\mathrm{GR}(\vec{\theta}), \\
    \delta\psi_{n(l)}(\vec{\theta}; \delta\varphi_{n(l)}) &\equiv \delta\varphi_{n(l)} \psi_{{n(l)}}^\mathrm{GR}(\vec{\theta}).
\end{align}
When the PN coefficient in GR vanishes (i.e., for $n=-2,1$), we consider the deviation parameter to be an absolute deviation instead. We do not include a deviation at $-0.5$PN ($n=-1$).

During the early inspiral stage (low frequencies), we model the phase of the frequency-domain waveform as
\begin{equation}
    \psi_{\ell m}(f, \vec{\theta}) = \psi_{\ell m}^\mathrm{GR}(f, \vec{\theta}) + \delta\psi_{\ell m}(f, \vec{\theta}; \delta\varphi_n, \delta\varphi_{n(l)}).
\end{equation}
During the post-inspiral stage (high frequencies), the phase must match the GR waveform phase, except for a constant shift due to accumulated dephasing during inspiral. Additionally, we require the waveform to be $C^2$ smooth over all frequencies. To achieve this, we apply a tapering function $W(f)$, to the phase corrections to transition smoothly between the early inspiral, which has corrections, and the unmodified post-inspiral. The chosen tapering function is~\cite{Mehta:2022pcn}
\begin{equation}
    W(f; v^\mathrm{tape}, \Delta v^\mathrm{tape}) \equiv \left[ 1 + \exp\left( \frac{v-v^\mathrm{tape}}{\Delta v^\mathrm{tape}} \right) \right]^{-1},
\end{equation}
which smoothly transitions between 0 and 1 around $v^\mathrm{tape}$ over the range of $\sim \Delta v^\mathrm{tape}$. This tapering function is applied to the second derivative of the phase correction with respect to frequency, which we denote as $\delta\psi_{\ell m}''(f)$. Integrating the tapered $\delta\psi_{\ell m}''(f)$ twice ensures $C^2$ smoothness of the phase, which leads us to
\begin{multline}
    \delta \psi_{\ell m}(f, \theta; \delta\psi_n, \delta\psi_{n(l)}; v^\mathrm{tape}, \Delta v^\mathrm{tape}) = \\
    \begin{aligned}
        \int_{f_{\ell m}^\mathrm{ref}}^f \mathrm{d}f' \int_{f_{22}^\mathrm{peak}}^{f'} \mathrm{d}f''  \delta\psi_{\ell m}''(f'', \theta; \delta\psi_n, \delta\psi_{n(l)})& \\
        \times W(f''; v^\mathrm{tape}, \Delta v^\mathrm{tape})&.
    \end{aligned}
\end{multline}
Using the reference frequency $f_{\ell m}^\mathrm{ref} = (m/2) f_{22}^\mathrm{ref}$ at which the phase of the $(\ell,m)$-mode vanishes as the first integration boundary ensures that the definition of the reference frequency does not change with respect to the GR waveform. The second integration boundary $f_{22}^\mathrm{peak}$, the frequency at which the $(2,2)$-mode peaks, ensures that the first derivative of $\delta\psi_{\ell m}(f)$ goes to zero at $f_{22}^\mathrm{peak}$ and thus that the GR waveform and non-GR waveform still align in the time domain.

After the phase corrections are computed, they can be added to the GR frequency-domain waveform $h_{\ell m}^{\mathrm{GR}}$ using
\begin{equation}
    h_{\ell m}^{\mathrm{FTI}}(f) = h_{\ell m}^{\mathrm{GR}}(f) e^{i\delta\psi_{\ell m}(f)},
\end{equation}
where the subscript $\ell m$ denotes the $(\ell,m)$-mode in the mode decomposition of the waveform. These modes can then be added together to get the plus and cross polarizations of the waveform using 
\begin{align}
    h_+^{\mathrm{FTI}}(f) &= \frac{1}{2} \sum_{\ell m} \left[ _{-2}Y_{\ell m}(\imath, \varphi_c ) + (-1)^\ell _{-2}Y^*_{\ell-m}(\imath, \varphi_c ) \right] h_{\ell m}^{\mathrm{FTI}}(f), \\
    h_\times^{\mathrm{FTI}}(f) &= \frac{i}{2} \sum_{\ell m} \left[ _{-2}Y_{\ell m}(\imath, \varphi_c ) - (-1)^\ell _{-2}Y^*_{\ell-m}(\imath, \varphi_c ) \right] h_{\ell m}^{\mathrm{FTI}}(f),
\end{align}
where $_{-2}Y_{\ell m}(\imath, \varphi_c )$ denotes the spin-weighted spherical harmonics of spin-weight $-2$, $\imath$ is the inclination angle, and $\varphi_c$ is the coalescence phase. For the FTI test with the \textsc{SEOBNRv5HM\_ROM} waveform model, the sum over $(\ell, m)$-modes here typically extends over $\{ (2, 2), (3, 3), (2, 1), (4, 4) \}$. 
The full list of modes available in the \textsc{SEOBNRv5HM\_ROM} model is $\{(2,2), (3,3), (2,1), (4,4), (5,5), (3,2), (4,3)\}$, but we only include the most important ones in order to limit the computational cost of waveform generation.

To make choosing the values for the tapering parameters $v^\mathrm{tape}$ and $\Delta v^\mathrm{tape}$ more natural, we are going to express them differently. We can write $v^\mathrm{tape}$ as a function of a tapering frequency $f^\mathrm{tape}$ using $v^\mathrm{tape} = (\pi f^\mathrm{tape} M)^{1/3}$. The frequency at which the inspiral ends is $\sim f_{22}^\mathrm{peak}$ and depends strongly on the masses. It is therefore easier to specify the tapering frequency as a fraction of the peak frequency so that $f^\mathrm{tape} = \alpha f_{22}^\mathrm{peak}$ with $\alpha$ a constant of order unity. Instead of specifying $\Delta v^\mathrm{tape}$, it is more useful to specify the number of GW cycles $\mathcal{N}_\mathrm{GW}$ over which the tapering function $W(f)$ changes from 0 to 1. The relation between these two is approximately given by
\begin{equation}
    \Delta v^\mathrm{tape} = \frac{128\eta}{3}\pi (v^\mathrm{tape})^6 \Gamma \mathcal{N}_\mathrm{GW},
\end{equation}
where $\Gamma = 1/50$. The choice for the values of the parameters $\alpha$ and $\mathcal{N}_\mathrm{GW}$ is completely phenomenological and they should be optimized. At first, LVK Collaboration analyses on GW signals used $f^\mathrm{tape} = 0.35 f_{22}^\mathrm{peak}$ and $\mathcal{N}_\mathrm{GW} = 1$~\cite{LIGOScientific:2019fpa, LIGOScientific:2020tif, LIGOScientific:2021sio}, while later this was changed to $f^\mathrm{tape} = 1.0 f_{22}^\mathrm{peak}$~\cite{Sanger:2024axs, LIGOScientific:2025cmm, LIGOScientific:2025wao, LIGOScientific:2025brd, LIGOScientific:2026fcf, LIGOScientific:2026qqv}.~\citet{Mehta:2022pcn} studied the effect of the tapering frequency on the constraints obtained for the deviation parameters and found that higher tapering frequencies improve the bounds up to a factor $\sim 7$, where the largest improvements were obtained for the higher PN deviation parameters. This is explained by the increase in the available signal-to-noise ratio (SNR) and the higher PN orders becoming more relevant at higher frequencies. They also found that changing $\mathcal{N}_\mathrm{GW}$ does not significantly influence their results.

\subsection{SIQM} \label{sec:SIQM}

The rotational deformation of spinning objects contributes to the multipole decomposition of their gravitational fields at quadrupolar and higher multipoles. Therefore, it affects the waveform of GWs from compact-object binaries. The spin-induced multipole moments of a Kerr BH depend solely on the mass and spin of the BH~\cite{Carter:1971zc, Hansen:1974zz}. The leading-order contribution comes from the SIQM, which can be represented as
\begin{equation}
    Q_i = -\kappa \chi_i^2 m_i^3,
\end{equation}
where $m_i$ and $\chi_i$ are the mass and dimensionless spin of the compact object.  Kerr BHs have $\kappa_i=1$, while other compact objects can have values of $\kappa_i$ that differ greatly from this. For example, spinning neutron stars have values between $\sim 2$ to $\sim 14$~\cite{Pappas:2012ns, Pappas:2012qg, Harry:2018hke}, while for spinning boson stars values for $\kappa_i$ as high as $\sim 10 - 150$ are possible~\cite{Ryan:1996nk}, and gravastars can even have negative values for $\kappa_i$~\cite{Uchikata:2015yma}. 

For compact binaries, the two additional SIQM parameters, $\kappa_1$ and $\kappa_2$, increase the total number of parameters required to describe waveforms from compact binaries to $13$. Attempts have been made to reduce the number of parameters required for accurate waveforms. Some parameterizations work well for searches (see, e.g.,~\cite{Chia:2022rwc}), but the match is not good enough for parameter estimation. Rather than inferring $\kappa_1$ and $\kappa_2$ directly, we use their symmetric and antisymmetric combinations, given by $\kappa_s = (\kappa_1 + \kappa_2)/2$ and $\kappa_a = (\kappa_1 - \kappa_2)/2$, respectively~\cite{Krishnendu:2017shb, LIGOScientific:2020tif}. For a BBH system, we have that $\kappa_s = 1$ and $\kappa_a = 0$. Thus, we parameterize deviations from BBHs as $\kappa_s = 1 + \delta\kappa_s$ and $\kappa_a = 0 + \delta\kappa_a$.

\begin{figure*}
    \centering
    \includegraphics[width=\textwidth]{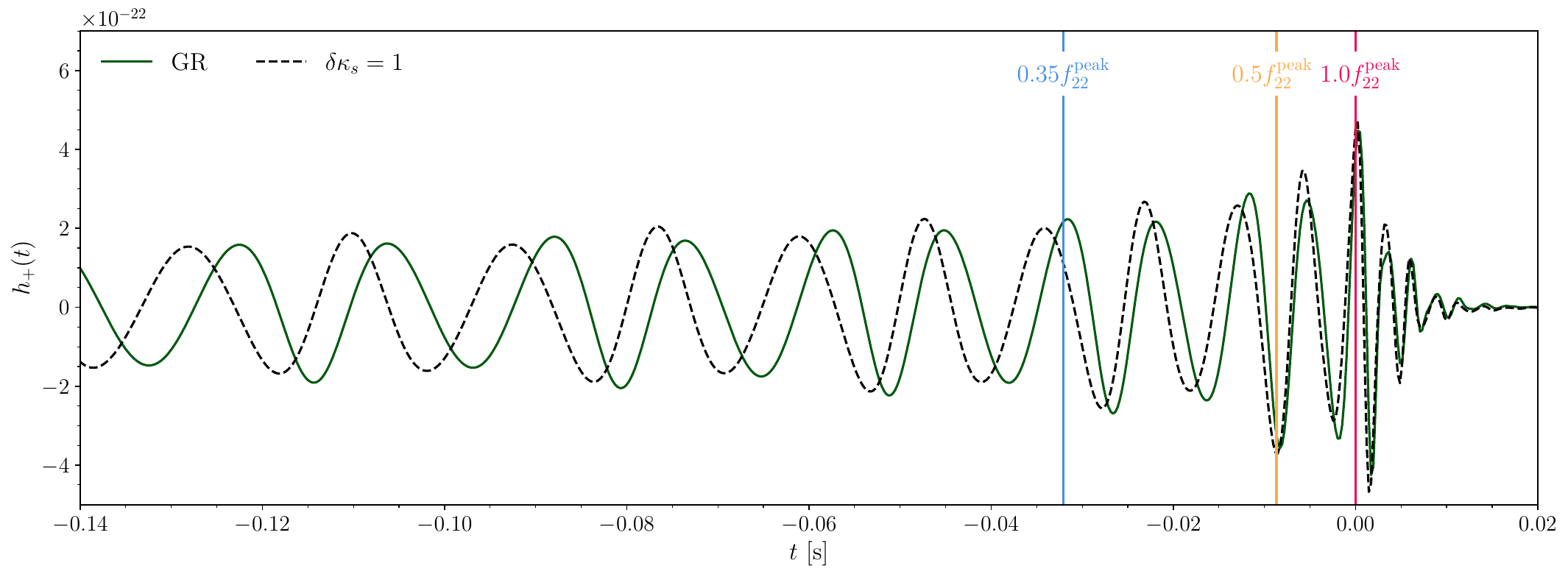}
    \caption{
    GW190412-like signal in GR (green solid line) and with a deviation in the SIQM of $\delta\kappa_s=1$ (black dashed line). The waveforms have been aligned at merger. The times to merger of the different tapering frequencies used in our analyses are indicated by the vertical lines.}
    \label{fig:waveform}
\end{figure*}

In the PN approximation, the SIQM first appears in the GW phase at 2PN order. We can write contributions from deviations in the SIQM in the same form as the generic deviations~\eqref{eq:generic-phase-correction}, which leads to
\begin{equation}
    \delta\psi_{\ell m}^\mathrm{SIQM}(f, \vec{\theta}; \delta\kappa_s, \delta\kappa_a) = \frac{3}{128\eta v^5}\frac{m}{2} \left[ \sum_{n=4}^7 \delta\psi_n^\mathrm{SIQM}(\vec{\theta}; \delta\kappa_s, \delta\kappa_a) v^n \right].
\end{equation}
The expressions for the coefficients that depend on the SIQM are~\cite{Krishnendu:2017shb}
\begin{widetext}
\begin{align}
    \delta\psi_4^{\rm SIQM}(\vec{\theta}) &= \left[ 50(2\eta-1) (\chi_s^2 + \chi_a^2) - 100d \chi_s\chi_a \right] \delta\kappa_s + \left[ 100(2\eta-1) \chi_s\chi_a -50d (\chi_s^2 + \chi_a^2) \right] \delta\kappa_a, \label{eq:dkappa2PN} \\
    \begin{split}
        \delta\psi_6^{\rm SIQM}(\vec{\theta}) &= \left[ \left( \frac{26015}{28} - \frac{44255}{21}\eta - 240\eta^2 \right) (\chi_s^2 + \chi_a^2) + d \left( \frac{26015}{14} - \frac{1495}{3}\eta \right) \chi_s\chi_a \right] \delta\kappa_s \\
        &+ \left[ d \left( \frac{26015}{28} - \frac{1495}{6}\eta \right) (\chi_s^2 + \chi_a^2) +
        \left( \frac{26015}{14} - \frac{88510}{21}\eta - 480\eta^2 \right) \chi_s \chi_a \right] \delta\kappa_a, \label{eq:dkappa3PN}
    \end{split} \\
    \begin{split}
        \delta\psi_7^{\rm SIQM}(\vec{\theta}) &= \left[ \left( \frac{3110}{3} - \frac{10250}{3}\eta + 40\eta^2 \right) \chi_s^3 + d \left( \frac{3110}{3} - 750\eta \right) \chi_a^3 + d \left( \frac{3110}{3} - \frac{10310}{3}\eta \right) \chi_s^2\chi_a + \left( 3110 - \frac{27190}{3}\eta + 40\eta^2 \right) \chi_s\chi_a^2 \right] \delta\kappa_s \\
        &+ \left[ d \left( \frac{3110}{3} - \frac{4030}{3}\eta \right) \chi_s^3 + \left( \frac{3110}{3} - \frac{8470}{3}\eta \right) \chi_a^3 + \left( 3110 - \frac{28970}{3}\eta + 80\eta^2 \right) \chi_s^2\chi_a + d \left( 3110 - \frac{8530}{3}\eta \right) \chi_s\chi_a^2 \right] \delta\kappa_a. \label{eq:dkappa3p5PN}
    \end{split}
\end{align}
\end{widetext}
Here
$d = (m_1-m_2)/M$ is the difference mass ratio and $\chi_s = (\chi_1+\chi_2)/2$ and $\chi_a = (\chi_1-\chi_2)/2$ are the symmetric and antisymmetric combinations of the spins of the objects. Figure~\ref{fig:waveform} illustrates how the modified SIQM alters the waveform.

The 3.5PN contribution from the SIQM is cubic in spin, so we expect it to be much smaller than contributions at lower PN orders. At 3.5PN, the SIOM begins to contribute and is proportional to spin cubed. Therefore, we also include SIOM contributions in the FTI model.

\subsection{SIOM} \label{sec:SIOM}

The SIOM can be represented as
\[
O_i = -\lambda_i \chi_i^3 m_i^4,
\]
where $\lambda_i$ is a dimensionless parameter that is 1 for Kerr BHs in GR~\cite{Hansen:1974zz}. For other compact objects, this value can be different, e.g., for neutron stars $\lambda_i$ is between $\sim 4 - 30$~\cite{Pappas:2012ns, Pappas:2012qg}, and for boson stars between $\sim 10 - 200$~\cite{Ryan:1996nk}.

We can again rewrite the component's SIOMs $\lambda_1$ and $\lambda_2$ into their symmetric and antisymmetric combinations $\lambda_s = (\lambda_1 + \lambda_2)/2$ and $\lambda_a = (\lambda_1 - \lambda_2)/2$~\cite{Krishnendu:2018nqa, Saini:2023gaw, Das:2026tel}. The deviations from the BBH value can then be expressed as $\lambda_s = 1 + \delta\lambda_s$ and $\lambda_a = 0 + \delta\lambda_a$. The contribution of deviations in the SIOM to the frequency-domain phase at 3.5PN is given by~\cite{Krishnendu:2017shb}

\begin{widetext}
\begin{align}
\delta\psi_7^{\rm SIOM}(\vec{\theta}) &= 440\left[ (3\eta-1)\chi_s^3 + d(\eta-1)\chi_a^3 + 3d(\eta-1)\chi_s^2\chi_a + 3(3\eta-1) \chi_s\chi_a^2 \right]\delta\lambda_s \\
&+ 440\left[ d(\eta-1)\chi_s^3 + (3\eta-1)\chi_a^3 + 3(3\eta-1)\chi_s^2\chi_a + 3d(\eta-1) \chi_s\chi_a^2 \right]\delta\lambda_a.
\end{align}
\end{widetext}

\subsection{Analysis setup}

As baseline GR model to which we add the FTI corrections we use the waveform approximants \textsc{SEOBNRv5HM\_ROM} and its $(2,2)$-mode only version \textsc{SEOBNRv5\_ROM}. \textsc{SEOBNRv5HM\_ROM} is the frequency-domain version of \textsc{SEOBNRv5HM}~\cite{Pompili:2023tna}, which is an effective-one-body waveform model, that assumes quasi-circular orbits and aligned spins, and includes higher order modes. The FTI waveforms are generated using \textsc{Bilby TGR}~\cite{ashton_2025_15676285}, an extension of \textsc{Bilby}~\cite{Ashton:2018jfp, Romero-Shaw:2020owr} to include waveforms for tests of GR. The baseline GR waveforms are generated using \textsc{LALSuite}~\cite{lalsuite, Wette:2020air}.

For parameter estimation, we use Bayesian analysis with the standard Gaussian likelihood function assuming stationary Gaussian noise characterized by its power spectral density (PSD).
The 1-dimensional posteriors for the deviation parameters are obtained by marginalizing over all other parameters. The sampling algorithm used is nested sampling with the \textsc{dynesty}~\cite{Speagle:2019ivv, sergey_koposov_2025_17268284} implementation in \textsc{Bilby}.

The priors used in our analyses are noninformative and generally wide enough to cover the full posterior support. We use uniform mass priors in the detector frame component masses. The spin priors are aligned-spin projections of isotropic spins with uniform spin magnitudes. The extrinsic parameters are chosen to be isotropic and uniform in comoving volume. The priors on the SIQM and SIOM parameters are chosen to be uniform with $\delta\kappa_s, \delta\kappa_a \in [-500, 500]$ and $\delta\lambda_s \in [-1000, 1000]$. We typically only vary one of $\left\{\delta\kappa_s, \delta\kappa_a, \delta\lambda_s\right\}$ at the same time, setting the other ones to their Kerr BH value of zero~\cite{Krishnendu:2017shb,Krishnendu:2018nqa, Saini:2023gaw}.

\section{Results from simulated signals} \label{sec:injections}

This section aims to evaluate the measurability of the SIQM and SIOM, and to adjust the tapering frequency and the number of higher modes employed. To this end, we employ the SIQM test within the FTI framework on simulated signals.

The simulated signals used in this analysis are similar to the BBH signals GW150914~\cite{LIGOScientific:2016aoc} and GW190412~\cite{LIGOScientific:2020stg}. We simulate signals with masses similar to these events; the injected parameters can be found in Table~\ref{tab:inj_param}. To be able to perform the SIQM test, a nonzero spin is required. We use four configurations for the spins: low spins with $\chi_1 = \chi_2 = 0.2$, intermediate spins with $\chi_1 = \chi_2 = 0.5$, high spins with $\chi_1 = \chi_2 = 0.8$, and negative spins $\chi_1 = \chi_2 = -0.2$. If not indicated otherwise, the intermediate spin configuration is used. The injections are done assuming zero noise (i.e., the data only contains the signal and no noise), but using a PSD corresponding to the design sensitivity of Advanced LIGO~\cite{KAGRA:2013rdx, psds_for_simulations} and Advanced Virgo~\cite{KAGRA:2013rdx, GWCommissionObserve}. We use the three-detector network consisting of the LIGO Hanford, LIGO Livingston, and Virgo detectors. Unless stated otherwise, we use the \textsc{SEOBNRv5HM\_ROM} waveform approximant~\cite{Pompili:2023tna} for injection (i.e., simulating the data) and recovery (i.e., Bayesian analysis).

\begin{table}
    \centering
    \begin{tabular}{lcc}
        \hline
        \hline
         & GW150914-like & GW190412-like \\
        \hline
        $M$ [$M_\odot$] & 70 & 40 \\
        $q$ & 0.8 & 0.4 \\
        $\chi_1, \chi_2$ & 0.5* & 0.5* \\
        $d_L$ [Mpc] & 1200 & 740 \\
        $\imath$ & 2.7 & 0.71 \\
        \hline
        Total SNR & 41.4 & 31.7 \\
        \hline
        \hline
    \end{tabular}
    \caption{Parameters for the simulated signals used in this work and their SNRs. The default value used for the spins is 0.5, but this is sometimes varied when indicated.}
    \label{tab:inj_param}
\end{table}

\subsection{Spin dependence of $\delta\kappa_s$}

\begin{figure*}
    \centering
    \includegraphics[width=\textwidth]{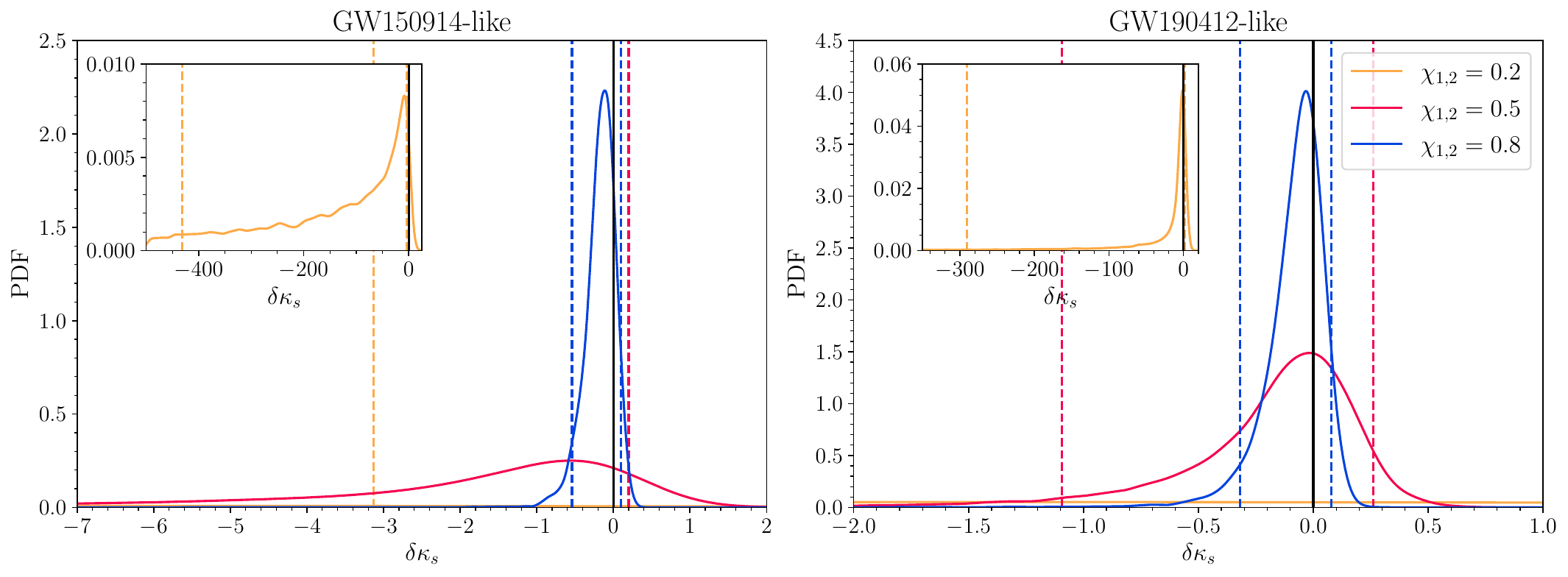}
    \caption{The posteriors on $\delta\kappa_s$ for the different spin cases for the GW150914-like (left) and GW190412-like (right) injections. The dashed lines indicate the 90\% credible intervals. We see that larger spins lead to tighter constraints on the SIQM.}
    \label{fig:spins}
\end{figure*}

We first study how well we can constrain the SIQM depending on the magnitude of the spins. The results for the different positive spin configurations are shown in Fig.~\ref{fig:spins}. We notice that the constraints on the SIQM are tighter for higher spins, which was already observed in Refs.~\cite{Krishnendu:2017shb, Krishnendu:2019tjp}. This is to be expected due to the spin dependence of the SIQM corrections, since the corrections are larger for larger spins.

\begin{figure}
    \centering
    \includegraphics[width=\columnwidth]{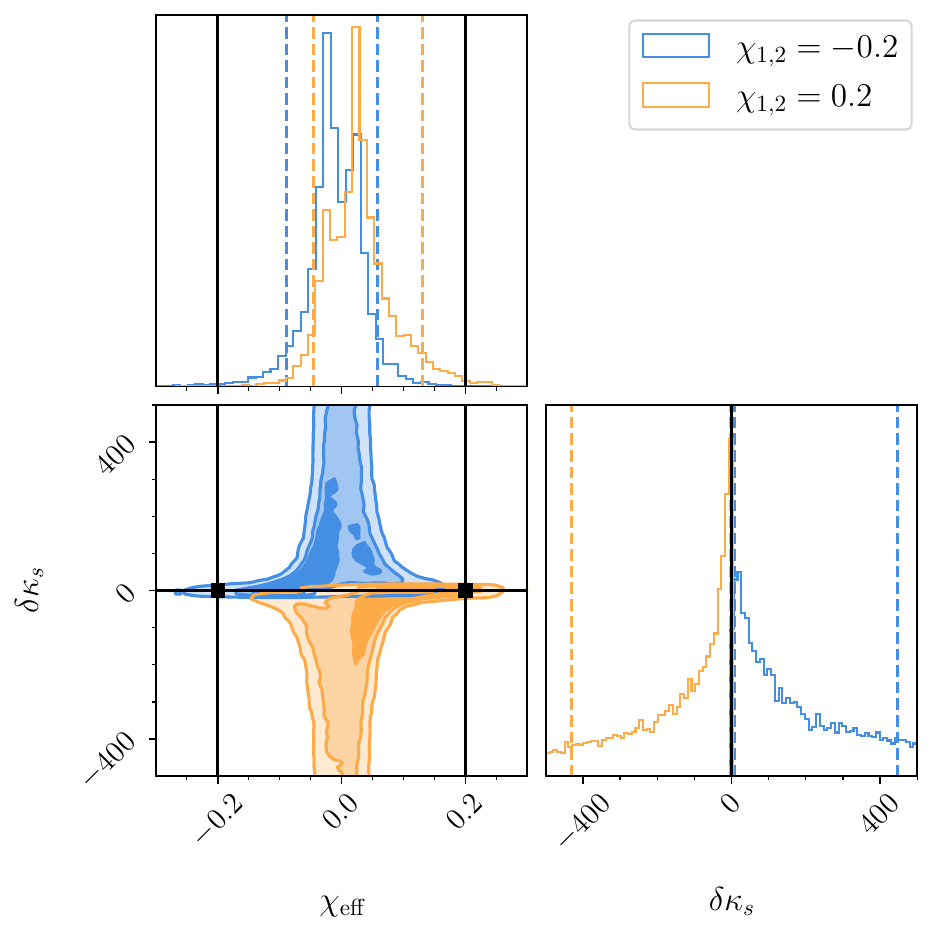}
    \caption{The joint posteriors on $\delta\kappa_s$ and $\chi_\mathrm{eff}$ for the positive (orange) and negative (blue) spin cases of the GW150914-like injections. The bold black lines indicate the injected values. We see that there is a correlation between these parameters and that the direction of the tail in $\delta\kappa_s$ depends on the sign of $\chi_\mathrm{eff}$.}
    \label{fig:neg_spin}
\end{figure}

We also notice that the posteriors for $\delta\kappa_s$ are asymmetric, with a heavier tail on the negative $\delta\kappa_s$ side due to correlations with the spins. This negative tail is especially prominent in the low-spin case. As shown in Fig.~\ref{fig:neg_spin}, the corner plot reveals the correlations between effective spin $\chi_\text{eff}$ and $\delta\kappa_s$ for the GW150914-like with $\chi_{1,2} = \pm 0.2$ cases. Hence the negative tail is due to the support for zero effective spin. Due to the spin dependence, the corrections remain zero for zero spins, whereas the SIQM parameter can take on any value. This leads to $\delta\kappa_s$ diverging close to $\chi_\mathrm{eff}=0$, giving it a long, heavy tail.

What is surprising though is that the tail only appears on the negative side for the $\chi_{1,2}=0.2$ case. For the negative spin case $\chi_{1,2}=-0.2$ on the other hand, the tail appears only on the positive side. This suggests that the side on which the tail appears is correlated with the sign of the spins. Looking at cases with larger spins reveals a similar, albeit weaker, correlation between the SIQM and spins, where smaller $\chi_\mathrm{eff}$ leads to a negative $\delta\kappa_s$. This correlation is amplified for small spins, leading to one-sided tails. Similar behavior was observed by \citet{Krishnendu:2019tjp}, who showed that this is indeed due to a degeneracy between $\chi_\text{eff}$ and $\delta\kappa_s$.

\subsection{The 3.5PN term} \label{sec:3p5PN}

The SIQM test as introduced in Ref.~\cite{Krishnendu:2017shb} and used by the LVK Collaboration in Refs.~\cite{LIGOScientific:2019fpa, LIGOScientific:2020tif, LIGOScientific:2021sio, LIGOScientific:2025brd, LIGOScientific:2026fcf}, does not contain the 3.5PN SIQM deviations. It was later added when including the SIOM at 3.5PN~\cite{Saini:2023gaw}. The 3.5PN term is not only higher PN order but also higher order of spin than the 2PN and 3PN terms. Therefore, we do not expect it to significantly impact the obtained constraints. Since the FTI method allows us to easily include or exclude the 3.5PN term, we would like to verify this.

\begin{figure}
    \centering
    \includegraphics[width=\columnwidth]{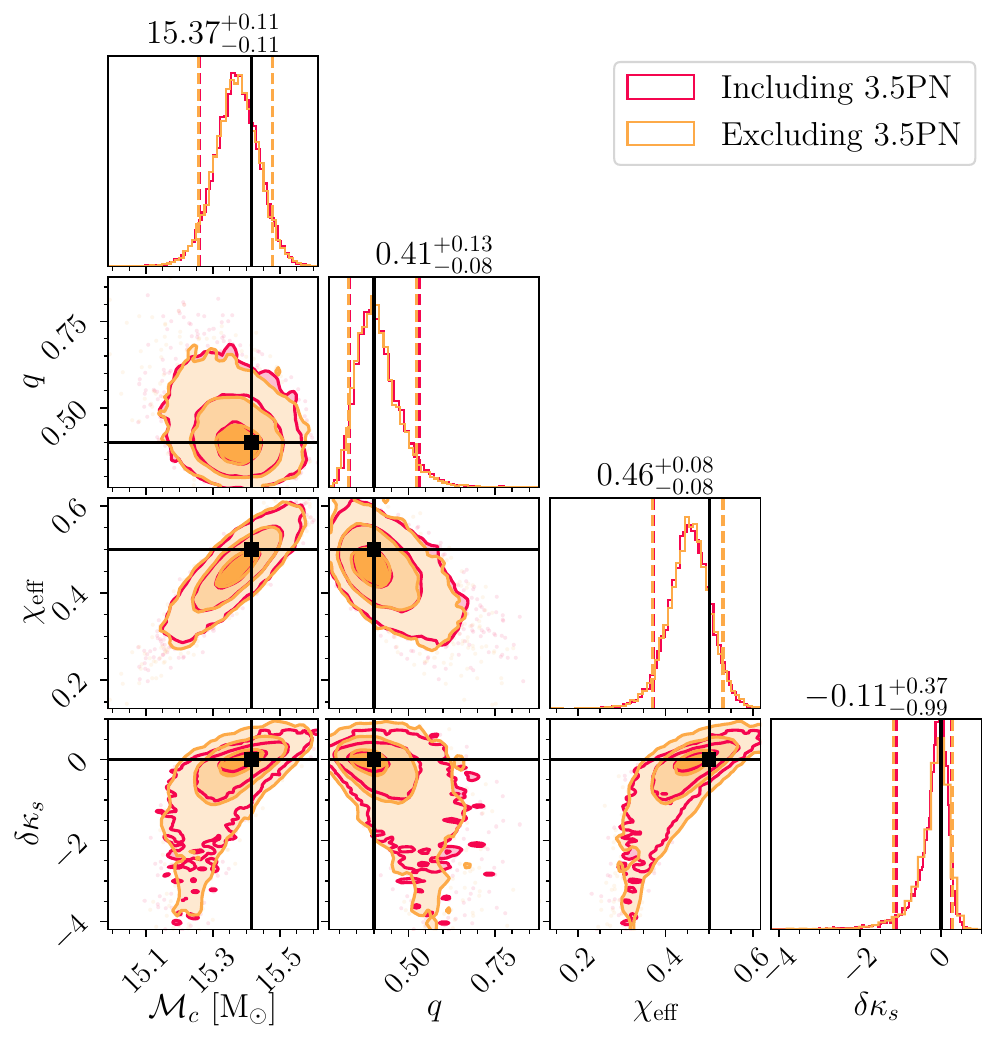}
    \caption{
    Results for the GW190412-like injection when including (red) or excluding (orange) the 3.5PN term in the SIQM corrections in the analysis. The bold black lines indicate the injected values. We see that there is no significant difference in the posteriors.}
    \label{fig:3.5PN}
\end{figure}

In Fig.~\ref{fig:3.5PN}, we compare the results obtained including and excluding the 3.5PN term in the SIQM contributions for the GW190412-like injection with spins of $\chi_1 = \chi_2 = 0.5$. We see that there is no significant difference between the posteriors, as expected. This is also true for the GW150914-like injection. From this we conclude that indeed the 3.5PN term is not important for these kind of systems. Inclusion of the 3.5PN term might become more important for extremal spins and higher SNRs.

\begin{figure*}
    \centering
    \includegraphics[width=\textwidth]{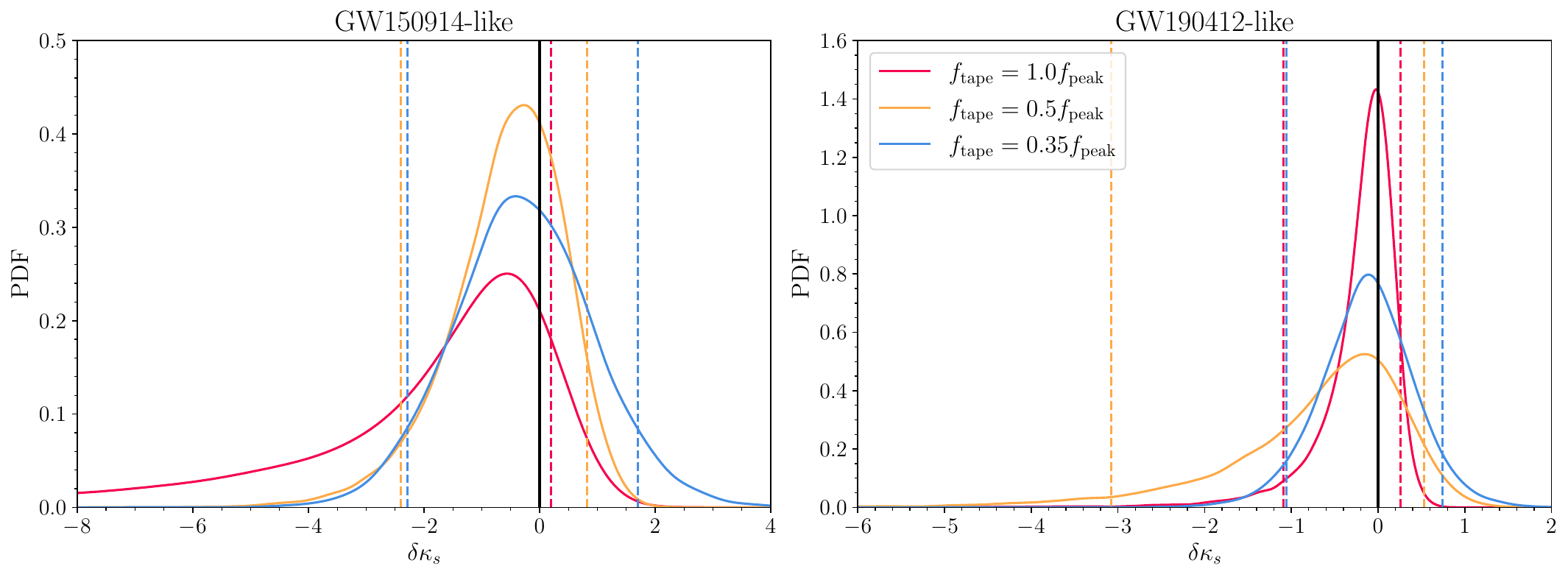}
    \caption{Results from varying the tapering frequency for the GW150914-like (left) and GW190412-like (right) injections.}
    \label{fig:ftape}
\end{figure*}

\subsection{Varying the tapering frequency}

As described in Sec.~\ref{sec:FTI}, the modifications to the inspiral are tapered off towards merger-ringdown. The tapering frequency $f^\mathrm{tape} = \alpha f_{22}^\mathrm{peak}$ is somewhat arbitrary. We would therefore like to investigate how the choice of tapering frequency influences the bounds found for the SIQM. For this, we analyze the signals with $\alpha=\{0.35, 0.5, 1.0\}$, where a larger value of $\alpha$ means we push the transition closer to the peak of the $(2,2)$-mode. The choice of $f^\mathrm{tape} = 0.35 f_{22}^\mathrm{peak}$, which still corresponds to few cycles before merger, was used in the past for FTI to approximately match the transition frequency of the Test Infrastructure for GEneral Relativity (TIGER)~\cite{Agathos:2013upa, Meidam:2017dgf, Roy:2025gzv}, the other generic parameterized inspiral test used by the LVK Collaboration. The choice of $f^\mathrm{tape} = f_{22}^\mathrm{peak}$, on the other hand, pushes the tapering to as late as possible to probe all of the inspiral regime. Figure~\ref{fig:waveform} illustrates where these different choices of tapering frequency are located for an example waveform.

The results for both GW150914-like and GW190412-like signals are shown in Fig.~\ref{fig:ftape}. We see that increasing the tapering frequency leads to tighter bounds on $\delta\kappa_s$ for the GW190412-like case. This is expected since the SIQM enters at high PN orders, meaning it does not significantly contribute to the phase until close to merger. Thus, increasing the tapering frequency probes more of the part of the signal where the SIQM is important. Increasing the tapering frequency also increases the SNR of the modified part of the signal, which leads to tighter bounds.

However, for the GW150914-like case we see that the constraints are less tight for $f^\mathrm{tape} = 1.0 f_{22}^\mathrm{peak}$ than for $f^\mathrm{tape} = 0.5 f_{22}^\mathrm{peak}$. This is likely due to increased correlations between $\delta\kappa_s$ and the other parameters of the binary. These correlations are typically not as strong for lower tapering frequencies because more of the waveform is left unmodified and can be used to better constrain the GR parameters.

\subsection{Higher modes and waveform approximant}

Including higher order modes in the waveform can improve the measurements of source parameters. This is especially the case for asymmetric binaries. \citet{Divyajyoti:2023izl} showed that this can also be the case for inference of the SIQM. In this section we investigate this effect on our SIQM test. We therefore analyze the same injections using the waveform models \textsc{SEOBNRv5\_ROM}~\cite{Pompili:2023tna}, which only contains the dominant $(2,2)$-mode, and \textsc{SEOBNRv5HM\_ROM}, which also includes the higher-order modes $(3,3), (2,1)$, and $(4,4)$. In both cases the injection was done with \textsc{SEOBNRv5HM\_ROM}.

\begin{figure*}
    \centering
    \includegraphics[width=\textwidth]{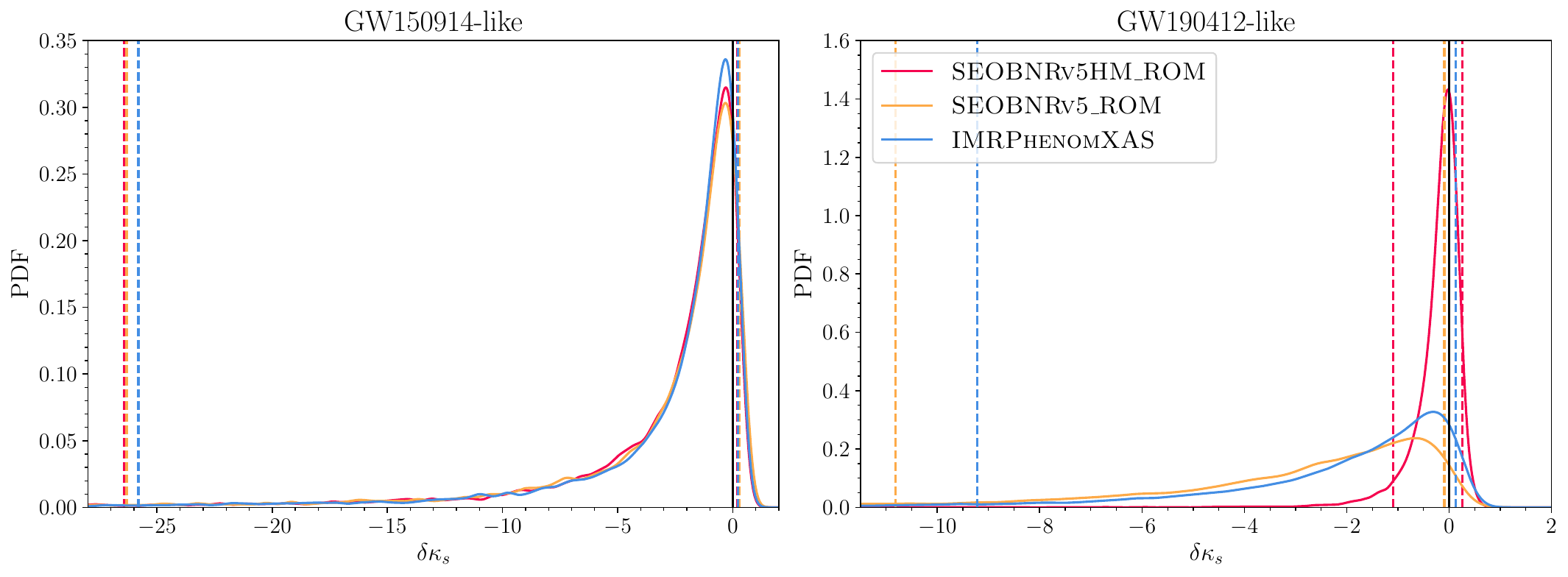}
    \caption{The posteriors on $\delta\kappa_s$ obtained using different waveform approximants for the GW150914-like (left) and GW190412-like (right) injections. We see that the inclusion of higher modes (red) greatly improves the results for the GW190412-like case.}
    \label{fig:approx}
\end{figure*}

Figure \ref{fig:approx} shows the posteriors on $\delta\kappa_s$ for both waveform models. For the GW150914-like injection, we see that the higher modes make no difference. This is expected because it is a relatively symmetric binary, meaning higher modes usually do not significantly contribute to the waveform unless the total SNR is very large. For the GW150914-like signal, the SNR of each higher mode peaks at approximately zero. Combining them yields a total SNR of less than two, indicating that no significant higher modes are present for this injection.

For the GW190412-like injection, we observe that the higher modes are important. Including the higher modes in the waveform model used for recovery yields a narrower posterior distribution for $\delta\kappa_s$. This is expected, as higher modes become important for more asymmetric binaries. In this case, the most significant higher mode is the $(3, 3)$ mode, which has an SNR of approximately $3.5$. This results in better measurement of the GR parameters, which leads to a narrower posterior on the TGR parameters due to correlations. Not including the higher modes could potentially introduce biases, but this is not the case here.

Imperfect modeling in GR waveform models can lead to systematic biases when using them for parameter estimation. These effects can be enhanced when doing tests of GR and can even lead to false deviations from GR. To see how sensitive the SIQM test is to biases due to mismodeling of the GR waveform, we perform analyses with both the \textsc{IMRPhenomXAS}~\cite{Pratten:2020fqn} and \textsc{SEOBNRv5\_ROM} waveform approximants. These both have only the dominant $(2,2)$-mode and use the aligned-spins approximation, so the physical effects included in the models are the same. However, the way the models are constructed is not the same and this can lead to small differences between the waveforms.

Figure \ref{fig:approx} shows the results obtained for $\delta\kappa_s$ using \textsc{IMRPhenomXAS} and \textsc{SEOBNRv5\_ROM}. The injection was in both cases done using \textsc{SEOBNRv5HM\_ROM}, which is also an aligned-spin model but includes higher order modes. We see that for both injections there are some minor differences between the \textsc{IMRPhenomXAS} and \textsc{SEOBNRv5\_ROM} posteriors. The posteriors obtained using \textsc{IMRPhenomXAS} are slightly tighter but the peak of the posterior is not shifted. The results are still comparable and consistent with the injected value of $\delta\kappa_s$. Although there is no significant impact from waveform systematics in these cases, this can become important for higher SNRs or systems in poorly modeled parts of the parameter space.

\subsection{Anti-symmetric combination of the SIQMs}

So far, we have assumed that the antisymmetric combination of the SIQMs $\delta\kappa_a$ vanishes (i.e., both objects have the same SIQM). However, this does not have to be the case. In this section, we study the measurability of $\delta\kappa_a$.

\begin{figure}
    \centering
    \includegraphics[width=\columnwidth]{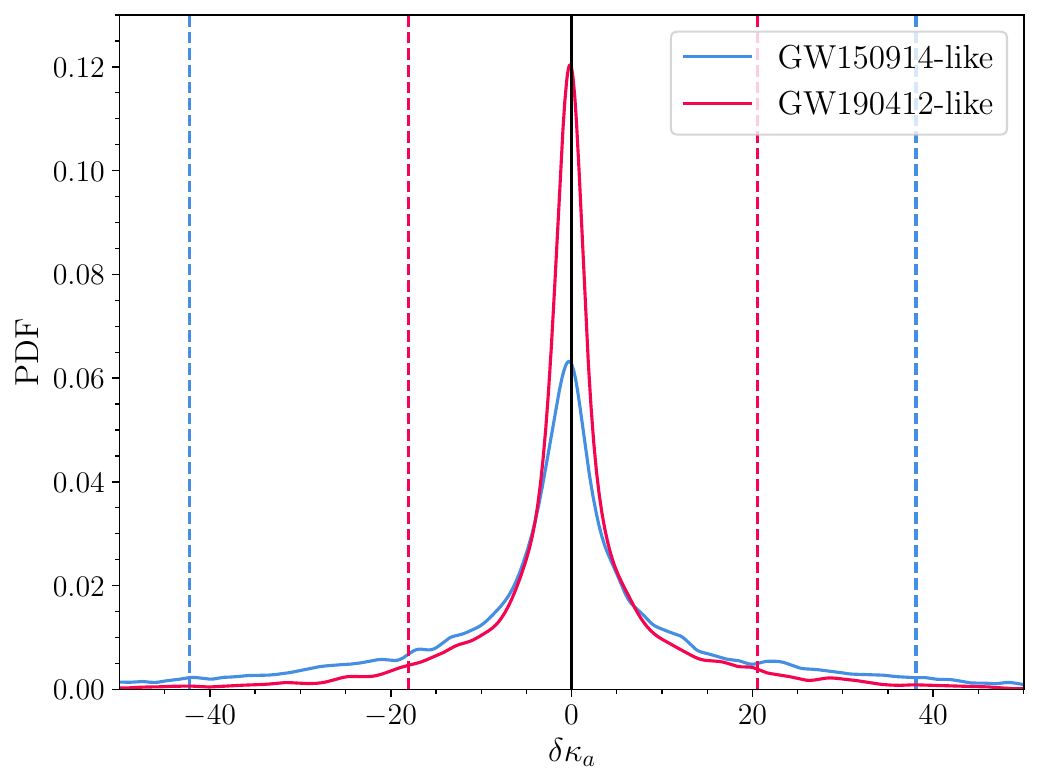}
    \caption{The posteriors on $\delta\kappa_a$ for the GW150914-like (blue) and GW190412-like (red) injections. Here, we keep the value of $\delta\kappa_s$ fixed to zero.}
    \label{fig:dkappaA}
\end{figure}

For an initial attempt to determine the measurability of $\delta\kappa_a$, we assume that the symmetric combination of the SIQMs $\delta\kappa_s=0$. This assumption is not physically realistic if there is a deviation from Kerr BHs in GR. Therefore, it should be seen purely as a consistency test of the model. The results are shown in Fig.~\ref{fig:dkappaA}, where we see that the $\delta\kappa_a$ posteriors peak nicely around zero and that the posteriors are tighter for the GW190412-like injection than the GW150914-like injection. We also note that these posteriors are symmetric around zero as opposed to the asymmetric posteriors observed for $\delta\kappa_s$.

\begin{figure}
    \centering
    \includegraphics[width=\columnwidth]{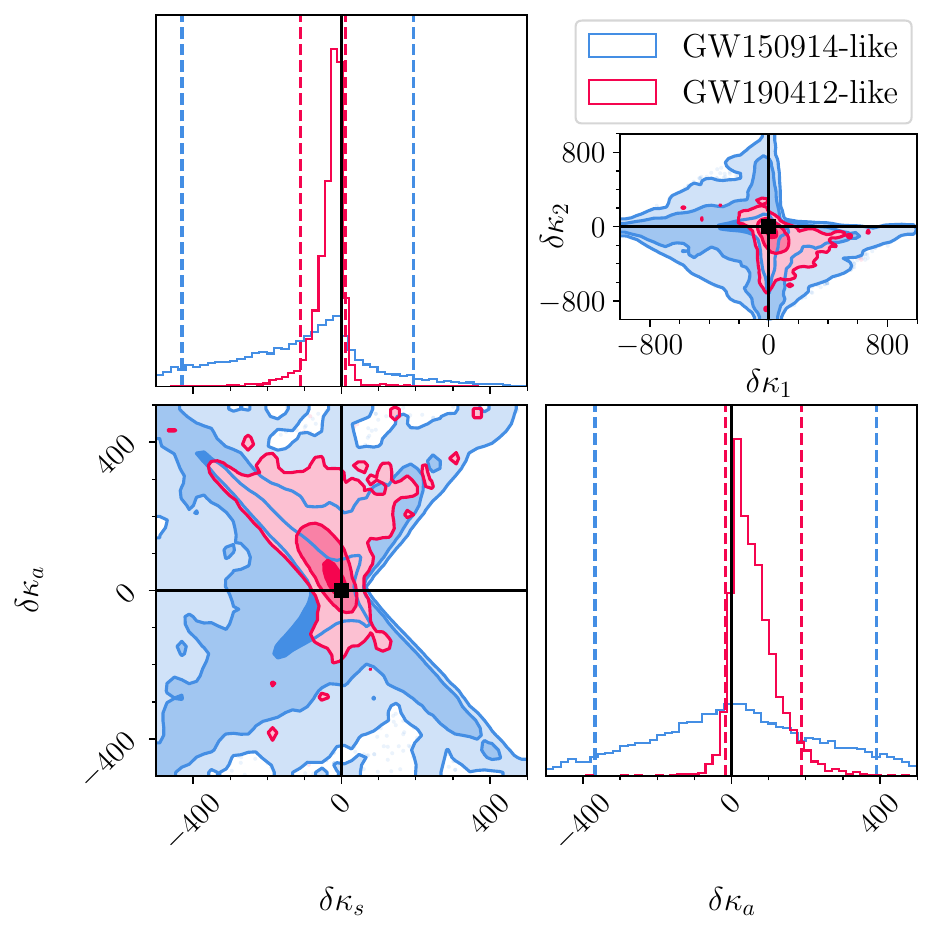}
    \caption{The joint posteriors on $\delta\kappa_s$ and $\delta\kappa_a$ when varying both at the same time for the GW150914-like (blue) and GW190412-like (red) injections. The inset in the top right show the posteriors converted to $\delta\kappa_1$ and $\delta\kappa_2$. The bold black lines indicate the injected values.}
    \label{fig:dkappaS_dkappaA}
\end{figure}

Next, we abandon the assumption that $\delta\kappa_s=0$ and vary both SIQM parameters simultaneously. The joint posteriors for $\delta\kappa_s$ and $\delta\kappa_a$ are shown in Fig.~\ref{fig:dkappaS_dkappaA}, where the top-right panel also shows the posteriors transformed into the individual component SIQMs $\delta\kappa_1$ and $\delta\kappa_2$. We notice that the posteriors for $\delta\kappa_s$ and $\delta\kappa_a$ become much wider than in the single-parameter test due to correlations between the two. For the GW150914-like injection, the SIQM parameters are not well constrained except for the fact that large positive values for both $\delta\kappa_1$ and $\delta\kappa_2$ at the same time are excluded. For the GW190412-like signal, on the other hand, the joint posterior is better constrained, mostly due to $\delta\kappa_1$ being well constrained. The SIQM for the secondary object is not as well constrained with $\delta\kappa_2$ having a tail towards negative values. For asymmetric binaries, the SIQM of the primary is expected to be more accurately determined than that of the secondary because the impact of the lower-mass object is suppressed by a factor of the mass ratio.

Overall, these results demonstrates that $\delta\kappa_a$ can be constrained in certain cases, but that higher SNRs are required in order to obtain meaningful constraints when allowing both SIQM parameters to vary.

\subsection{SIOM}

As discussed in Sec.~\ref{sec:SIOM}, the SIOM begins to contribute at 3.5PN. Since it enters at a high PN order and is proportional to the cube of the spin, it is difficult to constrain the SIOM. Here we study the measurability of the symmetric combination of the SIOMs $\delta\lambda_s$ and we assume that the antisymmetric combination $\delta\lambda_a$ vanishes.

\begin{figure}
    \centering
    \includegraphics[width=\columnwidth]{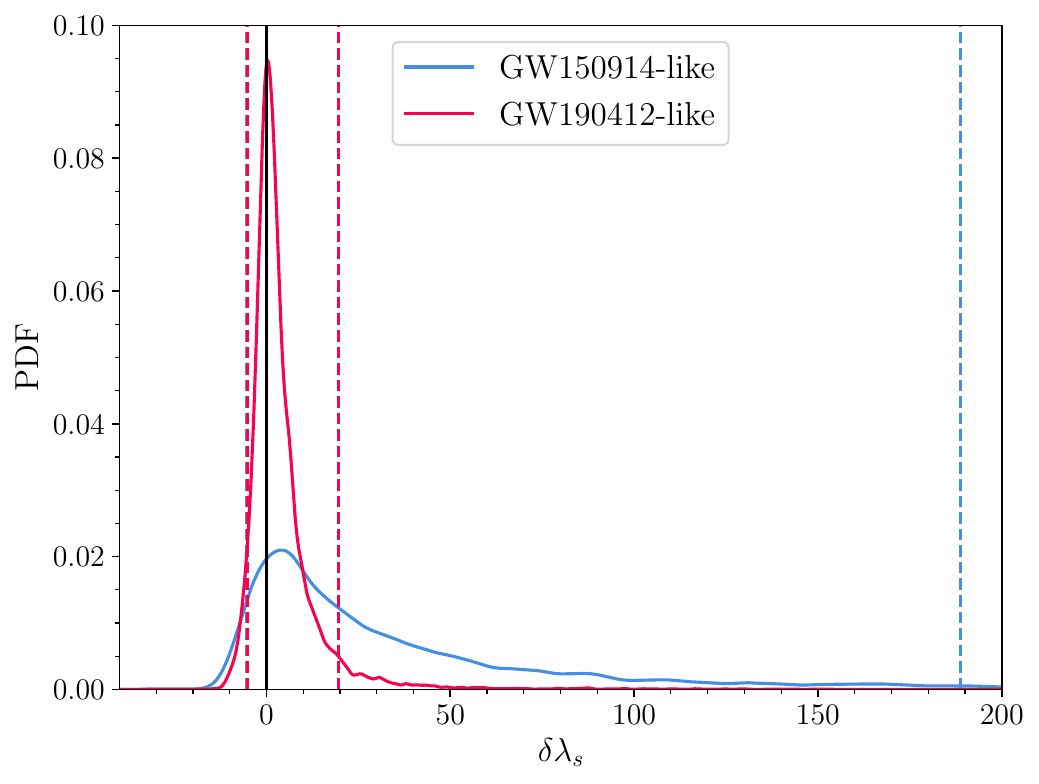}
    \caption{The posteriors on $\delta\lambda_s$ for the GW150914-like (blue) and GW190412-like (red) injections.}
    \label{fig:dlambdaS}
\end{figure}

The results are shown in Fig.~\ref{fig:dlambdaS}. We see that, similarly to $\delta\kappa_s$, the distribution is asymmetric with a longer tail on one side. However, for the SIOM, the tail is on the positive side for positive spins instead of the negative side, as is the case for the SIQM, due to$\delta\lambda_s$ and $\chi_{\rm eff}$ being anti-correlated while $\delta\kappa_s$ has a positive correlation with $\chi_{\rm eff}$. We also notice that the value of $\delta\lambda_s$ is not as well constrained as for $\delta\kappa_s$, most likely due to the corrections being higher order in spin and at a higher PN order, which makes the corrections smaller.

\subsection{Non-GR injections}

To see if the SIQM test can pick up on deviations from BBHs in GR, we inject signals with different values of $\delta\kappa_s$. We inject both positive and negative values, and keep the values of the other parameters the same as in Table~\ref{tab:inj_param}. We then analyze the injected signals leaving $\delta\kappa_s$ a free parameter but setting $\delta\kappa_a=0$.

\begin{figure*}
    \centering
    \includegraphics[width=\textwidth]{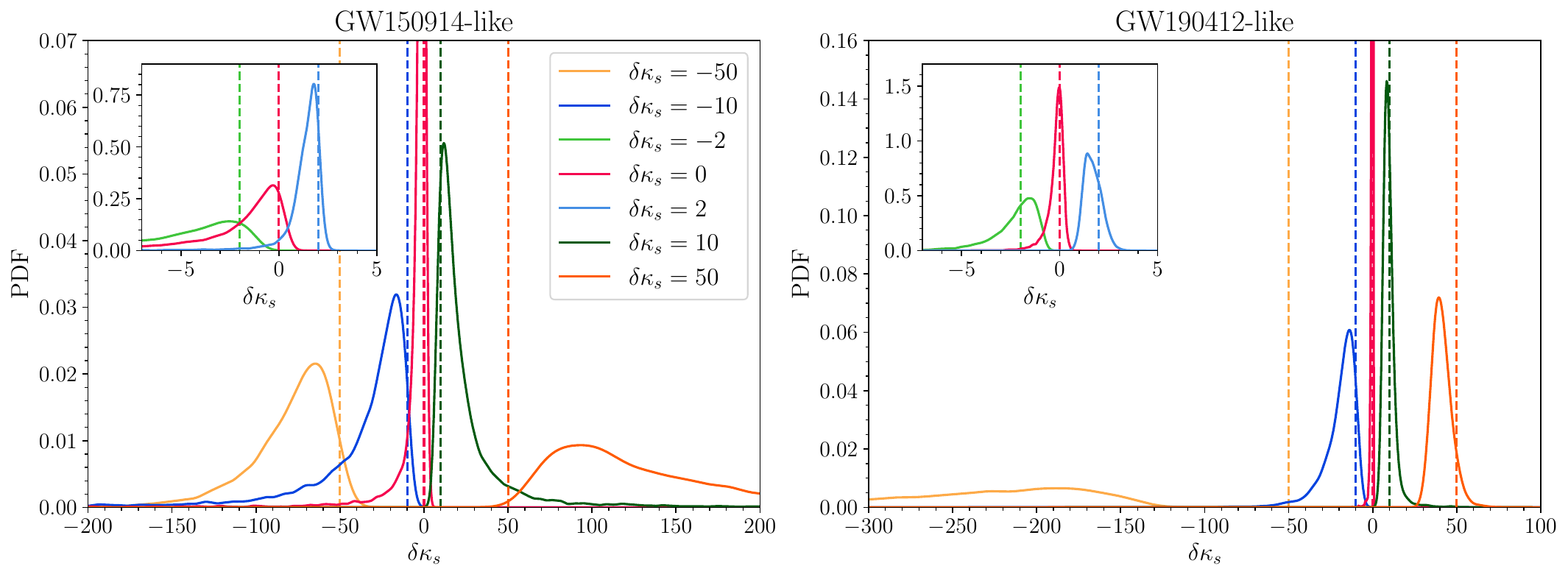}
    \caption{The solid lines show the posteriors obtained for injections with varying values for $\delta\kappa_s$; the injected values are indicated by the dashed lines.}
    \label{fig:nonGR}
\end{figure*}

The results from analyzing these injections with $\delta\kappa_s=\{\pm 2, \pm 10, \pm 50\}$ can be found in Fig.~\ref{fig:nonGR}. Each color corresponds to a different injection and the injected values are indicated with dashed vertical lines. The posteriors for each case are the solid lines with the same color. We also show the results for an injection with $\delta\kappa_s=0$ (red) in the same plots.

We notice that for the injections with the smallest values of $\delta\kappa_s = \pm 2$, the posteriors already clearly peak away from zero and can thus be confidently detected as a deviation from Kerr BHs in GR. For these and the injections with $\delta\kappa_s = \pm 10$, the posteriors also peak at the correct values. However, for the $\delta\kappa_s = \pm 50$ injections, the posteriors do not peak around the correct values, with the posterior for the GW190412-like case with $\delta\kappa_s = -50$ even completely excluding the injected value. There are also biases in other parameters, notably $\chi_{\rm eff}$ and $\mathcal{M}_c$ shift towards smaller values, while $q$ shifts towards more equal masses. One would expect that for zero-noise injections the true values are recovered, but this is not always the case due to correlations between parameters and the priors not being flat on all parameters.
FTI (as well as other inspiral-based null-tests of GR) is known to show such biases when the deviations become large~\cite{Johnson-McDaniel:2021yge}.

From these results we conclude that the SIQM test can be used for \textit{detecting} deviations in the SIQM away from the value for BHs in GR, but that it cannot be used for accurately \textit{measuring} the value of the SIQM of compact objects when the deviations become large.
It therefore merely serves as a null test of GR.

\section{Results from observed signals} \label{sec:real-events}

In this section, we apply the SIQM and SIOM test to a subset of the observed GW signals. We select the events from the first three GW Transient Catalogs (GWTC-3.0)~\cite{LIGOScientific:2018mvr,  LIGOScientific:2020ibl, LIGOScientific:2021usb, KAGRA:2021vkt} that meet the following selection criteria:
\begin{itemize}
    \item The signal is detected by at least two detectors.
    \item The false-alarm rate is less than 1 per 1000 years.
    \item The SNR of the inspiral part is at least 10, where the SNR is computed using the portion of the waveform that is between the minimum frequency used in the analysis $f_\mathrm{min}$, usually 20 Hz, and $f_{22}^\mathrm{peak}$, the frequency at the peak of the $(2,2)$-mode.
    \item The number of GW cycles during the inspiral is at least 5. The number of GW cycles is calculated in the frequency domain between $f_\mathrm{min}$, usually 20 Hz, and $f_{22}^\mathrm{peak}$.
    \item The effective spin $\chi_\mathrm{eff} = (m_1\chi_1 + m_2\chi_2)/M$ is nonzero at 90\% credible level. 
\end{itemize}
The seven GW events from GWTC-3.0 that meet these criteria and their main properties are listed in Table~\ref{tab:events}.

\begin{table}
    \centering
    \begin{tabular}{lcccc}
        \hline
        \hline
        Event & $\mathcal{M}_c$ [$\mathrm{M}_\odot$] & $\chi_\mathrm{eff}$ & Inspiral SNR & \# cycles \\
        \hline
        \textbf{GW190412} & $15.2^{+0.2}_{-0.2}$ & $0.25^{+0.08}_{-0.11}$ & 19.8 & 24 \\
        \textbf{GW190720}\_000836 & $10.4^{+0.2}_{-0.1}$ & $0.18^{+0.14}_{-0.12}$ & 11.6 & 51 \\
        \textbf{GW190728}\_064510 & $10.1^{+0.09}_{-0.08}$ & $0.12^{+0.20}_{-0.07}$ & 14.6 & 53 \\
        \textbf{GW190828}\_063405 & $34.5^{+2.9}_{-2.8}$ & $0.19^{+0.15}_{-0.16}$ & 14.5 & 6.3 \\
        \textbf{GW191204}\_171526 & $9.70^{+0.05}_{-0.05}$ & $0.16^{+0.08}_{-0.05}$ & 17.8 & 43 \\
        \textbf{GW191216}\_213338 & $8.94^{+0.05}_{-0.05}$ & $0.11^{+0.13}_{-0.06}$ & 18.9 & 48 \\
        \textbf{GW200316}\_215756 & $10.68^{+0.12}_{-0.12}$ & $0.13^{+0.27}_{-0.10}$ & 11.4 & 64 \\
        \hline
        \hline
    \end{tabular}
    \caption{The detector-frame chirp mass and effective spin of the events considered in this work. We also list the SNR of the inspiral and the number of GW cycles during inspiral.}
    \label{tab:events}
\end{table}

To stay as close as possible to previous analyses performed by the LVK Collaboration, we use the same data and settings wherever possible. More specifically, we use the strain data and PSDs released in Refs.~\cite{LIGOScientific:2019lzm, ligo_scientific_collaboration_and_virgo_2022_6513631, KAGRA:2023pio, ligo_scientific_collaboration_and_virgo_2021_5546663} and we use the same settings for duration, sampling rate, minimum and maximum frequency, priors, etc., as used in Refs.~\cite{LIGOScientific:2021usb, KAGRA:2021vkt}. The priors used for $\delta\kappa_s$ and $\delta\kappa_a$ are uniform in the range $[-500, 500]$, while the prior on $\delta\lambda_s$ is uniform between $[-1000, 1000]$. We only vary one of $\{\delta\kappa_s, \delta\kappa_a, \delta\lambda_s\}$ at a time while keeping the other ones fixed to their Kerr BH value of zero.

\begin{figure*}
    \centering
    \includegraphics[width=\textwidth]{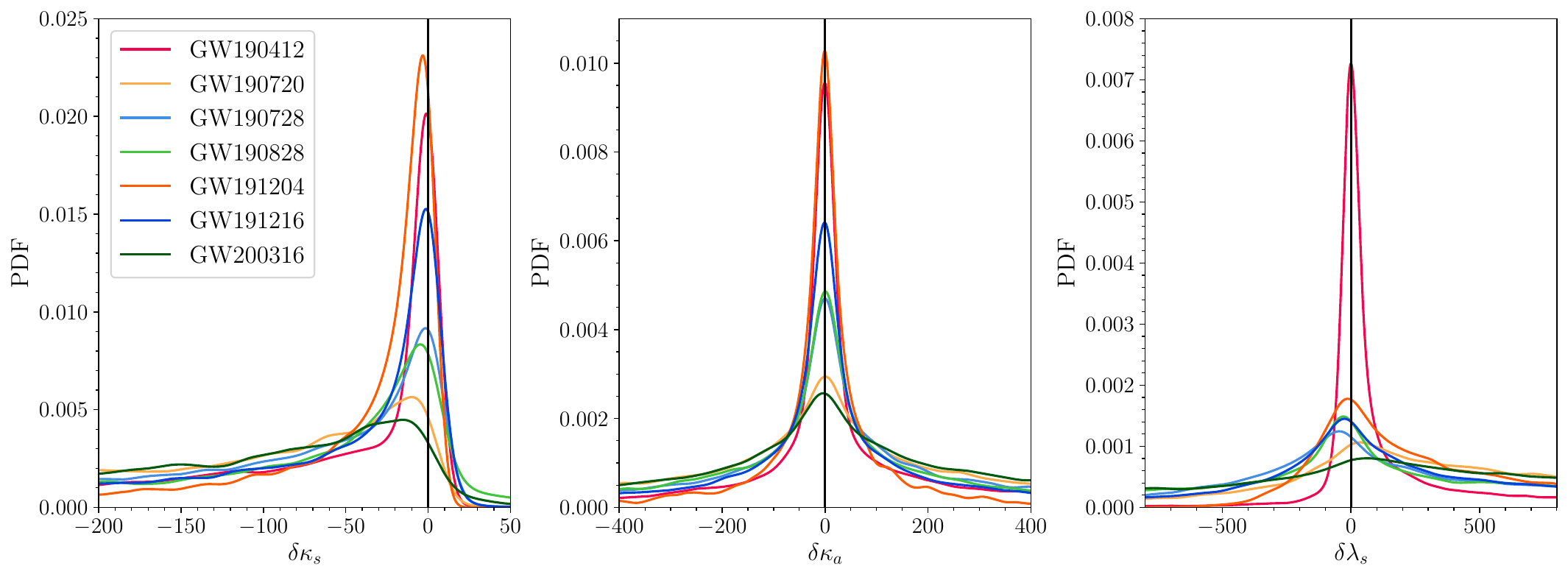}
    \caption{Results for $\delta\kappa_s$ (left), $\delta\kappa_a$ (middle), and $\delta\lambda_s$ (right) for the seven GWTC-3.0 events that meet our selection criteria. All results are consistent with Kerr BHs in GR.}
    \label{fig:events}
\end{figure*}

\subsection{Constraints on $\delta\kappa_s$, $\delta\kappa_a$, and $\delta\lambda_s$}

The posteriors on $\delta\kappa_s$ for the seven events are shown in the left panel of Fig.~\ref{fig:events}. All posteriors are consistent with the compact objects being Kerr BHs in GR. They also all have a tail towards negative values of $\delta\kappa_s$, which is consistent with all analyzed events having a small but positive $\chi_\mathrm{eff}$. The most constraining events are GW191204 and GW190412.

The middle panel of Fig.~\ref{fig:events} shows the posteriors on $\delta\kappa_a$ for the seven events. We once again see that all posteriors are consistent with Kerr BHs in GR. They are centered around zero with tails on both sides, which is consistent with what we saw in our injections. The most constraining events are again GW191204 and GW190412.

As can be seen in the right panel of Fig.~\ref{fig:events}, the posteriors on $\delta\lambda_s$ are mostly unconstrained, with only a small bump around zero and support over the full prior range. The only exception is GW190412, for which a clear peak around zero is obtained. Considering that it has the largest $\chi_\mathrm{eff}$ and inspiral SNR of the events considered, it is no surprise that this is the most constraining event.

We do not quote any bounds on the inferred individual SIQM and SIOM since all posteriors show tails that reach the edge of their respective priors. It is therefore not possible to obtain reliable 90\% credible intervals.

\subsection{Comparison with GWTC-3.0 LVK Collaboration results}

The LVK Collaboration previously presented results for a similar test of the SIQM (called SIM test) on GWTC\nobreakdash-3.0 events~\cite{LIGOScientific:2020tif, LIGOScientific:2021sio}, based on the \textsc{IMRPhenomPv2} waveform model~\cite{Hannam:2013oca, Bohe:PPv2}. This waveform model includes spin-precession effects and only models the dominant $(2,2)$-mode. The \textsc{SEOBNRv5HM\_ROM}~\cite{Pompili:2023tna} waveform model that we employ, on the other hand, does not include spin-precession effects but does include higher-order modes. Both tests include the same SIQM corrections in the inspiral regime, but the transition to the GR merger-ringdown is treated differently in the models. It is therefore interesting to compare the results from both tests to see how different choices in the modeling impact the constraints.

\begin{figure*}
    \centering
    \includegraphics[width=\textwidth]{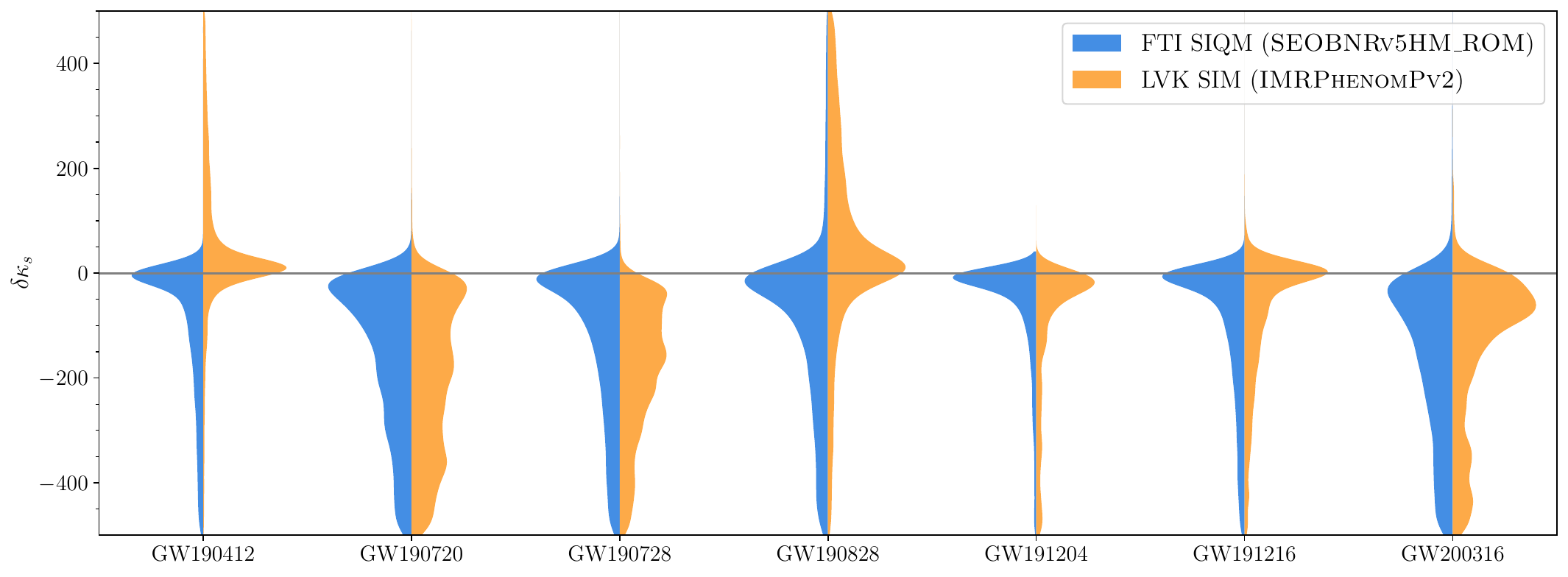}
    \caption{Comparison between our results (blue) and the results of the SIQM test performed by the LVK Collaboration (orange) for the GWTC-3.0~\cite{LIGOScientific:2020tif, LIGOScientific:2021sio}.}
    \label{fig:compare_SIM}
\end{figure*}

Figure~\ref{fig:compare_SIM} compares the posteriors from the LVK SIM test with the posteriors obtained with the FTI SIQM test for $\delta\kappa_s$ for the seven GW signals analyzed here. The results look mostly similar with FTI sometimes being more constraining. This is likely due to FTI turning off the SIQM corrections at a higher frequency, therefore probing more of the late inspiral where the SIQM corrections are more important. These differences in the results between FTI SIQM and LVK SIM should be seen as a measure of the systematic uncertainties in these SIQM tests.

\subsection{Combined GWTC-4.0 results}

We can combine the results from individual events to obtain more stringent constraints on the SIQM. When combining events, we also include results from events observed in first part of the fourth observing run of the LVK detectors (O4a)~\cite{LIGOScientific:2025slb}. The method used in this paper was applied to selected events from O4a by the LVK Collaboration in~\citet{LIGOScientific:2026fcf}, so we directly use those posteriors when combining events from GWTC-4.0~\cite{ligo_scientific_collaboration_2026_21403342}.

We use hierarchical inference to infer the underlying distribution of $\delta\kappa_s$ in the population. We assume that the values of the $\delta\kappa_s$ for the different events follow a Gaussian distribution characterized by a median $\mu$ and standard deviation $\sigma$~\cite{Isi:2019asy, LIGOScientific:2026qni}. We find when combining events from GWTC-4.0 that the distribution on the hyperparameters are consistent with GR ($\mu=\sigma=0$) at 90\% credible level, with $\mu = -29^{+25}_{-36}$ and $\sigma < 37$. The value of $\mu$ is shifted towards negative values due to the individual event posteriors all having a heavy tail at the negative side. The LVK Collaboration also applied the SIQM test based on the \textsc{IMRPhenomXPHM} waveform model~\cite{Divyajyoti:2023izl}, which includes spin-precession effects and higher-order multipoles, to events from GWTC-4.0, finding similar results~\cite{LIGOScientific:2026fcf}.

\begin{figure}
    \centering
    \includegraphics[width=\columnwidth]{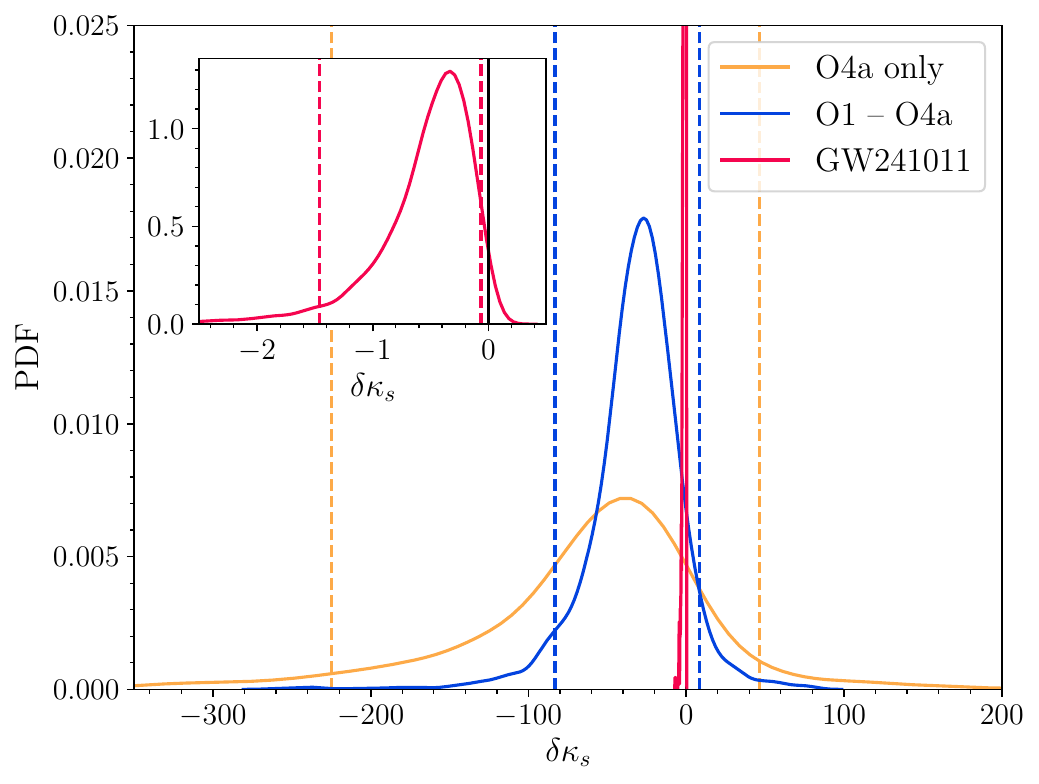}
    \caption{Combined constraints on $\delta\kappa_s$ obtained by hierarchically combining events only from O4a (orange,~\cite{LIGOScientific:2026fcf}) or including O1 -- O4a (blue). For comparison, we show the posterior on $\delta\kappa_s$ from GW241011\_233834 (red), an event observed during O4b and the best single event SIQM constraint to date. All results were obtained using FTI.}
    \label{fig:hier}
\end{figure}

The combined posterior on $\delta\kappa_s$ is shown in Fig.~\ref{fig:hier} and is consistent with GR with $\delta\kappa_s = -29^{+38}_{-54}$. For comparison, we also included the hierarchically combined posterior using only O4a events~\cite{LIGOScientific:2026fcf} and the posterior on $\delta\kappa_s$ for GW241011\_233834, an event from the second part of the fourth observing run of the LVK detectors (O4b) that has a highly spinning primary and therefore is able to put tight constraints on the SIQM~\cite{LIGOScientific:2025brd}. This event is not included in the combined results since it was observed later in O4b. This shows that including more events improves the constraints on the SIQM, but that a single highly spinning event provides much tighter bounds than can be obtained by combining multiple low-spin events. This means that population level constraints will likely be dominated by a few high-spin events.

\section{XG detectors} \label{sec:3G}

In this section, we make projections for the constraints on SIQM and SIOM deviations using XG detectors. To accomplish this, we examine two different configurations. The first configuration consists of a triangular Einstein Telescope (ET) with arm lengths of 10~km and located at the Virgo site. We use the ET-D sensitivity curve \cite{Hild:2010id} with a minimum frequency of 5~Hz in the analysis. For the second configuration, we add two Cosmic Explorer (CE) detectors, one with 40~km arms located at the LIGO Hanford site and the other with 20~km arms located at the LIGO Livingston site. They respectively use the CE40 and CE20 compact binary optimized configuration PSDs \cite{Srivastava:2022slt}, also using a minimum frequency of 5~Hz.

\subsection{Validating FIM error estimates}

The Fisher information matrix (FIM) can be used to estimate the measurement uncertainty of parameters~\cite{Cutler:1994ys, Poisson:1995ef}. Under the assumption of additive Gaussian noise, it is given by
\begin{equation}
    \mathbf{\Gamma}_{ij} = \sum_{k} \left( \partial_{\theta_i} h_k(f) \right. \left| \,\partial_{\theta_j} h_k(f) \right),
\end{equation}
where the sum is over the different detectors and the inner product is defined as
\begin{equation}
    (a|b) = 4 \Re \int_{f_{\rm min}}^{f_{\rm max}} \mathrm{d} f \frac{a^*(f) b(f)}{S_n(f)},
\end{equation}
where $S_n(f)$ denotes the PSD of the detector. The parameter errors are then determined by the covariance matrix $\mathbf{\Sigma} = \mathbf{\Gamma}^{-1}$, which is the inverse of the FIM. The standard deviations of the parameters $\theta_i$ are then given by the diagonal elements:
\begin{equation}
    \sigma_{\theta_i} = \sqrt{\mathbf{\Sigma}_{ii}}.
\end{equation}

The FIM assumes that the posteriors are Gaussian and unbiased. This is usually a reasonable assumption for the high SNRs expected in XG detectors. In some cases, however, we want to validate the FIM analysis by comparing it to a full Bayesian analysis, which is more computationally expensive. For both of the detector configurations described above, we inject the GW150914-like and GW190412-like signals from Table~\ref{tab:inj_param} into zero-noise using the \textsc{SEOBNRv5HM\_ROM} waveform model. The SNRs of these injections in both detector configurations are listed in Table~\ref{tab:snrs}. We then perform a FIM analysis to get an estimate of the expected errors on the parameters. For the FIMs, we use the \textsc{gwbench} package~\cite{Borhanian:2020ypi}.

\begin{table}[]
    \centering
    \begin{tabular}{llc}
        \hline
        \hline
        Detector configuration & Signal & SNR \\
        \hline
        ET & GW150914-like & 347 \\
        ET + CE40 + CE20 & GW150914-like & 1257 \\
        ET & GW190412-like & 269 \\
        ET + CE40 + CE20 & GW190412-like & 989 \\
        \hline
        \hline
    \end{tabular}
    \caption{SNR for the injected signals in XG detectors. The source parameters are the same as in Table~\ref{tab:inj_param}.}
    \label{tab:snrs}
\end{table}

We also perform a Bayesian analysis on the injected signal using \textsc{Bilby}. Since these signals are relatively long, requiring data segments of 64~s and 128~s, respectively, we use the multibanded likelihood approximation to speed up the analysis~\cite{Morisaki:2021ngj, Adhikari:2022mbj}. In this approximation, the likelihood is evaluated on a sparser frequency grid instead of the full, equally spaced one, which considerably speeds up the waveform generation and likelihood evaluation time for long signals without losing too much accuracy. We use the error estimation from the FIM to estimate the prior widths for the chirp mass and deviation parameters. This allows us to set reasonably tight priors, improving sampling convergence and time without having to rerun the analysis due to posteriors exceeding their prior bounds.

\begin{figure}
    \centering
    \includegraphics[width=\columnwidth]{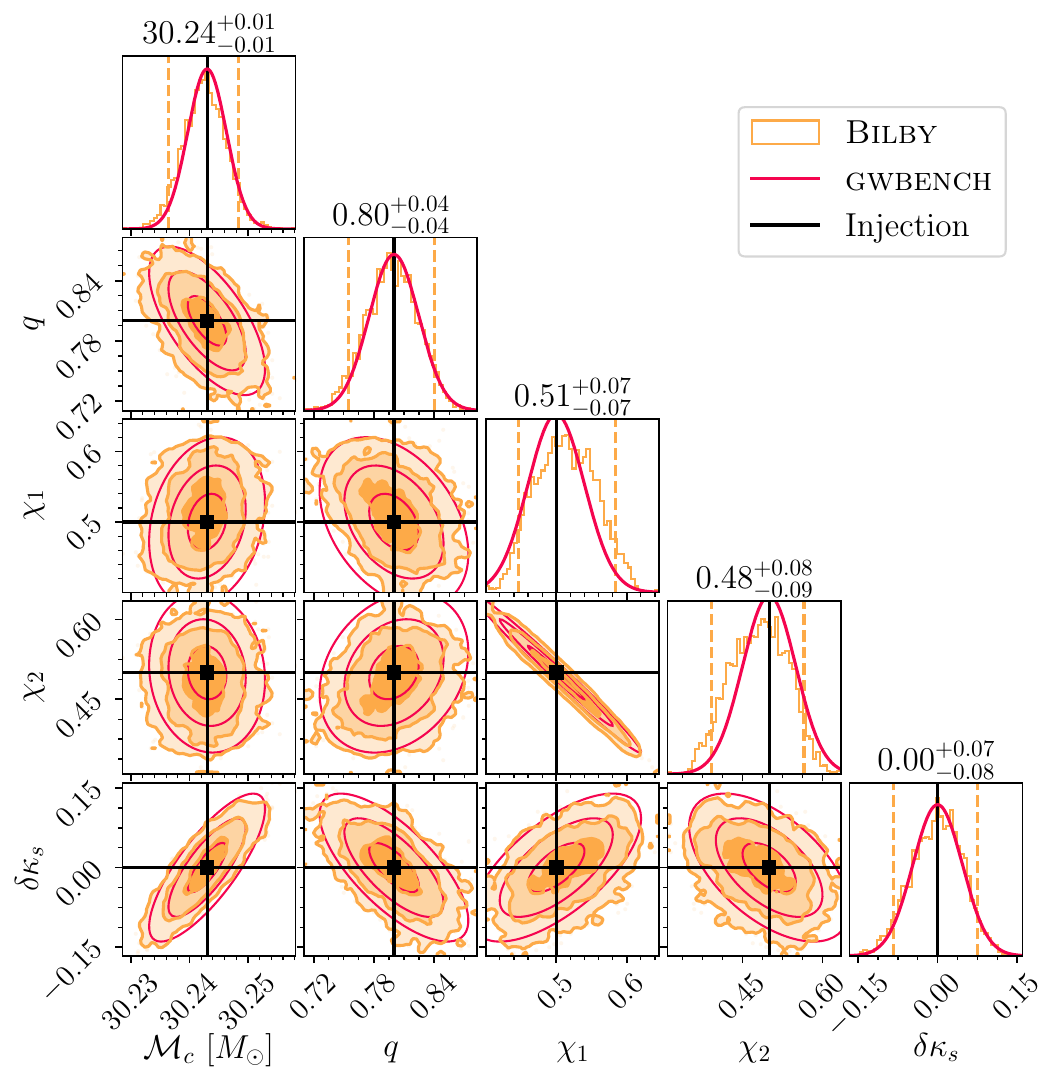}
    \caption{Bayesian- (\textsc{Bilby}, orange) and FIM-analysis (\textsc{gwbench}, red) results for $\delta\kappa_s$ for the GW150914-like injection (black) in the ET-only detector configuration. The 2d contours and ellipses correspond to $\{1,2,3\}$-sigma credible regions. The numbers at the top correspond to the medians and 1-sigma credible intervals recovered by \textsc{Bilby}, which is also indicated with vertical dashed lines in the 1d distributions. The bold black lines indicate the injected values.}
    \label{fig:fish-good}
\end{figure}

By comparing the FIM and Bayesian analyses, we can evaluate the effectiveness of the FIM approximation in estimating constraints on the SIQM and SIOM with XG detectors. For $\delta\kappa_s$ and $\delta\lambda_s$, the FIM and Bayesian analyses agree well for all cases, like for the example shown in Fig.~\ref{fig:fish-good}. For $\delta\kappa_a$, the GW190412-like injections also give comparable results. Minor differences sometimes arise between the Bayesian and FIM analyses due to correlations between the parameters. However, the estimated width of the posteriors on the deviation parameters consistently aligns well between the FIM and Bayesian analyses.

\begin{figure}
    \centering
    \includegraphics[width=\columnwidth]{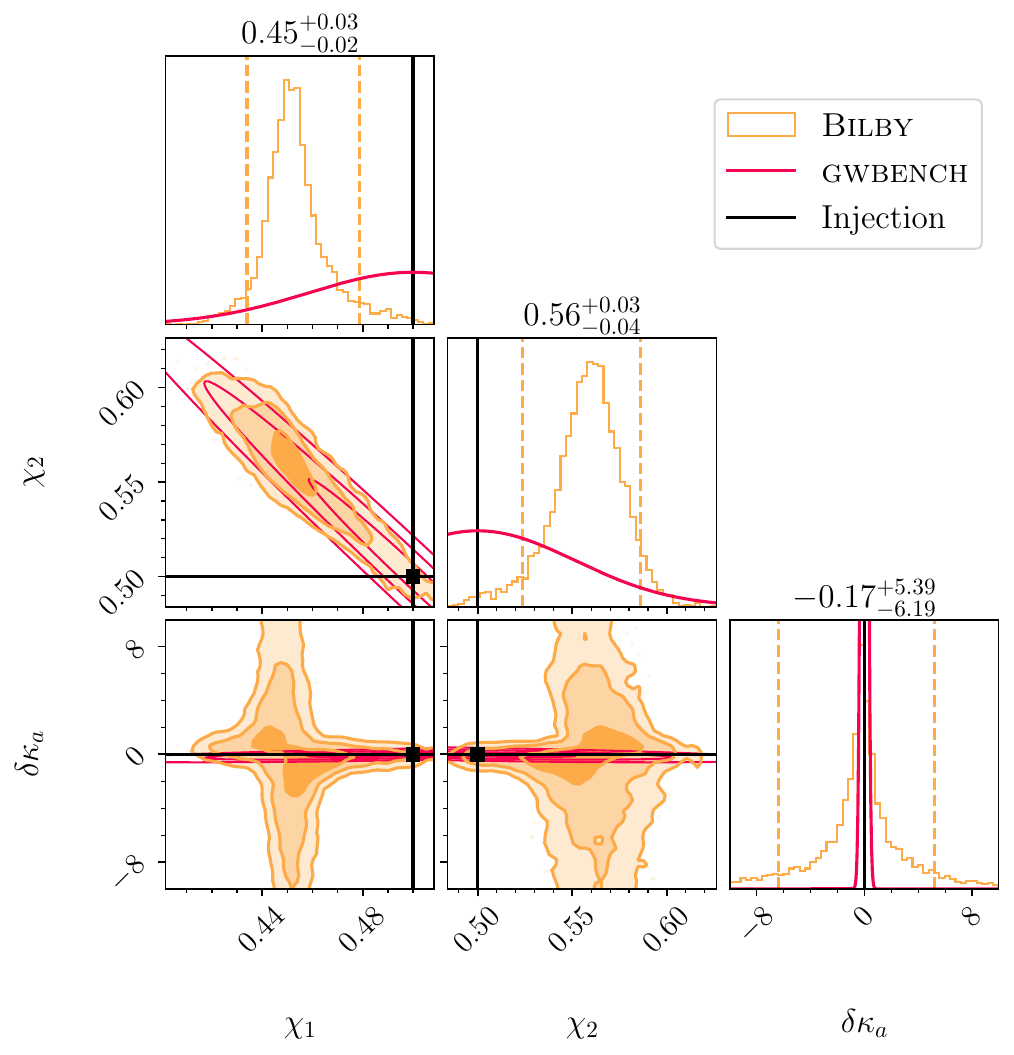}
    \caption{Bayesian- (\textsc{Bilby}, orange) and FIM-analysis (\textsc{gwbench}, red) results for $\delta\kappa_a$ for the GW150914-like injection (black) in the ET-only detector configuration. The bold black lines indicate the injected values. There is a pole in the SIQM corrections close to the injected values of the masses and spins, leading to a much wider posterior for $\delta\kappa_a$ than the FIM provides.}
    \label{fig:fish-bad-dkappaA}
\end{figure}

Conversely, the FIM analysis for the GW150914-like case of $\delta\kappa_a$ does not agree with the \textsc{Bilby} results. The former provides much tighter constraints on $\delta\kappa_a$ than the latter. As Fig.~\ref{fig:fish-bad-dkappaA} shows, the posteriors for $\delta\kappa_a$ and the spins $\chi_1$ and $\chi_2$ are non-Gaussian. Therefore, the FIM is not expected to work. These posteriors are non-Gaussian because for certain values of the masses and spins the SIQM corrections in Eqs. \eqref{eq:dkappa2PN}--\eqref{eq:dkappa3p5PN} become zero for the antisymmetric combination of the SIQMs. This occurs close enough to the injected values of the mass ratio and spins that the posterior supports these values to some extent. This then allows for the possibility of large values of $\delta\kappa_a$, resulting in heavy tails in $\delta\kappa_a$ and biased spins. Thus, the FIM underestimates the error on $\delta\kappa_a$.
This only occurs when the prefactor for $\delta\kappa_a$ in Eq. \eqref{eq:dkappa2PN} is small (i.e., when a tight bound on $\delta\kappa_a$ can typically not be achieved). Therefore, when searching for the best constraints, we do not need to worry about the FIM underestimating the statistical errors for $\delta\kappa_a$; the systems that provide the best constraints should not suffer from this issue.

\begin{figure}
    \centering
    \includegraphics[width=\columnwidth]{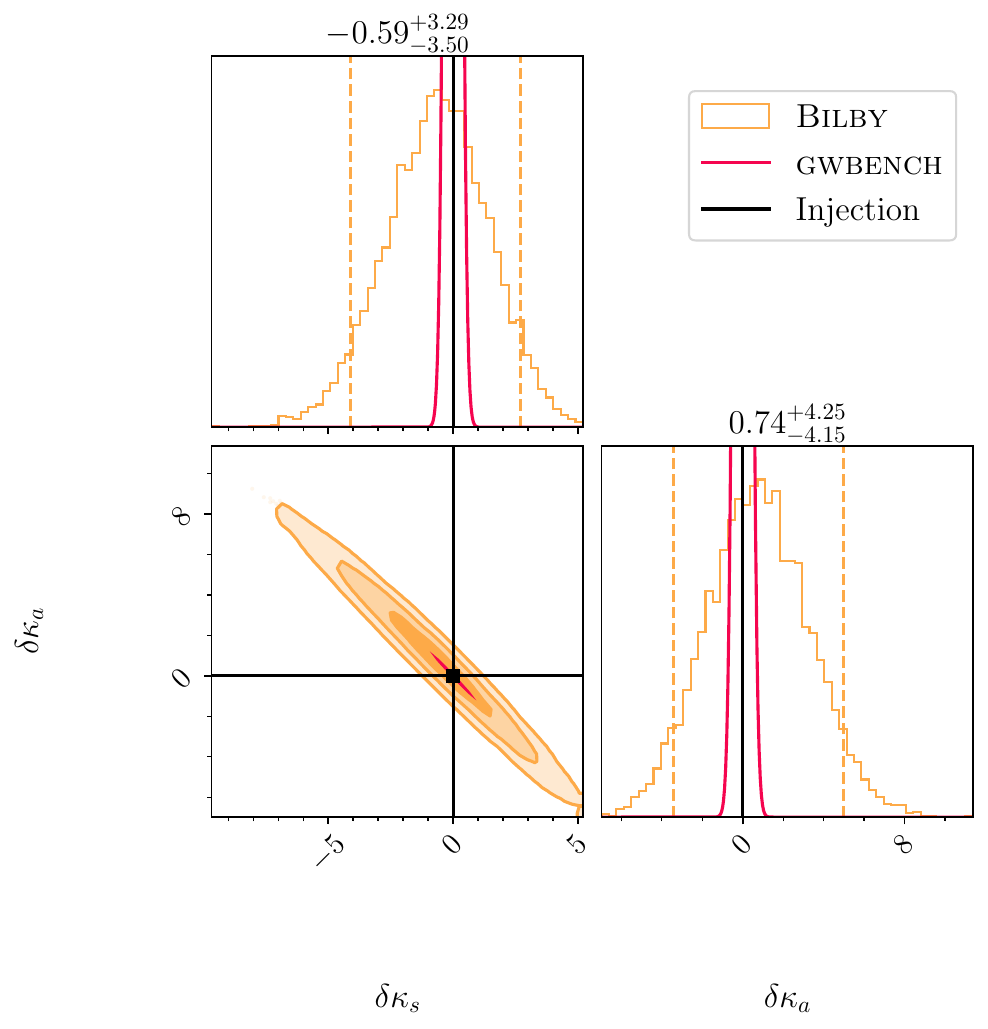}
    \caption{Bayesian- (\textsc{Bilby}, orange) and FIM-analysis (\textsc{gwbench}, red) results when varying both $\delta\kappa_s$ and $\delta\kappa_a$ at the same time for the GW190412-like injection (black) in the 2 CEs + ET detector configuration. The bold black lines indicate the injected values. The FIM does not adequately capture the strength of the degeneracy between $\delta\kappa_s$ and $\delta\kappa_a$, resulting in overly tight error estimates.}
    \label{fig:fish-bad-dkappaS-dkappaA}
\end{figure}

We also notice that the FIM analysis for cases in which we allow both the symmetric and antisymmetric combinations of the SIQMs to vary simultaneously does not match the \textsc{Bilby} results. 
As illustrated in Fig.~\ref{fig:fish-bad-dkappaS-dkappaA}, the FIM has difficulty capturing the strong correlation between $\delta\kappa_s$ and $\delta\kappa_a$. Consequently, it underestimates the errors on both parameters. This is caused by numerical errors when inverting the FIM due to eigenvalues of the FIM being of vastly different orders of magnitude. This can happen when there are strongly correlated parameters. Therefore, we conclude that using the FIM is ineffective in this case because it tends to provide unrealistically narrow error estimates.

There is one small caveat to the results of the ET-only configuration presented above. A challenge of a single triangular ET detector network is sky localization. For long signals, the rotation of the Earth can be used to accurately measure the sky location because it induces amplitude and phase modulations. However, this is not the case for short signals, such as those analyzed here. This often leads to multimodal posteriors for the extrinsic parameters. The FIM cannot handle this and instead provides an estimate of the posterior width of only the mode containing the injected sky location. We do not expect this to significantly impact the error estimates for the intrinsic parameters of the binary. \textsc{Bilby} should be able to handle these multimodal posteriors. However, we notice that the sky location was not recovered correctly in our analyses. This occurs because the error in the likelihood estimate when using the multibanded likelihood is similar to the difference in likelihood between the different sky-location modes. Thus, the sampler picks the incorrect mode. Again, we do not expect this to greatly affect the posteriors of the intrinsic parameters. Therefore, we believe that the above conclusions should still be valid. This problem does not occur with the triangular ET plus two CE detector network because the sky location is accurately measured and becomes unimodal with three detectors.

\subsection{FIM analysis for a population of BBHs}

In order to estimate how well XG detectors can constrain the spin-induced multipole moments, we first need to generate a population of merging BBHs. For this, we use the \textsc{gwforge} package~\cite{Chandra:2024dhf}. We use the Madau-Dickinson model for the redshift~\cite{Madau:2014bja}, the power-law + peak model for the masses, and the `default' spin model. We take the merger rate, mass, and spin population parameters from the inferred GWTC-3.0 population~\cite{KAGRA:2021duu}. Assuming the binaries consist entirely of Kerr BHs in GR, the deviation parameters are set to zero. With these population settings, we obtain $71\,419$ merging binaries in one year of observation time.

\begin{figure}
    \centering
    \includegraphics[width=\columnwidth]{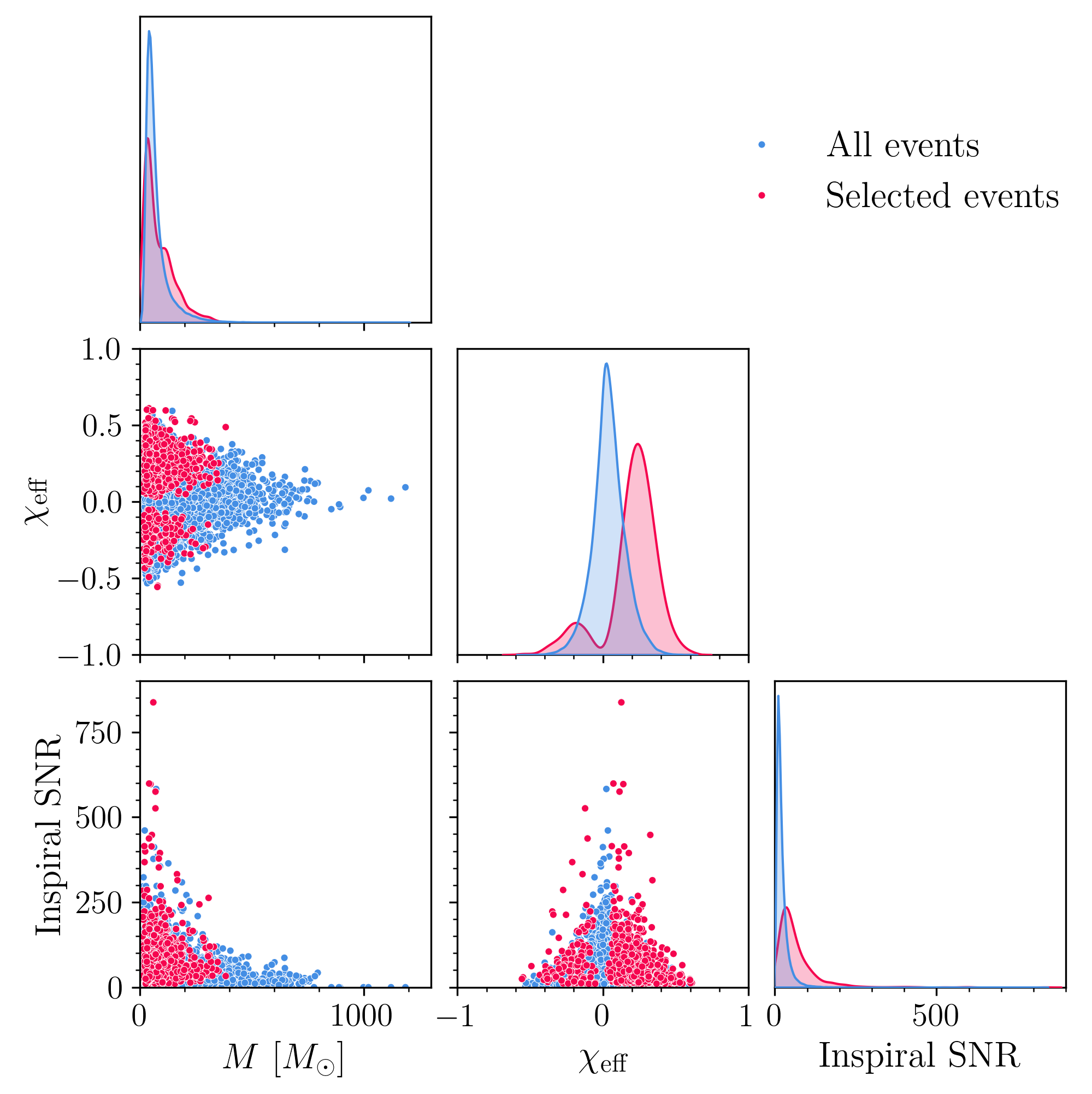}
    \caption{The total mass, effective spin, and inspiral SNR distributions for all events in the generated population (blue) and the selected events (red) that pass our selection criteria for the ET-only detector configuration.}
    \label{fig:population}
\end{figure}

Next, we calculate the inspiral SNR and the number of inspiral cycles for the merging binaries in the two detector configurations. We compute these quantities starting from 5~Hz and up to the frequency at the peak of the $(2,2)$-mode. We then select only the events that have inspiral SNR larger than 10 and at least 5 GW cycles during the inspiral. For the events that pass this selection criteria, we then perform a FIM analysis assuming Kerr BHs in GR to select the events that have $\chi_\mathrm{eff}\neq0$ at 90\% credible level. This leaves us with 1249 events for the ET-only configuration and 4090 events for the 2 CEs + ET configuration. Figure~\ref{fig:population} shows the total mass, effective spin, and inspiral SNR for all events in the generated population and the selected events for the ET-only configuration.  We notice that all events with a large total mass are discarded because they do not have sufficient inspiral signal in the detectors' frequency band, and events with small spins are excluded as well. Interesting enough, there are less events with $\chi_\mathrm{eff}<0$ selected than with $\chi_\mathrm{eff}>0$. Assuming all other parameters remain the same, it is reasonable to conclude that fewer binaries with negative spins pass the SNR selection criterion, since SNRs are generally lower for negative spins and they have lesser inspiral cycles.

\begin{figure}
    \centering
    \includegraphics[width=\columnwidth]{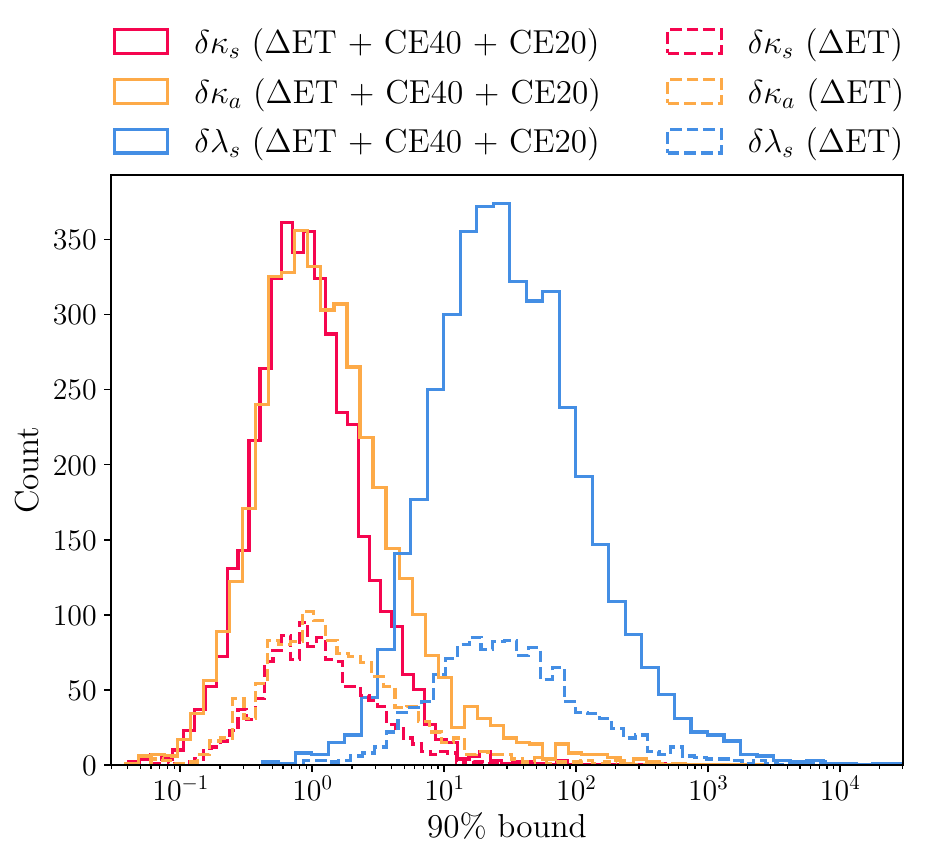}
    \caption{The distributions of 90\% upper bounds on the magnitude of $\delta\kappa_s$ (red), $\delta\kappa_a$ (orange), and $\delta\lambda_s$ (blue) for a population of BBHs detected by a (2 CEs + ET) detector network (solid lines) or triangular ET-only detector network (dashed lines).}
    \label{fig:next-gen}
\end{figure}

We perform FIM analyses on the single-parameter tests for $\delta\kappa_s$, $\delta\kappa_a$, and $\delta\lambda_s$, for all the selected events. While it would be more physically realistic to make predictions by varying $\delta\kappa_s$ and $\delta\kappa_a$ simultaneously, the aforementioned disagreement between the FIM and Bayesian analyses for this case limits us to single-parameter tests. Next, we convert the error estimates from the FIM to 90\% upper bounds on the magnitude of the deviation parameters. The distributions of these bounds are shown in Fig.~\ref{fig:next-gen}. The distribution of bounds is similar for $\delta\kappa_s$ and $\delta\kappa_a$, but $\delta\lambda_s$ is less well constrained. This is expected because it is a higher-order-in-spin effect and contributes at a higher PN order. We also observe that the distribution of bounds for ET-only and 2 CEs + ET detector networks are similar. The most noticeable difference is that the ET-only detector network observes fewer events. Adding the two CE detectors to the network improves the best bounds by only a factor $\lesssim 2$, due to the tail of the distribution extending slightly toward smaller values. The best single event bounds for $\delta\kappa_s$ are comparable to those found in earlier works~\cite{Krishnendu:2018nqa, Branchesi:2023mws, ET:2025xjr}.

We estimate the bounds by combining the constraints from all events in the population, assuming they share a common value. The combined bound is then given by
\begin{equation}
    \frac{1}{\sigma_N^2} = \sum_i \frac{1}{\sigma_i^2},
\end{equation}
where $\sigma_i$ are the individual event bounds and the sum goes over all events. This gives
\begin{align*}
    &|\delta\kappa_s| < 0.013, \\
    &|\delta\kappa_a| < 0.014, \\
    &|\delta\lambda_s| < 0.24,
\end{align*}
for the ET-only detector configuration and
\begin{align*}
    &|\delta\kappa_s| < 6.5\times10^{-3}, \\
    &|\delta\kappa_a| < 6.9\times10^{-3}, \\
    &|\delta\lambda_s| < 0.12,
\end{align*}
for the 2 CEs + ET detector configuration.
The combined bounds are a factor $\sim 2$ tighter when adding the two CE detectors to the network, consistent with the increase in number of events by a factor of $\sim 4$.

These estimates show that it will be possible to constrain the values of $|\delta\kappa_s|$, $|\delta\kappa_a|$, and $|\delta\lambda_s|$ to $\ll 1$, meaning it will be possible to determine with confidence whether these effects are present. Therefore, XG detectors will be able to routinely test the BH nature of compact binary coalescences and distinguish BHs from neutron stars and exotic compact objects.

\section{Conclusion}

In this study, we developed a test for the SIQM and SIOM of compact objects using the FTI framework~\cite{Mehta:2022pcn}. Using simulated BBH signals, we analyzed the performance of a detector network consisting of two Advanced LIGO detectors~\cite{LIGOScientific:2014pky} and one Advanced Virgo detector~\cite{VIRGO:2014yos}. As expected, we found that BBHs with large spins provide better constraints on the SIQM. Additionally, we observed that small spins can lead to long tails in the posteriors due to poles in the SIQM corrections. While we could constrain $\delta\kappa_a$ by holding $\delta\kappa_s$ at $0$, when we varied both SIQM parameters simultaneously, they were not well constrained. Furthermore, we demonstrated that it is possible to obtain SIOM constraints, though they are not as tight as SIQM constraints.

The flexibility of the FTI framework allowed us to examine the effects of certain modeling choices on the SIQM constraints. Including the 3.5PN-term in the corrections, which was not included in previous studies~\cite{Krishnendu:2017shb, LIGOScientific:2019fpa, LIGOScientific:2020tif, LIGOScientific:2021sio, LIGOScientific:2025brd, LIGOScientific:2026fcf}, does not affect the results at the considered SNRs. We demonstrated that the tapering frequency of the corrections affects the constraints; larger tapering frequencies tend to produce tighter constraints. Additionally, we demonstrated that incorporating higher modes into the waveform model is sometimes crucial, especially for asymmetric binaries.

We also checked whether the test could detect deviations from a Kerr BH in GR by simulating signals with a nonzero $\delta\kappa_s$. We detected deviations as small as $\delta\kappa_s = \pm2$. However, the test does not accurately measure large deviations. Therefore, we conclude that, while this test can be used as a null test to detect deviations from Kerr BHs in GR, it is generally not suitable for accurately measuring the SIQM value when deviations are large.

Next, we applied our test to selected GW signals observed by the LVK Collaboration 
during their first three observing runs~\cite{KAGRA:2021vkt}. We found that all of these signals have SIQM posteriors consistent with Kerr BHs in GR. Our results are consistent with the constraints on $\delta\kappa_s$ obtained by the LVK Collaboration for GWTC-3.0~\cite{LIGOScientific:2021sio}. This SIQM test has also been used by the LVK Collaboration in GWTC-4.0, finding consistency with GR~\cite{LIGOScientific:2026fcf}. We presented here updated joint constraints by combining those results obtained with FTI with the events analyzed in this work, giving $\delta\kappa_s = -29^{+38}_{-54}$. We were unable to constrain the SIOM for any of the signals except for GW190412.

Lastly, we made projections on how effectively XG detectors~\cite{Hild:2010id, Srivastava:2022slt} will constrain the spin-induced multipole moments. To accomplish this, we performed FIM analyses on a population of BBHs. We validated our results using the FIM by performing full Bayesian analyses on a few selected cases. Our results showed that XG detectors will be able to constrain $|\delta\kappa_s|$, $|\delta\kappa_a|$, and $|\delta\lambda_s|$ to values much less than one. This means that it will be possible to distinguish BHs from neutron stars and exotic compact objects. We compared the performance of a triangular ET-only detector network with that of a triangular ET plus two CEs network. We found that adding the two CE detectors increases the number of events by a factor of about $4$ and improves the bounds by a factor of about $2$.

Overall, our results show that, although it is possible to constrain the SIQM and SIOM using the LIGO-Virgo detector network, obtaining tight bounds requires some fortune in detecting an event with large spins and a long inspiral. Indeed, the FTI-based test using \textsc{SEOBNRv5HM}, presented in this work, (and the IMRPhenomX-based SIQM test~\cite{Divyajyoti:2023izl}) has already been employed by the LVK Collaboration to constrain the value of $\delta\kappa_1$ for the highly-spinning event GW241011\_233834~\cite{LIGOScientific:2025brd}, providing stringent constraints of around unity. GW241011\_233834 has also been used to place constraints on the SIOM of $\mathcal{O}(10)$~\cite{Das:2026tel}. The increased SNR and signal duration in XG detectors will help overcome this problem, making it possible to routinely test the BH nature of compact objects.

\vspace{1em}

\section*{Acknowledgments}

The authors would like to thank Rimo Das and N.V. Krishnendu for reviewing the manuscript and providing useful comments.
The authors are grateful for computational resources provided by the LIGO Laboratory and supported by National Science Foundation Grants PHY-0757058 and PHY-0823459, as well as computational resources from the Max Planck Institute for Gravitational Physics in Potsdam.
This research has made use of data or software obtained from the Gravitational Wave Open Science Center (\href{https://www.gwosc.org}{https://www.gwosc.org}), a service of the LIGO Scientific Collaboration, the Virgo Collaboration, and KAGRA. This material is based upon work supported by NSF's LIGO Laboratory which is a major facility fully funded by the National Science Foundation, as well as the Science and Technology Facilities Council (STFC) of the United Kingdom, the Max-Planck-Society (MPS), and the State of Niedersachsen/Germany for support of the construction of Advanced LIGO and construction and operation of the GEO600 detector. Additional support for Advanced LIGO was provided by the Australian Research Council. Virgo is funded, through the European Gravitational Observatory (EGO), by the French Centre National de Recherche Scientifique (CNRS), the Italian Istituto Nazionale di Fisica Nucleare (INFN) and the Dutch Nikhef, with contributions by institutions from Belgium, Germany, Greece, Hungary, Ireland, Japan, Monaco, Poland, Portugal, Spain. KAGRA is supported by Ministry of Education, Culture, Sports, Science and Technology (MEXT), Japan Society for the Promotion of Science (JSPS) in Japan; National Research Foundation (NRF) and Ministry of Science and ICT (MSIT) in Korea; Academia Sinica (AS) and National Science and Technology Council (NSTC) in Taiwan.

\bibliography{biblio}

%apsrev4-2.bst 2019-01-14 (MD) hand-edited version of apsrev4-1.bst
%Control: key (0)
%Control: author (8) initials jnrlst
%Control: editor formatted (1) identically to author
%Control: production of article title (0) allowed
%Control: page (0) single
%Control: year (1) truncated
%Control: production of eprint (0) enabled
\begin{thebibliography}{86}%
\makeatletter
\providecommand \@ifxundefined [1]{%
 \@ifx{#1\undefined}
}%
\providecommand \@ifnum [1]{%
 \ifnum #1\expandafter \@firstoftwo
 \else \expandafter \@secondoftwo
 \fi
}%
\providecommand \@ifx [1]{%
 \ifx #1\expandafter \@firstoftwo
 \else \expandafter \@secondoftwo
 \fi
}%
\providecommand \natexlab [1]{#1}%
\providecommand \enquote  [1]{``#1''}%
\providecommand \bibnamefont  [1]{#1}%
\providecommand \bibfnamefont [1]{#1}%
\providecommand \citenamefont [1]{#1}%
\providecommand \href@noop [0]{\@secondoftwo}%
\providecommand \href [0]{\begingroup \@sanitize@url \@href}%
\providecommand \@href[1]{\@@startlink{#1}\@@href}%
\providecommand \@@href[1]{\endgroup#1\@@endlink}%
\providecommand \@sanitize@url [0]{\catcode `\\12\catcode `\$12\catcode
  `\&12\catcode `\#12\catcode `\^12\catcode `\_12\catcode `\%12\relax}%
\providecommand \@@startlink[1]{}%
\providecommand \@@endlink[0]{}%
\providecommand \url  [0]{\begingroup\@sanitize@url \@url }%
\providecommand \@url [1]{\endgroup\@href {#1}{\urlprefix }}%
\providecommand \urlprefix  [0]{URL }%
\providecommand \Eprint [0]{\href }%
\providecommand \doibase [0]{https://doi.org/}%
\providecommand \selectlanguage [0]{\@gobble}%
\providecommand \bibinfo  [0]{\@secondoftwo}%
\providecommand \bibfield  [0]{\@secondoftwo}%
\providecommand \translation [1]{[#1]}%
\providecommand \BibitemOpen [0]{}%
\providecommand \bibitemStop [0]{}%
\providecommand \bibitemNoStop [0]{.\EOS\space}%
\providecommand \EOS [0]{\spacefactor3000\relax}%
\providecommand \BibitemShut  [1]{\csname bibitem#1\endcsname}%
\let\auto@bib@innerbib\@empty
%</preamble>
\bibitem [{\citenamefont {Carter}(1971)}]{Carter:1971zc}%
  \BibitemOpen
  \bibfield  {author} {\bibinfo {author} {\bibfnamefont {B.}~\bibnamefont
  {Carter}},\ }\bibfield  {title} {\bibinfo {title} {{Axisymmetric Black Hole
  Has Only Two Degrees of Freedom}},\ }\href
  {https://doi.org/10.1103/PhysRevLett.26.331} {\bibfield  {journal} {\bibinfo
  {journal} {Phys. Rev. Lett.}\ }\textbf {\bibinfo {volume} {26}},\ \bibinfo
  {pages} {331} (\bibinfo {year} {1971})}\BibitemShut {NoStop}%
\bibitem [{\citenamefont {Hansen}(1974)}]{Hansen:1974zz}%
  \BibitemOpen
  \bibfield  {author} {\bibinfo {author} {\bibfnamefont {R.~O.}\ \bibnamefont
  {Hansen}},\ }\bibfield  {title} {\bibinfo {title} {{Multipole moments of
  stationary space-times}},\ }\href {https://doi.org/10.1063/1.1666501}
  {\bibfield  {journal} {\bibinfo  {journal} {J. Math. Phys.}\ }\textbf
  {\bibinfo {volume} {15}},\ \bibinfo {pages} {46} (\bibinfo {year}
  {1974})}\BibitemShut {NoStop}%
\bibitem [{\citenamefont {Krishnendu}\ \emph {et~al.}(2017)\citenamefont
  {Krishnendu}, \citenamefont {Arun},\ and\ \citenamefont
  {Mishra}}]{Krishnendu:2017shb}%
  \BibitemOpen
  \bibfield  {author} {\bibinfo {author} {\bibfnamefont {N.~V.}\ \bibnamefont
  {Krishnendu}}, \bibinfo {author} {\bibfnamefont {K.~G.}\ \bibnamefont
  {Arun}},\ and\ \bibinfo {author} {\bibfnamefont {C.~K.}\ \bibnamefont
  {Mishra}},\ }\bibfield  {title} {\bibinfo {title} {{Testing the binary black
  hole nature of a compact binary coalescence}},\ }\href
  {https://doi.org/10.1103/PhysRevLett.119.091101} {\bibfield  {journal}
  {\bibinfo  {journal} {Phys. Rev. Lett.}\ }\textbf {\bibinfo {volume} {119}},\
  \bibinfo {pages} {091101} (\bibinfo {year} {2017})},\ \Eprint
  {https://arxiv.org/abs/1701.06318} {arXiv:1701.06318 [gr-qc]} \BibitemShut
  {NoStop}%
\bibitem [{\citenamefont {Lyu}\ \emph {et~al.}(2024)\citenamefont {Lyu},
  \citenamefont {LaHaye}, \citenamefont {Yang},\ and\ \citenamefont
  {Bonga}}]{Lyu:2023zxv}%
  \BibitemOpen
  \bibfield  {author} {\bibinfo {author} {\bibfnamefont {Z.}~\bibnamefont
  {Lyu}}, \bibinfo {author} {\bibfnamefont {M.}~\bibnamefont {LaHaye}},
  \bibinfo {author} {\bibfnamefont {H.}~\bibnamefont {Yang}},\ and\ \bibinfo
  {author} {\bibfnamefont {B.}~\bibnamefont {Bonga}},\ }\bibfield  {title}
  {\bibinfo {title} {{Probing spin-induced quadrupole moments in precessing
  compact binaries}},\ }\href {https://doi.org/10.1103/PhysRevD.109.064081}
  {\bibfield  {journal} {\bibinfo  {journal} {Phys. Rev. D}\ }\textbf {\bibinfo
  {volume} {109}},\ \bibinfo {pages} {064081} (\bibinfo {year} {2024})},\
  \Eprint {https://arxiv.org/abs/2308.09032} {arXiv:2308.09032 [gr-qc]}
  \BibitemShut {NoStop}%
\bibitem [{\citenamefont {Arun}\ \emph {et~al.}(2009)\citenamefont {Arun},
  \citenamefont {Buonanno}, \citenamefont {Faye},\ and\ \citenamefont
  {Ochsner}}]{Arun:2008kb}%
  \BibitemOpen
  \bibfield  {author} {\bibinfo {author} {\bibfnamefont {K.~G.}\ \bibnamefont
  {Arun}}, \bibinfo {author} {\bibfnamefont {A.}~\bibnamefont {Buonanno}},
  \bibinfo {author} {\bibfnamefont {G.}~\bibnamefont {Faye}},\ and\ \bibinfo
  {author} {\bibfnamefont {E.}~\bibnamefont {Ochsner}},\ }\bibfield  {title}
  {\bibinfo {title} {{Higher-order spin effects in the amplitude and phase of
  gravitational waveforms emitted by inspiraling compact binaries: Ready-to-use
  gravitational waveforms}},\ }\href
  {https://doi.org/10.1103/PhysRevD.79.104023} {\bibfield  {journal} {\bibinfo
  {journal} {Phys. Rev. D}\ }\textbf {\bibinfo {volume} {79}},\ \bibinfo
  {pages} {104023} (\bibinfo {year} {2009})},\ \bibinfo {note} {[Erratum:
  Phys.Rev.D 84, 049901 (2011)]},\ \Eprint {https://arxiv.org/abs/0810.5336}
  {arXiv:0810.5336 [gr-qc]} \BibitemShut {NoStop}%
\bibitem [{\citenamefont {Mishra}\ \emph {et~al.}(2016)\citenamefont {Mishra},
  \citenamefont {Kela}, \citenamefont {Arun},\ and\ \citenamefont
  {Faye}}]{Mishra:2016whh}%
  \BibitemOpen
  \bibfield  {author} {\bibinfo {author} {\bibfnamefont {C.~K.}\ \bibnamefont
  {Mishra}}, \bibinfo {author} {\bibfnamefont {A.}~\bibnamefont {Kela}},
  \bibinfo {author} {\bibfnamefont {K.~G.}\ \bibnamefont {Arun}},\ and\
  \bibinfo {author} {\bibfnamefont {G.}~\bibnamefont {Faye}},\ }\bibfield
  {title} {\bibinfo {title} {{Ready-to-use post-Newtonian gravitational
  waveforms for binary black holes with nonprecessing spins: An update}},\
  }\href {https://doi.org/10.1103/PhysRevD.93.084054} {\bibfield  {journal}
  {\bibinfo  {journal} {Phys. Rev. D}\ }\textbf {\bibinfo {volume} {93}},\
  \bibinfo {pages} {084054} (\bibinfo {year} {2016})},\ \Eprint
  {https://arxiv.org/abs/1601.05588} {arXiv:1601.05588 [gr-qc]} \BibitemShut
  {NoStop}%
\bibitem [{\citenamefont {Aasi}\ \emph {et~al.}(2015)\citenamefont {Aasi} \emph
  {et~al.}}]{LIGOScientific:2014pky}%
  \BibitemOpen
  \bibfield  {author} {\bibinfo {author} {\bibfnamefont {J.}~\bibnamefont
  {Aasi}} \emph {et~al.} (\bibinfo {collaboration} {LIGO Scientific}),\
  }\bibfield  {title} {\bibinfo {title} {{Advanced LIGO}},\ }\href
  {https://doi.org/10.1088/0264-9381/32/7/074001} {\bibfield  {journal}
  {\bibinfo  {journal} {Class. Quant. Grav.}\ }\textbf {\bibinfo {volume}
  {32}},\ \bibinfo {pages} {074001} (\bibinfo {year} {2015})},\ \Eprint
  {https://arxiv.org/abs/1411.4547} {arXiv:1411.4547 [gr-qc]} \BibitemShut
  {NoStop}%
\bibitem [{\citenamefont {Acernese}\ \emph {et~al.}(2015)\citenamefont
  {Acernese} \emph {et~al.}}]{VIRGO:2014yos}%
  \BibitemOpen
  \bibfield  {author} {\bibinfo {author} {\bibfnamefont {F.}~\bibnamefont
  {Acernese}} \emph {et~al.} (\bibinfo {collaboration} {VIRGO}),\ }\bibfield
  {title} {\bibinfo {title} {{Advanced Virgo: a second-generation
  interferometric gravitational wave detector}},\ }\href
  {https://doi.org/10.1088/0264-9381/32/2/024001} {\bibfield  {journal}
  {\bibinfo  {journal} {Class. Quant. Grav.}\ }\textbf {\bibinfo {volume}
  {32}},\ \bibinfo {pages} {024001} (\bibinfo {year} {2015})},\ \Eprint
  {https://arxiv.org/abs/1408.3978} {arXiv:1408.3978 [gr-qc]} \BibitemShut
  {NoStop}%
\bibitem [{\citenamefont {Akutsu}\ \emph {et~al.}(2021)\citenamefont {Akutsu}
  \emph {et~al.}}]{KAGRA:2020tym}%
  \BibitemOpen
  \bibfield  {author} {\bibinfo {author} {\bibfnamefont {T.}~\bibnamefont
  {Akutsu}} \emph {et~al.} (\bibinfo {collaboration} {KAGRA}),\ }\bibfield
  {title} {\bibinfo {title} {{Overview of KAGRA: Detector design and
  construction history}},\ }\href {https://doi.org/10.1093/ptep/ptaa125}
  {\bibfield  {journal} {\bibinfo  {journal} {PTEP}\ }\textbf {\bibinfo
  {volume} {2021}},\ \bibinfo {pages} {05A101} (\bibinfo {year} {2021})},\
  \Eprint {https://arxiv.org/abs/2005.05574} {arXiv:2005.05574
  [physics.ins-det]} \BibitemShut {NoStop}%
\bibitem [{\citenamefont {Krishnendu}\ \emph
  {et~al.}(2019{\natexlab{a}})\citenamefont {Krishnendu}, \citenamefont
  {Saleem}, \citenamefont {Samajdar}, \citenamefont {Arun}, \citenamefont
  {Del~Pozzo},\ and\ \citenamefont {Mishra}}]{Krishnendu:2019tjp}%
  \BibitemOpen
  \bibfield  {author} {\bibinfo {author} {\bibfnamefont {N.~V.}\ \bibnamefont
  {Krishnendu}}, \bibinfo {author} {\bibfnamefont {M.}~\bibnamefont {Saleem}},
  \bibinfo {author} {\bibfnamefont {A.}~\bibnamefont {Samajdar}}, \bibinfo
  {author} {\bibfnamefont {K.~G.}\ \bibnamefont {Arun}}, \bibinfo {author}
  {\bibfnamefont {W.}~\bibnamefont {Del~Pozzo}},\ and\ \bibinfo {author}
  {\bibfnamefont {C.~K.}\ \bibnamefont {Mishra}},\ }\bibfield  {title}
  {\bibinfo {title} {{Constraints on the binary black hole nature of GW151226
  and GW170608 from the measurement of spin-induced quadrupole moments}},\
  }\href {https://doi.org/10.1103/PhysRevD.100.104019} {\bibfield  {journal}
  {\bibinfo  {journal} {Phys. Rev. D}\ }\textbf {\bibinfo {volume} {100}},\
  \bibinfo {pages} {104019} (\bibinfo {year} {2019}{\natexlab{a}})},\ \Eprint
  {https://arxiv.org/abs/1908.02247} {arXiv:1908.02247 [gr-qc]} \BibitemShut
  {NoStop}%
\bibitem [{\citenamefont {Abbott}\ \emph
  {et~al.}(2020{\natexlab{a}})\citenamefont {Abbott} \emph
  {et~al.}}]{LIGOScientific:2020zkf}%
  \BibitemOpen
  \bibfield  {author} {\bibinfo {author} {\bibfnamefont {R.}~\bibnamefont
  {Abbott}} \emph {et~al.} (\bibinfo {collaboration} {LIGO Scientific,
  Virgo}),\ }\bibfield  {title} {\bibinfo {title} {{GW190814: Gravitational
  Waves from the Coalescence of a 23 Solar Mass Black Hole with a 2.6 Solar
  Mass Compact Object}},\ }\href {https://doi.org/10.3847/2041-8213/ab960f}
  {\bibfield  {journal} {\bibinfo  {journal} {Astrophys. J. Lett.}\ }\textbf
  {\bibinfo {volume} {896}},\ \bibinfo {pages} {L44} (\bibinfo {year}
  {2020}{\natexlab{a}})},\ \Eprint {https://arxiv.org/abs/2006.12611}
  {arXiv:2006.12611 [astro-ph.HE]} \BibitemShut {NoStop}%
\bibitem [{\citenamefont {Abbott}\ \emph
  {et~al.}(2021{\natexlab{a}})\citenamefont {Abbott} \emph
  {et~al.}}]{LIGOScientific:2020tif}%
  \BibitemOpen
  \bibfield  {author} {\bibinfo {author} {\bibfnamefont {R.}~\bibnamefont
  {Abbott}} \emph {et~al.} (\bibinfo {collaboration} {LIGO Scientific,
  Virgo}),\ }\bibfield  {title} {\bibinfo {title} {{Tests of general relativity
  with binary black holes from the second LIGO-Virgo gravitational-wave
  transient catalog}},\ }\href {https://doi.org/10.1103/PhysRevD.103.122002}
  {\bibfield  {journal} {\bibinfo  {journal} {Phys. Rev. D}\ }\textbf {\bibinfo
  {volume} {103}},\ \bibinfo {pages} {122002} (\bibinfo {year}
  {2021}{\natexlab{a}})},\ \Eprint {https://arxiv.org/abs/2010.14529}
  {arXiv:2010.14529 [gr-qc]} \BibitemShut {NoStop}%
\bibitem [{\citenamefont {Abbott}\ \emph {et~al.}(2025)\citenamefont {Abbott}
  \emph {et~al.}}]{LIGOScientific:2021sio}%
  \BibitemOpen
  \bibfield  {author} {\bibinfo {author} {\bibfnamefont {R.}~\bibnamefont
  {Abbott}} \emph {et~al.} (\bibinfo {collaboration} {LIGO Scientific, VIRGO,
  KAGRA}),\ }\bibfield  {title} {\bibinfo {title} {{Tests of General Relativity
  with GWTC-3}},\ }\href {https://doi.org/10.1103/PhysRevD.112.084080}
  {\bibfield  {journal} {\bibinfo  {journal} {Phys. Rev. D}\ }\textbf {\bibinfo
  {volume} {112}},\ \bibinfo {pages} {084080} (\bibinfo {year} {2025})},\
  \Eprint {https://arxiv.org/abs/2112.06861} {arXiv:2112.06861 [gr-qc]}
  \BibitemShut {NoStop}%
\bibitem [{\citenamefont {Abac}\ \emph {et~al.}(2025)\citenamefont {Abac} \emph
  {et~al.}}]{LIGOScientific:2025brd}%
  \BibitemOpen
  \bibfield  {author} {\bibinfo {author} {\bibfnamefont {A.~G.}\ \bibnamefont
  {Abac}} \emph {et~al.} (\bibinfo {collaboration} {LIGO Scientific, Virgo,
  KAGRA}),\ }\bibfield  {title} {\bibinfo {title} {{GW241011 and GW241110:
  Exploring Binary Formation and Fundamental Physics with Asymmetric, High-spin
  Black Hole Coalescences}},\ }\href {https://doi.org/10.3847/2041-8213/ae0d54}
  {\bibfield  {journal} {\bibinfo  {journal} {Astrophys. J. Lett.}\ }\textbf
  {\bibinfo {volume} {993}},\ \bibinfo {pages} {L21} (\bibinfo {year}
  {2025})},\ \Eprint {https://arxiv.org/abs/2510.26931} {arXiv:2510.26931
  [astro-ph.HE]} \BibitemShut {NoStop}%
\bibitem [{\citenamefont {Saleem}\ \emph {et~al.}(2022)\citenamefont {Saleem},
  \citenamefont {Krishnendu}, \citenamefont {Ghosh}, \citenamefont {Gupta},
  \citenamefont {Del~Pozzo}, \citenamefont {Ghosh},\ and\ \citenamefont
  {Arun}}]{Saleem:2021vph}%
  \BibitemOpen
  \bibfield  {author} {\bibinfo {author} {\bibfnamefont {M.}~\bibnamefont
  {Saleem}}, \bibinfo {author} {\bibfnamefont {N.~V.}\ \bibnamefont
  {Krishnendu}}, \bibinfo {author} {\bibfnamefont {A.}~\bibnamefont {Ghosh}},
  \bibinfo {author} {\bibfnamefont {A.}~\bibnamefont {Gupta}}, \bibinfo
  {author} {\bibfnamefont {W.}~\bibnamefont {Del~Pozzo}}, \bibinfo {author}
  {\bibfnamefont {A.}~\bibnamefont {Ghosh}},\ and\ \bibinfo {author}
  {\bibfnamefont {K.~G.}\ \bibnamefont {Arun}},\ }\bibfield  {title} {\bibinfo
  {title} {{Population inference of spin-induced quadrupole moments as a probe
  for nonblack hole compact binaries}},\ }\href
  {https://doi.org/10.1103/PhysRevD.105.104066} {\bibfield  {journal} {\bibinfo
   {journal} {Phys. Rev. D}\ }\textbf {\bibinfo {volume} {105}},\ \bibinfo
  {pages} {104066} (\bibinfo {year} {2022})},\ \Eprint
  {https://arxiv.org/abs/2111.04135} {arXiv:2111.04135 [gr-qc]} \BibitemShut
  {NoStop}%
\bibitem [{\citenamefont {Divyajyoti}\ \emph {et~al.}(2024)\citenamefont
  {Divyajyoti}, \citenamefont {Krishnendu}, \citenamefont {Saleem},
  \citenamefont {Colleoni}, \citenamefont {Vijaykumar}, \citenamefont {Arun},\
  and\ \citenamefont {Mishra}}]{Divyajyoti:2023izl}%
  \BibitemOpen
  \bibfield  {author} {\bibinfo {author} {\bibnamefont {Divyajyoti}}, \bibinfo
  {author} {\bibfnamefont {N.~V.}\ \bibnamefont {Krishnendu}}, \bibinfo
  {author} {\bibfnamefont {M.}~\bibnamefont {Saleem}}, \bibinfo {author}
  {\bibfnamefont {M.}~\bibnamefont {Colleoni}}, \bibinfo {author}
  {\bibfnamefont {A.}~\bibnamefont {Vijaykumar}}, \bibinfo {author}
  {\bibfnamefont {K.~G.}\ \bibnamefont {Arun}},\ and\ \bibinfo {author}
  {\bibfnamefont {C.~K.}\ \bibnamefont {Mishra}},\ }\bibfield  {title}
  {\bibinfo {title} {{Effect of double spin-precession and higher harmonics on
  spin-induced quadrupole moment measurements}},\ }\href
  {https://doi.org/10.1103/PhysRevD.109.023016} {\bibfield  {journal} {\bibinfo
   {journal} {Phys. Rev. D}\ }\textbf {\bibinfo {volume} {109}},\ \bibinfo
  {pages} {023016} (\bibinfo {year} {2024})},\ \Eprint
  {https://arxiv.org/abs/2311.05506} {arXiv:2311.05506 [gr-qc]} \BibitemShut
  {NoStop}%
\bibitem [{\citenamefont {Krishnendu}\ \emph
  {et~al.}(2019{\natexlab{b}})\citenamefont {Krishnendu}, \citenamefont
  {Mishra},\ and\ \citenamefont {Arun}}]{Krishnendu:2018nqa}%
  \BibitemOpen
  \bibfield  {author} {\bibinfo {author} {\bibfnamefont {N.~V.}\ \bibnamefont
  {Krishnendu}}, \bibinfo {author} {\bibfnamefont {C.~K.}\ \bibnamefont
  {Mishra}},\ and\ \bibinfo {author} {\bibfnamefont {K.~G.}\ \bibnamefont
  {Arun}},\ }\bibfield  {title} {\bibinfo {title} {{Spin-induced deformations
  and tests of binary black hole nature using third-generation detectors}},\
  }\href {https://doi.org/10.1103/PhysRevD.99.064008} {\bibfield  {journal}
  {\bibinfo  {journal} {Phys. Rev. D}\ }\textbf {\bibinfo {volume} {99}},\
  \bibinfo {pages} {064008} (\bibinfo {year} {2019}{\natexlab{b}})},\ \Eprint
  {https://arxiv.org/abs/1811.00317} {arXiv:1811.00317 [gr-qc]} \BibitemShut
  {NoStop}%
\bibitem [{\citenamefont {Saini}\ and\ \citenamefont
  {Krishnendu}(2024)}]{Saini:2023gaw}%
  \BibitemOpen
  \bibfield  {author} {\bibinfo {author} {\bibfnamefont {P.}~\bibnamefont
  {Saini}}\ and\ \bibinfo {author} {\bibfnamefont {N.~V.}\ \bibnamefont
  {Krishnendu}},\ }\bibfield  {title} {\bibinfo {title} {{Constraining the
  nature of dark compact objects with spin-induced octupole moment
  measurement}},\ }\href {https://doi.org/10.1103/PhysRevD.109.024009}
  {\bibfield  {journal} {\bibinfo  {journal} {Phys. Rev. D}\ }\textbf {\bibinfo
  {volume} {109}},\ \bibinfo {pages} {024009} (\bibinfo {year} {2024})},\
  \Eprint {https://arxiv.org/abs/2308.01309} {arXiv:2308.01309 [gr-qc]}
  \BibitemShut {NoStop}%
\bibitem [{\citenamefont {Naqvi}\ and\ \citenamefont
  {Mishra}(2025)}]{Naqvi:2025gly}%
  \BibitemOpen
  \bibfield  {author} {\bibinfo {author} {\bibfnamefont {S.~U.}\ \bibnamefont
  {Naqvi}}\ and\ \bibinfo {author} {\bibfnamefont {C.~K.}\ \bibnamefont
  {Mishra}},\ }\bibfield  {title} {\bibinfo {title} {{Spin-induced Quadrupole
  Moment (SIQM) Test for Eccentric Compact Binaries}}\ }(\bibinfo {year}
  {2025})\ \Eprint {https://arxiv.org/abs/2509.10675} {arXiv:2509.10675
  [gr-qc]} \BibitemShut {NoStop}%
\bibitem [{\citenamefont {Krishnendu}(2026)}]{Krishnendu:2025awt}%
  \BibitemOpen
  \bibfield  {author} {\bibinfo {author} {\bibfnamefont {N.~V.}\ \bibnamefont
  {Krishnendu}},\ }\bibfield  {title} {\bibinfo {title} {{Test for eccentric
  binaries based on the spin-induced quadrupole moment}},\ }\href
  {https://doi.org/10.1103/xmv7-sxll} {\bibfield  {journal} {\bibinfo
  {journal} {Phys. Rev. D}\ }\textbf {\bibinfo {volume} {113}},\ \bibinfo
  {pages} {124008} (\bibinfo {year} {2026})},\ \Eprint
  {https://arxiv.org/abs/2512.20579} {arXiv:2512.20579 [gr-qc]} \BibitemShut
  {NoStop}%
\bibitem [{\citenamefont {Branchesi}\ \emph {et~al.}(2023)\citenamefont
  {Branchesi} \emph {et~al.}}]{Branchesi:2023mws}%
  \BibitemOpen
  \bibfield  {author} {\bibinfo {author} {\bibfnamefont {M.}~\bibnamefont
  {Branchesi}} \emph {et~al.},\ }\bibfield  {title} {\bibinfo {title} {{Science
  with the Einstein Telescope: a comparison of different designs}},\ }\href
  {https://doi.org/10.1088/1475-7516/2023/07/068} {\bibfield  {journal}
  {\bibinfo  {journal} {JCAP}\ }\textbf {\bibinfo {volume} {07}},\ \bibinfo
  {pages} {068}},\ \Eprint {https://arxiv.org/abs/2303.15923} {arXiv:2303.15923
  [gr-qc]} \BibitemShut {NoStop}%
\bibitem [{\citenamefont {Abac}\ \emph
  {et~al.}(2026{\natexlab{a}})\citenamefont {Abac} \emph
  {et~al.}}]{ET:2025xjr}%
  \BibitemOpen
  \bibfield  {author} {\bibinfo {author} {\bibfnamefont {A.}~\bibnamefont
  {Abac}} \emph {et~al.} (\bibinfo {collaboration} {ET}),\ }\bibfield  {title}
  {\bibinfo {title} {{The Science of the Einstein Telescope}},\ }\href
  {https://doi.org/10.1088/1475-7516/2026/03/081} {\bibfield  {journal}
  {\bibinfo  {journal} {JCAP}\ }\textbf {\bibinfo {volume} {03}},\ \bibinfo
  {pages} {081}},\ \Eprint {https://arxiv.org/abs/2503.12263} {arXiv:2503.12263
  [gr-qc]} \BibitemShut {NoStop}%
\bibitem [{\citenamefont {Krishnendu}\ and\ \citenamefont
  {Yelikar}(2020)}]{Krishnendu:2019ebd}%
  \BibitemOpen
  \bibfield  {author} {\bibinfo {author} {\bibfnamefont {N.~V.}\ \bibnamefont
  {Krishnendu}}\ and\ \bibinfo {author} {\bibfnamefont {A.~B.}\ \bibnamefont
  {Yelikar}},\ }\bibfield  {title} {\bibinfo {title} {{Testing the Kerr nature
  of supermassive and intermediate-mass black hole binaries using spin-induced
  multipole moment measurements}},\ }\href
  {https://doi.org/10.1088/1361-6382/ababb1} {\bibfield  {journal} {\bibinfo
  {journal} {Class. Quant. Grav.}\ }\textbf {\bibinfo {volume} {37}},\ \bibinfo
  {pages} {205019} (\bibinfo {year} {2020})},\ \Eprint
  {https://arxiv.org/abs/1904.12712} {arXiv:1904.12712 [gr-qc]} \BibitemShut
  {NoStop}%
\bibitem [{\citenamefont {Kong}\ and\ \citenamefont
  {Zhang}(2024)}]{Kong:2024ssa}%
  \BibitemOpen
  \bibfield  {author} {\bibinfo {author} {\bibfnamefont {Y.-L.}\ \bibnamefont
  {Kong}}\ and\ \bibinfo {author} {\bibfnamefont {J.-d.}\ \bibnamefont
  {Zhang}},\ }\bibfield  {title} {\bibinfo {title} {{Probing the spin-induced
  quadrupole moment of massive black holes with the inspiral of binary black
  holes}},\ }\href {https://doi.org/10.1103/PhysRevD.110.024059} {\bibfield
  {journal} {\bibinfo  {journal} {Phys. Rev. D}\ }\textbf {\bibinfo {volume}
  {110}},\ \bibinfo {pages} {024059} (\bibinfo {year} {2024})},\ \Eprint
  {https://arxiv.org/abs/2401.12066} {arXiv:2401.12066 [gr-qc]} \BibitemShut
  {NoStop}%
\bibitem [{\citenamefont {Piarulli}\ \emph {et~al.}(2025)\citenamefont
  {Piarulli}, \citenamefont {Marsat}, \citenamefont {S{\"a}nger}, \citenamefont
  {Buonanno}, \citenamefont {Steinhoff},\ and\ \citenamefont
  {Tamanini}}]{Piarulli:2025rvr}%
  \BibitemOpen
  \bibfield  {author} {\bibinfo {author} {\bibfnamefont {M.}~\bibnamefont
  {Piarulli}}, \bibinfo {author} {\bibfnamefont {S.}~\bibnamefont {Marsat}},
  \bibinfo {author} {\bibfnamefont {E.~M.}\ \bibnamefont {S{\"a}nger}},
  \bibinfo {author} {\bibfnamefont {A.}~\bibnamefont {Buonanno}}, \bibinfo
  {author} {\bibfnamefont {J.}~\bibnamefont {Steinhoff}},\ and\ \bibinfo
  {author} {\bibfnamefont {N.}~\bibnamefont {Tamanini}},\ }\bibfield  {title}
  {\bibinfo {title} {{Parametrized test of general relativity for LISA massive
  black hole binary inspirals}},\ }\href {https://doi.org/10.1103/59zd-qvbd}
  {\bibfield  {journal} {\bibinfo  {journal} {Phys. Rev. D}\ }\textbf {\bibinfo
  {volume} {112}},\ \bibinfo {pages} {124044} (\bibinfo {year} {2025})},\
  \Eprint {https://arxiv.org/abs/2510.06330} {arXiv:2510.06330 [gr-qc]}
  \BibitemShut {NoStop}%
\bibitem [{\citenamefont {Mehta}\ \emph {et~al.}(2023)\citenamefont {Mehta},
  \citenamefont {Buonanno}, \citenamefont {Cotesta}, \citenamefont {Ghosh},
  \citenamefont {Sennett},\ and\ \citenamefont {Steinhoff}}]{Mehta:2022pcn}%
  \BibitemOpen
  \bibfield  {author} {\bibinfo {author} {\bibfnamefont {A.~K.}\ \bibnamefont
  {Mehta}}, \bibinfo {author} {\bibfnamefont {A.}~\bibnamefont {Buonanno}},
  \bibinfo {author} {\bibfnamefont {R.}~\bibnamefont {Cotesta}}, \bibinfo
  {author} {\bibfnamefont {A.}~\bibnamefont {Ghosh}}, \bibinfo {author}
  {\bibfnamefont {N.}~\bibnamefont {Sennett}},\ and\ \bibinfo {author}
  {\bibfnamefont {J.}~\bibnamefont {Steinhoff}},\ }\bibfield  {title} {\bibinfo
  {title} {{Tests of general relativity with gravitational-wave observations
  using a flexible theory-independent method}},\ }\href
  {https://doi.org/10.1103/PhysRevD.107.044020} {\bibfield  {journal} {\bibinfo
   {journal} {Phys. Rev. D}\ }\textbf {\bibinfo {volume} {107}},\ \bibinfo
  {pages} {044020} (\bibinfo {year} {2023})},\ \Eprint
  {https://arxiv.org/abs/2203.13937} {arXiv:2203.13937 [gr-qc]} \BibitemShut
  {NoStop}%
\bibitem [{\citenamefont {Abbott}\ \emph
  {et~al.}(2019{\natexlab{a}})\citenamefont {Abbott} \emph
  {et~al.}}]{LIGOScientific:2018dkp}%
  \BibitemOpen
  \bibfield  {author} {\bibinfo {author} {\bibfnamefont {B.~P.}\ \bibnamefont
  {Abbott}} \emph {et~al.} (\bibinfo {collaboration} {LIGO Scientific,
  Virgo}),\ }\bibfield  {title} {\bibinfo {title} {{Tests of General Relativity
  with GW170817}},\ }\href {https://doi.org/10.1103/PhysRevLett.123.011102}
  {\bibfield  {journal} {\bibinfo  {journal} {Phys. Rev. Lett.}\ }\textbf
  {\bibinfo {volume} {123}},\ \bibinfo {pages} {011102} (\bibinfo {year}
  {2019}{\natexlab{a}})},\ \Eprint {https://arxiv.org/abs/1811.00364}
  {arXiv:1811.00364 [gr-qc]} \BibitemShut {NoStop}%
\bibitem [{\citenamefont {Abbott}\ \emph
  {et~al.}(2019{\natexlab{b}})\citenamefont {Abbott} \emph
  {et~al.}}]{LIGOScientific:2019fpa}%
  \BibitemOpen
  \bibfield  {author} {\bibinfo {author} {\bibfnamefont {B.~P.}\ \bibnamefont
  {Abbott}} \emph {et~al.} (\bibinfo {collaboration} {LIGO Scientific,
  Virgo}),\ }\bibfield  {title} {\bibinfo {title} {{Tests of General Relativity
  with the Binary Black Hole Signals from the LIGO-Virgo Catalog GWTC-1}},\
  }\href {https://doi.org/10.1103/PhysRevD.100.104036} {\bibfield  {journal}
  {\bibinfo  {journal} {Phys. Rev. D}\ }\textbf {\bibinfo {volume} {100}},\
  \bibinfo {pages} {104036} (\bibinfo {year} {2019}{\natexlab{b}})},\ \Eprint
  {https://arxiv.org/abs/1903.04467} {arXiv:1903.04467 [gr-qc]} \BibitemShut
  {NoStop}%
\bibitem [{\citenamefont {Abbott}\ \emph
  {et~al.}(2020{\natexlab{b}})\citenamefont {Abbott} \emph
  {et~al.}}]{LIGOScientific:2020aai}%
  \BibitemOpen
  \bibfield  {author} {\bibinfo {author} {\bibfnamefont {B.~P.}\ \bibnamefont
  {Abbott}} \emph {et~al.} (\bibinfo {collaboration} {LIGO Scientific,
  Virgo}),\ }\bibfield  {title} {\bibinfo {title} {{GW190425: Observation of a
  Compact Binary Coalescence with Total Mass $\sim 3.4 M_{\odot}$}},\ }\href
  {https://doi.org/10.3847/2041-8213/ab75f5} {\bibfield  {journal} {\bibinfo
  {journal} {Astrophys. J. Lett.}\ }\textbf {\bibinfo {volume} {892}},\
  \bibinfo {pages} {L3} (\bibinfo {year} {2020}{\natexlab{b}})},\ \Eprint
  {https://arxiv.org/abs/2001.01761} {arXiv:2001.01761 [astro-ph.HE]}
  \BibitemShut {NoStop}%
\bibitem [{\citenamefont {S{\"a}nger}\ \emph {et~al.}(2026)\citenamefont
  {S{\"a}nger} \emph {et~al.}}]{Sanger:2024axs}%
  \BibitemOpen
  \bibfield  {author} {\bibinfo {author} {\bibfnamefont {E.~M.}\ \bibnamefont
  {S{\"a}nger}} \emph {et~al.},\ }\bibfield  {title} {\bibinfo {title} {{Tests
  of general relativity with GW230529: A neutron star merging with a lower
  mass-gap compact object}},\ }\href {https://doi.org/10.1103/r43k-51yq}
  {\bibfield  {journal} {\bibinfo  {journal} {Phys. Rev. D}\ }\textbf {\bibinfo
  {volume} {113}},\ \bibinfo {pages} {084070} (\bibinfo {year} {2026})},\
  \Eprint {https://arxiv.org/abs/2406.03568} {arXiv:2406.03568 [gr-qc]}
  \BibitemShut {NoStop}%
\bibitem [{\citenamefont {Abac}\ \emph
  {et~al.}(2026{\natexlab{b}})\citenamefont {Abac} \emph
  {et~al.}}]{LIGOScientific:2025cmm}%
  \BibitemOpen
  \bibfield  {author} {\bibinfo {author} {\bibfnamefont {A.~G.}\ \bibnamefont
  {Abac}} \emph {et~al.} (\bibinfo {collaboration} {LIGO Scientific, Virgo,
  KAGRA}),\ }\bibfield  {title} {\bibinfo {title} {{GW230814: Investigation of
  a Loud Gravitational-wave Signal Observed with a Single Detector}},\ }\href
  {https://doi.org/10.3847/2041-8213/ae2ad3} {\bibfield  {journal} {\bibinfo
  {journal} {Astrophys. J. Lett.}\ }\textbf {\bibinfo {volume} {1004}},\
  \bibinfo {pages} {L23} (\bibinfo {year} {2026}{\natexlab{b}})},\ \Eprint
  {https://arxiv.org/abs/2509.07348} {arXiv:2509.07348 [gr-qc]} \BibitemShut
  {NoStop}%
\bibitem [{\citenamefont {Abac}\ \emph
  {et~al.}(2026{\natexlab{c}})\citenamefont {Abac} \emph
  {et~al.}}]{LIGOScientific:2025wao}%
  \BibitemOpen
  \bibfield  {author} {\bibinfo {author} {\bibfnamefont {A.~G.}\ \bibnamefont
  {Abac}} \emph {et~al.} (\bibinfo {collaboration} {LIGO Scientific, Virgo,
  KAGRA}),\ }\bibfield  {title} {\bibinfo {title} {{Black Hole Spectroscopy and
  Tests of General Relativity with GW250114}},\ }\href
  {https://doi.org/10.1103/6c61-fm1n} {\bibfield  {journal} {\bibinfo
  {journal} {Phys. Rev. Lett.}\ }\textbf {\bibinfo {volume} {136}},\ \bibinfo
  {pages} {041403} (\bibinfo {year} {2026}{\natexlab{c}})},\ \Eprint
  {https://arxiv.org/abs/2509.08099} {arXiv:2509.08099 [gr-qc]} \BibitemShut
  {NoStop}%
\bibitem [{\citenamefont {Abac}\ \emph
  {et~al.}(2026{\natexlab{d}})\citenamefont {Abac} \emph
  {et~al.}}]{LIGOScientific:2026fcf}%
  \BibitemOpen
  \bibfield  {author} {\bibinfo {author} {\bibfnamefont {A.~G.}\ \bibnamefont
  {Abac}} \emph {et~al.} (\bibinfo {collaboration} {LIGO Scientific, VIRGO,
  KAGRA}),\ }\bibfield  {title} {\bibinfo {title} {{GWTC-4.0: Tests of General
  Relativity. II. Parameterized Tests}},\ }\href@noop {} {\  (\bibinfo {year}
  {2026}{\natexlab{d}})},\ \Eprint {https://arxiv.org/abs/2603.19020}
  {arXiv:2603.19020 [gr-qc]} \BibitemShut {NoStop}%
\bibitem [{\citenamefont {Abac}\ \emph
  {et~al.}(2026{\natexlab{e}})\citenamefont {Abac} \emph
  {et~al.}}]{LIGOScientific:2026qqv}%
  \BibitemOpen
  \bibfield  {author} {\bibinfo {author} {\bibfnamefont {A.~G.}\ \bibnamefont
  {Abac}} \emph {et~al.} (\bibinfo {collaboration} {LIGO Scientific, Virgo,
  KAGRA}),\ }\bibfield  {title} {\bibinfo {title} {{GW240925 and GW250207:
  Astrophysical Calibration of Gravitational-wave Detectors}}\ }\href
  {https://doi.org/10.1103/gzrj-mwv3} {10.1103/gzrj-mwv3} (\bibinfo {year}
  {2026}{\natexlab{e}}),\ \Eprint {https://arxiv.org/abs/2605.11703}
  {arXiv:2605.11703 [gr-qc]} \BibitemShut {NoStop}%
\bibitem [{\citenamefont {Blanchet}(2024)}]{Blanchet:2013haa}%
  \BibitemOpen
  \bibfield  {author} {\bibinfo {author} {\bibfnamefont {L.}~\bibnamefont
  {Blanchet}},\ }\bibfield  {title} {\bibinfo {title} {{Post-Newtonian Theory
  for Gravitational Waves}},\ }\href
  {https://doi.org/10.1007/s41114-024-00050-z} {\bibfield  {journal} {\bibinfo
  {journal} {Living Rev. Rel.}\ }\textbf {\bibinfo {volume} {27}},\ \bibinfo
  {pages} {4} (\bibinfo {year} {2024})},\ \Eprint
  {https://arxiv.org/abs/1310.1528} {arXiv:1310.1528 [gr-qc]} \BibitemShut
  {NoStop}%
\bibitem [{\citenamefont {Sathyaprakash}\ and\ \citenamefont
  {Dhurandhar}(1991)}]{Sathyaprakash:1991mt}%
  \BibitemOpen
  \bibfield  {author} {\bibinfo {author} {\bibfnamefont {B.~S.}\ \bibnamefont
  {Sathyaprakash}}\ and\ \bibinfo {author} {\bibfnamefont {S.~V.}\ \bibnamefont
  {Dhurandhar}},\ }\bibfield  {title} {\bibinfo {title} {{Choice of filters for
  the detection of gravitational waves from coalescing binaries}},\ }\href
  {https://doi.org/10.1103/PhysRevD.44.3819} {\bibfield  {journal} {\bibinfo
  {journal} {Phys. Rev. D}\ }\textbf {\bibinfo {volume} {44}},\ \bibinfo
  {pages} {3819} (\bibinfo {year} {1991})}\BibitemShut {NoStop}%
\bibitem [{\citenamefont {Cutler}\ and\ \citenamefont
  {Flanagan}(1994)}]{Cutler:1994ys}%
  \BibitemOpen
  \bibfield  {author} {\bibinfo {author} {\bibfnamefont {C.}~\bibnamefont
  {Cutler}}\ and\ \bibinfo {author} {\bibfnamefont {E.~E.}\ \bibnamefont
  {Flanagan}},\ }\bibfield  {title} {\bibinfo {title} {{Gravitational waves
  from merging compact binaries: How accurately can one extract the binary's
  parameters from the inspiral wave form?}},\ }\href
  {https://doi.org/10.1103/PhysRevD.49.2658} {\bibfield  {journal} {\bibinfo
  {journal} {Phys. Rev. D}\ }\textbf {\bibinfo {volume} {49}},\ \bibinfo
  {pages} {2658} (\bibinfo {year} {1994})},\ \Eprint
  {https://arxiv.org/abs/gr-qc/9402014} {arXiv:gr-qc/9402014} \BibitemShut
  {NoStop}%
\bibitem [{\citenamefont {Buonanno}\ \emph {et~al.}(2009)\citenamefont
  {Buonanno}, \citenamefont {Iyer}, \citenamefont {Ochsner}, \citenamefont
  {Pan},\ and\ \citenamefont {Sathyaprakash}}]{Buonanno:2009zt}%
  \BibitemOpen
  \bibfield  {author} {\bibinfo {author} {\bibfnamefont {A.}~\bibnamefont
  {Buonanno}}, \bibinfo {author} {\bibfnamefont {B.}~\bibnamefont {Iyer}},
  \bibinfo {author} {\bibfnamefont {E.}~\bibnamefont {Ochsner}}, \bibinfo
  {author} {\bibfnamefont {Y.}~\bibnamefont {Pan}},\ and\ \bibinfo {author}
  {\bibfnamefont {B.~S.}\ \bibnamefont {Sathyaprakash}},\ }\bibfield  {title}
  {\bibinfo {title} {{Comparison of post-Newtonian templates for compact binary
  inspiral signals in gravitational-wave detectors}},\ }\href
  {https://doi.org/10.1103/PhysRevD.80.084043} {\bibfield  {journal} {\bibinfo
  {journal} {Phys. Rev. D}\ }\textbf {\bibinfo {volume} {80}},\ \bibinfo
  {pages} {084043} (\bibinfo {year} {2009})},\ \Eprint
  {https://arxiv.org/abs/0907.0700} {arXiv:0907.0700 [gr-qc]} \BibitemShut
  {NoStop}%
\bibitem [{\citenamefont {Pappas}\ and\ \citenamefont
  {Apostolatos}(2012{\natexlab{a}})}]{Pappas:2012ns}%
  \BibitemOpen
  \bibfield  {author} {\bibinfo {author} {\bibfnamefont {G.}~\bibnamefont
  {Pappas}}\ and\ \bibinfo {author} {\bibfnamefont {T.~A.}\ \bibnamefont
  {Apostolatos}},\ }\bibfield  {title} {\bibinfo {title} {{Revising the
  multipole moments of numerical spacetimes, and its consequences}},\ }\href
  {https://doi.org/10.1103/PhysRevLett.108.231104} {\bibfield  {journal}
  {\bibinfo  {journal} {Phys. Rev. Lett.}\ }\textbf {\bibinfo {volume} {108}},\
  \bibinfo {pages} {231104} (\bibinfo {year} {2012}{\natexlab{a}})},\ \Eprint
  {https://arxiv.org/abs/1201.6067} {arXiv:1201.6067 [gr-qc]} \BibitemShut
  {NoStop}%
\bibitem [{\citenamefont {Pappas}\ and\ \citenamefont
  {Apostolatos}(2012{\natexlab{b}})}]{Pappas:2012qg}%
  \BibitemOpen
  \bibfield  {author} {\bibinfo {author} {\bibfnamefont {G.}~\bibnamefont
  {Pappas}}\ and\ \bibinfo {author} {\bibfnamefont {T.~A.}\ \bibnamefont
  {Apostolatos}},\ }\bibfield  {title} {\bibinfo {title} {{Multipole Moments of
  numerical spacetimes}},\ }\href@noop {} {\  (\bibinfo {year}
  {2012}{\natexlab{b}})},\ \Eprint {https://arxiv.org/abs/1211.6299}
  {arXiv:1211.6299 [gr-qc]} \BibitemShut {NoStop}%
\bibitem [{\citenamefont {Harry}\ and\ \citenamefont
  {Hinderer}(2018)}]{Harry:2018hke}%
  \BibitemOpen
  \bibfield  {author} {\bibinfo {author} {\bibfnamefont {I.}~\bibnamefont
  {Harry}}\ and\ \bibinfo {author} {\bibfnamefont {T.}~\bibnamefont
  {Hinderer}},\ }\bibfield  {title} {\bibinfo {title} {{Observing and measuring
  the neutron-star equation-of-state in spinning binary neutron star
  systems}},\ }\href {https://doi.org/10.1088/1361-6382/aac7e3} {\bibfield
  {journal} {\bibinfo  {journal} {Class. Quant. Grav.}\ }\textbf {\bibinfo
  {volume} {35}},\ \bibinfo {pages} {145010} (\bibinfo {year} {2018})},\
  \Eprint {https://arxiv.org/abs/1801.09972} {arXiv:1801.09972 [gr-qc]}
  \BibitemShut {NoStop}%
\bibitem [{\citenamefont {Ryan}(1997)}]{Ryan:1996nk}%
  \BibitemOpen
  \bibfield  {author} {\bibinfo {author} {\bibfnamefont {F.~D.}\ \bibnamefont
  {Ryan}},\ }\bibfield  {title} {\bibinfo {title} {{Spinning boson stars with
  large selfinteraction}},\ }\href {https://doi.org/10.1103/PhysRevD.55.6081}
  {\bibfield  {journal} {\bibinfo  {journal} {Phys. Rev. D}\ }\textbf {\bibinfo
  {volume} {55}},\ \bibinfo {pages} {6081} (\bibinfo {year}
  {1997})}\BibitemShut {NoStop}%
\bibitem [{\citenamefont {Uchikata}\ and\ \citenamefont
  {Yoshida}(2016)}]{Uchikata:2015yma}%
  \BibitemOpen
  \bibfield  {author} {\bibinfo {author} {\bibfnamefont {N.}~\bibnamefont
  {Uchikata}}\ and\ \bibinfo {author} {\bibfnamefont {S.}~\bibnamefont
  {Yoshida}},\ }\bibfield  {title} {\bibinfo {title} {{Slowly rotating thin
  shell gravastars}},\ }\href {https://doi.org/10.1088/0264-9381/33/2/025005}
  {\bibfield  {journal} {\bibinfo  {journal} {Class. Quant. Grav.}\ }\textbf
  {\bibinfo {volume} {33}},\ \bibinfo {pages} {025005} (\bibinfo {year}
  {2016})},\ \Eprint {https://arxiv.org/abs/1506.06485} {arXiv:1506.06485
  [gr-qc]} \BibitemShut {NoStop}%
\bibitem [{\citenamefont {Chia}\ \emph {et~al.}(2022)\citenamefont {Chia},
  \citenamefont {Edwards}, \citenamefont {George}, \citenamefont {Zimmerman},
  \citenamefont {Coogan}, \citenamefont {Freese}, \citenamefont {Messick},\
  and\ \citenamefont {Setzer}}]{Chia:2022rwc}%
  \BibitemOpen
  \bibfield  {author} {\bibinfo {author} {\bibfnamefont {H.~S.}\ \bibnamefont
  {Chia}}, \bibinfo {author} {\bibfnamefont {T.~D.~P.}\ \bibnamefont
  {Edwards}}, \bibinfo {author} {\bibfnamefont {R.~N.}\ \bibnamefont {George}},
  \bibinfo {author} {\bibfnamefont {A.}~\bibnamefont {Zimmerman}}, \bibinfo
  {author} {\bibfnamefont {A.}~\bibnamefont {Coogan}}, \bibinfo {author}
  {\bibfnamefont {K.}~\bibnamefont {Freese}}, \bibinfo {author} {\bibfnamefont
  {C.}~\bibnamefont {Messick}},\ and\ \bibinfo {author} {\bibfnamefont {C.~N.}\
  \bibnamefont {Setzer}},\ }\bibfield  {title} {\bibinfo {title}
  {{Dimensionally Reduced Waveforms for Spin-Induced Quadrupole Searches}},\
  }\href@noop {} {\  (\bibinfo {year} {2022})},\ \Eprint
  {https://arxiv.org/abs/2211.00039} {arXiv:2211.00039 [gr-qc]} \BibitemShut
  {NoStop}%
\bibitem [{\citenamefont {Das}\ \emph {et~al.}(2026)\citenamefont {Das},
  \citenamefont {Krishnendu}, \citenamefont {Saleem}, \citenamefont {Mishra},\
  and\ \citenamefont {Arun}}]{Das:2026tel}%
  \BibitemOpen
  \bibfield  {author} {\bibinfo {author} {\bibfnamefont {R.}~\bibnamefont
  {Das}}, \bibinfo {author} {\bibfnamefont {N.~V.}\ \bibnamefont {Krishnendu}},
  \bibinfo {author} {\bibfnamefont {M.}~\bibnamefont {Saleem}}, \bibinfo
  {author} {\bibfnamefont {C.~K.}\ \bibnamefont {Mishra}},\ and\ \bibinfo
  {author} {\bibfnamefont {K.~G.}\ \bibnamefont {Arun}},\ }\bibfield  {title}
  {\bibinfo {title} {{Testing the Kerr hypothesis beyond the quadrupole with
  GW241011}},\ }\href@noop {} {\  (\bibinfo {year} {2026})},\ \Eprint
  {https://arxiv.org/abs/2604.09828} {arXiv:2604.09828 [gr-qc]} \BibitemShut
  {NoStop}%
\bibitem [{\citenamefont {Pompili}\ \emph {et~al.}(2023)\citenamefont {Pompili}
  \emph {et~al.}}]{Pompili:2023tna}%
  \BibitemOpen
  \bibfield  {author} {\bibinfo {author} {\bibfnamefont {L.}~\bibnamefont
  {Pompili}} \emph {et~al.},\ }\bibfield  {title} {\bibinfo {title} {{Laying
  the foundation of the effective-one-body waveform models SEOBNRv5: Improved
  accuracy and efficiency for spinning nonprecessing binary black holes}},\
  }\href {https://doi.org/10.1103/PhysRevD.108.124035} {\bibfield  {journal}
  {\bibinfo  {journal} {Phys. Rev. D}\ }\textbf {\bibinfo {volume} {108}},\
  \bibinfo {pages} {124035} (\bibinfo {year} {2023})},\ \Eprint
  {https://arxiv.org/abs/2303.18039} {arXiv:2303.18039 [gr-qc]} \BibitemShut
  {NoStop}%
\bibitem [{\citenamefont {Ashton}\ \emph {et~al.}(2025)\citenamefont {Ashton},
  \citenamefont {Talbot}, \citenamefont {Roy}, \citenamefont {Pratten},
  \citenamefont {Pang}, \citenamefont {Agathos}, \citenamefont {Baka},
  \citenamefont {Sänger}, \citenamefont {Mehta}, \citenamefont {Steinhoff},
  \citenamefont {Maggio}, \citenamefont {Ghosh}, \citenamefont {Vijaykumar},
  \citenamefont {Enficiaud},\ and\ \citenamefont
  {Pompili}}]{ashton_2025_15676285}%
  \BibitemOpen
  \bibfield  {author} {\bibinfo {author} {\bibfnamefont {G.}~\bibnamefont
  {Ashton}}, \bibinfo {author} {\bibfnamefont {C.}~\bibnamefont {Talbot}},
  \bibinfo {author} {\bibfnamefont {S.}~\bibnamefont {Roy}}, \bibinfo {author}
  {\bibfnamefont {G.}~\bibnamefont {Pratten}}, \bibinfo {author} {\bibfnamefont
  {T.-H.}\ \bibnamefont {Pang}}, \bibinfo {author} {\bibfnamefont
  {M.}~\bibnamefont {Agathos}}, \bibinfo {author} {\bibfnamefont
  {T.}~\bibnamefont {Baka}}, \bibinfo {author} {\bibfnamefont {E.}~\bibnamefont
  {Sänger}}, \bibinfo {author} {\bibfnamefont {A.}~\bibnamefont {Mehta}},
  \bibinfo {author} {\bibfnamefont {J.}~\bibnamefont {Steinhoff}}, \bibinfo
  {author} {\bibfnamefont {E.}~\bibnamefont {Maggio}}, \bibinfo {author}
  {\bibfnamefont {A.}~\bibnamefont {Ghosh}}, \bibinfo {author} {\bibfnamefont
  {A.}~\bibnamefont {Vijaykumar}}, \bibinfo {author} {\bibfnamefont
  {R.}~\bibnamefont {Enficiaud}},\ and\ \bibinfo {author} {\bibfnamefont
  {L.}~\bibnamefont {Pompili}},\ }\href
  {https://doi.org/10.5281/zenodo.15676285} {\bibinfo {title} {{Bilby TGR}}}
  (\bibinfo {year} {2025})\BibitemShut {NoStop}%
\bibitem [{\citenamefont {Ashton}\ \emph {et~al.}(2019)\citenamefont {Ashton}
  \emph {et~al.}}]{Ashton:2018jfp}%
  \BibitemOpen
  \bibfield  {author} {\bibinfo {author} {\bibfnamefont {G.}~\bibnamefont
  {Ashton}} \emph {et~al.},\ }\bibfield  {title} {\bibinfo {title} {{BILBY: A
  user-friendly Bayesian inference library for gravitational-wave astronomy}},\
  }\href {https://doi.org/10.3847/1538-4365/ab06fc} {\bibfield  {journal}
  {\bibinfo  {journal} {Astrophys. J. Suppl.}\ }\textbf {\bibinfo {volume}
  {241}},\ \bibinfo {pages} {27} (\bibinfo {year} {2019})},\ \Eprint
  {https://arxiv.org/abs/1811.02042} {arXiv:1811.02042 [astro-ph.IM]}
  \BibitemShut {NoStop}%
\bibitem [{\citenamefont {Romero-Shaw}\ \emph {et~al.}(2020)\citenamefont
  {Romero-Shaw} \emph {et~al.}}]{Romero-Shaw:2020owr}%
  \BibitemOpen
  \bibfield  {author} {\bibinfo {author} {\bibfnamefont {I.~M.}\ \bibnamefont
  {Romero-Shaw}} \emph {et~al.},\ }\bibfield  {title} {\bibinfo {title}
  {{Bayesian inference for compact binary coalescences with bilby: validation
  and application to the first LIGO{\textendash}Virgo gravitational-wave
  transient catalogue}},\ }\href {https://doi.org/10.1093/mnras/staa2850}
  {\bibfield  {journal} {\bibinfo  {journal} {Mon. Not. Roy. Astron. Soc.}\
  }\textbf {\bibinfo {volume} {499}},\ \bibinfo {pages} {3295} (\bibinfo {year}
  {2020})},\ \Eprint {https://arxiv.org/abs/2006.00714} {arXiv:2006.00714
  [astro-ph.IM]} \BibitemShut {NoStop}%
\bibitem [{\citenamefont {{LIGO Scientific Collaboration}}\ \emph
  {et~al.}(2018)\citenamefont {{LIGO Scientific Collaboration}}, \citenamefont
  {{Virgo Collaboration}},\ and\ \citenamefont {{KAGRA
  Collaboration}}}]{lalsuite}%
  \BibitemOpen
  \bibfield  {author} {\bibinfo {author} {\bibnamefont {{LIGO Scientific
  Collaboration}}}, \bibinfo {author} {\bibnamefont {{Virgo Collaboration}}},\
  and\ \bibinfo {author} {\bibnamefont {{KAGRA Collaboration}}},\ }\href
  {https://doi.org/10.7935/GT1W-FZ16} {\bibinfo {title} {{LVK} {A}lgorithm
  {L}ibrary - {LALS}uite}},\ \bibinfo {howpublished} {Free software (GPL)}
  (\bibinfo {year} {2018})\BibitemShut {NoStop}%
\bibitem [{\citenamefont {Wette}(2020)}]{Wette:2020air}%
  \BibitemOpen
  \bibfield  {author} {\bibinfo {author} {\bibfnamefont {K.}~\bibnamefont
  {Wette}},\ }\bibfield  {title} {\bibinfo {title} {{SWIGLAL: Python and Octave
  interfaces to the LALSuite gravitational-wave data analysis libraries}},\
  }\href {https://doi.org/10.1016/j.softx.2020.100634} {\bibfield  {journal}
  {\bibinfo  {journal} {SoftwareX}\ }\textbf {\bibinfo {volume} {12}},\
  \bibinfo {pages} {100634} (\bibinfo {year} {2020})},\ \Eprint
  {https://arxiv.org/abs/2012.09552} {arXiv:2012.09552 [astro-ph.IM]}
  \BibitemShut {NoStop}%
\bibitem [{\citenamefont {Speagle}(2020)}]{Speagle:2019ivv}%
  \BibitemOpen
  \bibfield  {author} {\bibinfo {author} {\bibfnamefont {J.~S.}\ \bibnamefont
  {Speagle}},\ }\bibfield  {title} {\bibinfo {title} {{dynesty: a dynamic
  nested sampling package for estimating Bayesian posteriors and evidences}},\
  }\href {https://doi.org/10.1093/mnras/staa278} {\bibfield  {journal}
  {\bibinfo  {journal} {Mon. Not. Roy. Astron. Soc.}\ }\textbf {\bibinfo
  {volume} {493}},\ \bibinfo {pages} {3132} (\bibinfo {year} {2020})},\ \Eprint
  {https://arxiv.org/abs/1904.02180} {arXiv:1904.02180 [astro-ph.IM]}
  \BibitemShut {NoStop}%
\bibitem [{\citenamefont {Koposov}\ \emph {et~al.}(2025)\citenamefont
  {Koposov}, \citenamefont {Speagle}, \citenamefont {Barbary}, \citenamefont
  {Ashton}, \citenamefont {Bennett}, \citenamefont {Buchner}, \citenamefont
  {Scheffler}, \citenamefont {Talbot}, \citenamefont {Cook}, \citenamefont
  {Guillochon}, \citenamefont {Cubillos}, \citenamefont {Ramos}, \citenamefont
  {Dartiailh}, \citenamefont {Ilya}, \citenamefont {Tollerud}, \citenamefont
  {Lang}, \citenamefont {Johnson}, \citenamefont {jtmendel}, \citenamefont
  {Higson}, \citenamefont {Vandal}, \citenamefont {Daylan}, \citenamefont
  {Angus}, \citenamefont {patelR}, \citenamefont {Cargile}, \citenamefont
  {Sheehan}, \citenamefont {Pitkin}, \citenamefont {Kirk}, \citenamefont {Xu},
  \citenamefont {Leja},\ and\ \citenamefont
  {joezuntz}}]{sergey_koposov_2025_17268284}%
  \BibitemOpen
  \bibfield  {author} {\bibinfo {author} {\bibfnamefont {S.}~\bibnamefont
  {Koposov}}, \bibinfo {author} {\bibfnamefont {J.}~\bibnamefont {Speagle}},
  \bibinfo {author} {\bibfnamefont {K.}~\bibnamefont {Barbary}}, \bibinfo
  {author} {\bibfnamefont {G.}~\bibnamefont {Ashton}}, \bibinfo {author}
  {\bibfnamefont {E.}~\bibnamefont {Bennett}}, \bibinfo {author} {\bibfnamefont
  {J.}~\bibnamefont {Buchner}}, \bibinfo {author} {\bibfnamefont
  {C.}~\bibnamefont {Scheffler}}, \bibinfo {author} {\bibfnamefont
  {C.}~\bibnamefont {Talbot}}, \bibinfo {author} {\bibfnamefont
  {B.}~\bibnamefont {Cook}}, \bibinfo {author} {\bibfnamefont {J.}~\bibnamefont
  {Guillochon}}, \bibinfo {author} {\bibfnamefont {P.}~\bibnamefont
  {Cubillos}}, \bibinfo {author} {\bibfnamefont {A.~A.}\ \bibnamefont {Ramos}},
  \bibinfo {author} {\bibfnamefont {M.}~\bibnamefont {Dartiailh}}, \bibinfo
  {author} {\bibnamefont {Ilya}}, \bibinfo {author} {\bibfnamefont
  {E.}~\bibnamefont {Tollerud}}, \bibinfo {author} {\bibfnamefont
  {D.}~\bibnamefont {Lang}}, \bibinfo {author} {\bibfnamefont {B.}~\bibnamefont
  {Johnson}}, \bibinfo {author} {\bibnamefont {jtmendel}}, \bibinfo {author}
  {\bibfnamefont {E.}~\bibnamefont {Higson}}, \bibinfo {author} {\bibfnamefont
  {T.}~\bibnamefont {Vandal}}, \bibinfo {author} {\bibfnamefont
  {T.}~\bibnamefont {Daylan}}, \bibinfo {author} {\bibfnamefont
  {R.}~\bibnamefont {Angus}}, \bibinfo {author} {\bibnamefont {patelR}},
  \bibinfo {author} {\bibfnamefont {P.}~\bibnamefont {Cargile}}, \bibinfo
  {author} {\bibfnamefont {P.}~\bibnamefont {Sheehan}}, \bibinfo {author}
  {\bibfnamefont {M.}~\bibnamefont {Pitkin}}, \bibinfo {author} {\bibfnamefont
  {M.}~\bibnamefont {Kirk}}, \bibinfo {author} {\bibfnamefont {L.}~\bibnamefont
  {Xu}}, \bibinfo {author} {\bibfnamefont {J.}~\bibnamefont {Leja}},\ and\
  \bibinfo {author} {\bibnamefont {joezuntz}},\ }\href
  {https://doi.org/10.5281/zenodo.17268284} {\bibinfo {title}
  {joshspeagle/dynesty: v3.0.0}} (\bibinfo {year} {2025})\BibitemShut {NoStop}%
\bibitem [{\citenamefont {Abbott}\ \emph {et~al.}(2016)\citenamefont {Abbott}
  \emph {et~al.}}]{LIGOScientific:2016aoc}%
  \BibitemOpen
  \bibfield  {author} {\bibinfo {author} {\bibfnamefont {B.~P.}\ \bibnamefont
  {Abbott}} \emph {et~al.} (\bibinfo {collaboration} {LIGO Scientific,
  Virgo}),\ }\bibfield  {title} {\bibinfo {title} {{Observation of
  Gravitational Waves from a Binary Black Hole Merger}},\ }\href
  {https://doi.org/10.1103/PhysRevLett.116.061102} {\bibfield  {journal}
  {\bibinfo  {journal} {Phys. Rev. Lett.}\ }\textbf {\bibinfo {volume} {116}},\
  \bibinfo {pages} {061102} (\bibinfo {year} {2016})},\ \Eprint
  {https://arxiv.org/abs/1602.03837} {arXiv:1602.03837 [gr-qc]} \BibitemShut
  {NoStop}%
\bibitem [{\citenamefont {Abbott}\ \emph
  {et~al.}(2020{\natexlab{c}})\citenamefont {Abbott} \emph
  {et~al.}}]{LIGOScientific:2020stg}%
  \BibitemOpen
  \bibfield  {author} {\bibinfo {author} {\bibfnamefont {R.}~\bibnamefont
  {Abbott}} \emph {et~al.} (\bibinfo {collaboration} {LIGO Scientific,
  Virgo}),\ }\bibfield  {title} {\bibinfo {title} {{GW190412: Observation of a
  Binary-Black-Hole Coalescence with Asymmetric Masses}},\ }\href
  {https://doi.org/10.1103/PhysRevD.102.043015} {\bibfield  {journal} {\bibinfo
   {journal} {Phys. Rev. D}\ }\textbf {\bibinfo {volume} {102}},\ \bibinfo
  {pages} {043015} (\bibinfo {year} {2020}{\natexlab{c}})},\ \Eprint
  {https://arxiv.org/abs/2004.08342} {arXiv:2004.08342 [astro-ph.HE]}
  \BibitemShut {NoStop}%
\bibitem [{\citenamefont {Abbott}\ \emph
  {et~al.}(2020{\natexlab{d}})\citenamefont {Abbott} \emph
  {et~al.}}]{KAGRA:2013rdx}%
  \BibitemOpen
  \bibfield  {author} {\bibinfo {author} {\bibfnamefont {B.~P.}\ \bibnamefont
  {Abbott}} \emph {et~al.} (\bibinfo {collaboration} {KAGRA, LIGO Scientific,
  Virgo}),\ }\bibfield  {title} {\bibinfo {title} {{Prospects for observing and
  localizing gravitational-wave transients with Advanced LIGO, Advanced Virgo
  and KAGRA}},\ }\href {https://doi.org/10.1007/s41114-020-00026-9} {\bibfield
  {journal} {\bibinfo  {journal} {Living Rev. Rel.}\ }\textbf {\bibinfo
  {volume} {23}},\ \bibinfo {pages} {3} (\bibinfo {year}
  {2020}{\natexlab{d}})},\ \Eprint {https://arxiv.org/abs/1304.0670}
  {arXiv:1304.0670 [gr-qc]} \BibitemShut {NoStop}%
\bibitem [{\citenamefont {O'Reilly}\ \emph {et~al.}(2022)\citenamefont
  {O'Reilly} \emph {et~al.}}]{psds_for_simulations}%
  \BibitemOpen
  \bibfield  {author} {\bibinfo {author} {\bibfnamefont {B.}~\bibnamefont
  {O'Reilly}} \emph {et~al.},\ }\href {https://dcc.ligo.org/LIGO-T2000012}
  {\emph {\bibinfo {title} {Noise curves used for Simulations in the update of
  the Observing Scenarios Paper}}},\ \bibinfo {type} {Tech. Rep.}\ \bibinfo
  {number} {{LIGO}-T2000012}\ (\bibinfo  {institution} {{LIGO} Project},\
  \bibinfo {year} {2022})\BibitemShut {NoStop}%
\bibitem [{\citenamefont {{The LIGO Scientific Collaboration and the Virgo
  Collaboration and the KAGRA Collaboration}}(2013)}]{GWCommissionObserve}%
  \BibitemOpen
  \bibfield  {author} {\bibinfo {author} {\bibnamefont {{The LIGO Scientific
  Collaboration and the Virgo Collaboration and the KAGRA Collaboration}}},\
  }\href {https://dcc.ligo.org/LIGO-P1200087-v42/public} {\emph {\bibinfo
  {title} {{Prospects for Observing and Localizing Gravitational-Wave
  Transients with Advanced LIGO, Advanced Virgo and KAGRA}}}},\ \bibinfo {type}
  {Tech. Rep.}\ \bibinfo {number} {LIGO-P1200087-v42}\ (\bibinfo  {institution}
  {{The LIGO Scientific Collaboration and the Virgo Collaboration and the KAGRA
  Collaboration}},\ \bibinfo {year} {2013})\BibitemShut {NoStop}%
\bibitem [{\citenamefont {Agathos}\ \emph {et~al.}(2014)\citenamefont
  {Agathos}, \citenamefont {Del~Pozzo}, \citenamefont {Li}, \citenamefont {Van
  Den~Broeck}, \citenamefont {Veitch},\ and\ \citenamefont
  {Vitale}}]{Agathos:2013upa}%
  \BibitemOpen
  \bibfield  {author} {\bibinfo {author} {\bibfnamefont {M.}~\bibnamefont
  {Agathos}}, \bibinfo {author} {\bibfnamefont {W.}~\bibnamefont {Del~Pozzo}},
  \bibinfo {author} {\bibfnamefont {T.~G.~F.}\ \bibnamefont {Li}}, \bibinfo
  {author} {\bibfnamefont {C.}~\bibnamefont {Van Den~Broeck}}, \bibinfo
  {author} {\bibfnamefont {J.}~\bibnamefont {Veitch}},\ and\ \bibinfo {author}
  {\bibfnamefont {S.}~\bibnamefont {Vitale}},\ }\bibfield  {title} {\bibinfo
  {title} {{TIGER: A data analysis pipeline for testing the strong-field
  dynamics of general relativity with gravitational wave signals from
  coalescing compact binaries}},\ }\href
  {https://doi.org/10.1103/PhysRevD.89.082001} {\bibfield  {journal} {\bibinfo
  {journal} {Phys. Rev. D}\ }\textbf {\bibinfo {volume} {89}},\ \bibinfo
  {pages} {082001} (\bibinfo {year} {2014})},\ \Eprint
  {https://arxiv.org/abs/1311.0420} {arXiv:1311.0420 [gr-qc]} \BibitemShut
  {NoStop}%
\bibitem [{\citenamefont {Meidam}\ \emph {et~al.}(2018)\citenamefont {Meidam}
  \emph {et~al.}}]{Meidam:2017dgf}%
  \BibitemOpen
  \bibfield  {author} {\bibinfo {author} {\bibfnamefont {J.}~\bibnamefont
  {Meidam}} \emph {et~al.},\ }\bibfield  {title} {\bibinfo {title}
  {{Parametrized tests of the strong-field dynamics of general relativity using
  gravitational wave signals from coalescing binary black holes: Fast
  likelihood calculations and sensitivity of the method}},\ }\href
  {https://doi.org/10.1103/PhysRevD.97.044033} {\bibfield  {journal} {\bibinfo
  {journal} {Phys. Rev. D}\ }\textbf {\bibinfo {volume} {97}},\ \bibinfo
  {pages} {044033} (\bibinfo {year} {2018})},\ \Eprint
  {https://arxiv.org/abs/1712.08772} {arXiv:1712.08772 [gr-qc]} \BibitemShut
  {NoStop}%
\bibitem [{\citenamefont {Roy}\ \emph {et~al.}(2026)\citenamefont {Roy},
  \citenamefont {Haney}, \citenamefont {Pratten}, \citenamefont {T.~H.~Pang},\
  and\ \citenamefont {Van Den~Broeck}}]{Roy:2025gzv}%
  \BibitemOpen
  \bibfield  {author} {\bibinfo {author} {\bibfnamefont {S.}~\bibnamefont
  {Roy}}, \bibinfo {author} {\bibfnamefont {M.}~\bibnamefont {Haney}}, \bibinfo
  {author} {\bibfnamefont {G.}~\bibnamefont {Pratten}}, \bibinfo {author}
  {\bibfnamefont {P.}~\bibnamefont {T.~H.~Pang}},\ and\ \bibinfo {author}
  {\bibfnamefont {C.}~\bibnamefont {Van Den~Broeck}},\ }\bibfield  {title}
  {\bibinfo {title} {{Improved parametrized test of general relativity using
  the IMRPhenomX waveform family: Including higher harmonics and precession}},\
  }\href {https://doi.org/10.1103/855k-sys5} {\bibfield  {journal} {\bibinfo
  {journal} {Phys. Rev. D}\ }\textbf {\bibinfo {volume} {113}},\ \bibinfo
  {pages} {024016} (\bibinfo {year} {2026})},\ \Eprint
  {https://arxiv.org/abs/2504.21147} {arXiv:2504.21147 [gr-qc]} \BibitemShut
  {NoStop}%
\bibitem [{\citenamefont {Pratten}\ \emph {et~al.}(2020)\citenamefont
  {Pratten}, \citenamefont {Husa}, \citenamefont {Garcia-Quiros}, \citenamefont
  {Colleoni}, \citenamefont {Ramos-Buades}, \citenamefont {Estelles},\ and\
  \citenamefont {Jaume}}]{Pratten:2020fqn}%
  \BibitemOpen
  \bibfield  {author} {\bibinfo {author} {\bibfnamefont {G.}~\bibnamefont
  {Pratten}}, \bibinfo {author} {\bibfnamefont {S.}~\bibnamefont {Husa}},
  \bibinfo {author} {\bibfnamefont {C.}~\bibnamefont {Garcia-Quiros}}, \bibinfo
  {author} {\bibfnamefont {M.}~\bibnamefont {Colleoni}}, \bibinfo {author}
  {\bibfnamefont {A.}~\bibnamefont {Ramos-Buades}}, \bibinfo {author}
  {\bibfnamefont {H.}~\bibnamefont {Estelles}},\ and\ \bibinfo {author}
  {\bibfnamefont {R.}~\bibnamefont {Jaume}},\ }\bibfield  {title} {\bibinfo
  {title} {{Setting the cornerstone for a family of models for gravitational
  waves from compact binaries: The dominant harmonic for nonprecessing
  quasicircular black holes}},\ }\href
  {https://doi.org/10.1103/PhysRevD.102.064001} {\bibfield  {journal} {\bibinfo
   {journal} {Phys. Rev. D}\ }\textbf {\bibinfo {volume} {102}},\ \bibinfo
  {pages} {064001} (\bibinfo {year} {2020})},\ \Eprint
  {https://arxiv.org/abs/2001.11412} {arXiv:2001.11412 [gr-qc]} \BibitemShut
  {NoStop}%
\bibitem [{\citenamefont {Johnson-McDaniel}\ \emph {et~al.}(2022)\citenamefont
  {Johnson-McDaniel}, \citenamefont {Ghosh}, \citenamefont {Ghonge},
  \citenamefont {Saleem}, \citenamefont {Krishnendu},\ and\ \citenamefont
  {Clark}}]{Johnson-McDaniel:2021yge}%
  \BibitemOpen
  \bibfield  {author} {\bibinfo {author} {\bibfnamefont {N.~K.}\ \bibnamefont
  {Johnson-McDaniel}}, \bibinfo {author} {\bibfnamefont {A.}~\bibnamefont
  {Ghosh}}, \bibinfo {author} {\bibfnamefont {S.}~\bibnamefont {Ghonge}},
  \bibinfo {author} {\bibfnamefont {M.}~\bibnamefont {Saleem}}, \bibinfo
  {author} {\bibfnamefont {N.~V.}\ \bibnamefont {Krishnendu}},\ and\ \bibinfo
  {author} {\bibfnamefont {J.~A.}\ \bibnamefont {Clark}},\ }\bibfield  {title}
  {\bibinfo {title} {{Investigating the relation between gravitational wave
  tests of general relativity}},\ }\href
  {https://doi.org/10.1103/PhysRevD.105.044020} {\bibfield  {journal} {\bibinfo
   {journal} {Phys. Rev. D}\ }\textbf {\bibinfo {volume} {105}},\ \bibinfo
  {pages} {044020} (\bibinfo {year} {2022})},\ \Eprint
  {https://arxiv.org/abs/2109.06988} {arXiv:2109.06988 [gr-qc]} \BibitemShut
  {NoStop}%
\bibitem [{\citenamefont {Abbott}\ \emph
  {et~al.}(2019{\natexlab{c}})\citenamefont {Abbott} \emph
  {et~al.}}]{LIGOScientific:2018mvr}%
  \BibitemOpen
  \bibfield  {author} {\bibinfo {author} {\bibfnamefont {B.~P.}\ \bibnamefont
  {Abbott}} \emph {et~al.} (\bibinfo {collaboration} {LIGO Scientific,
  Virgo}),\ }\bibfield  {title} {\bibinfo {title} {{GWTC-1: A
  Gravitational-Wave Transient Catalog of Compact Binary Mergers Observed by
  LIGO and Virgo during the First and Second Observing Runs}},\ }\href
  {https://doi.org/10.1103/PhysRevX.9.031040} {\bibfield  {journal} {\bibinfo
  {journal} {Phys. Rev. X}\ }\textbf {\bibinfo {volume} {9}},\ \bibinfo {pages}
  {031040} (\bibinfo {year} {2019}{\natexlab{c}})},\ \Eprint
  {https://arxiv.org/abs/1811.12907} {arXiv:1811.12907 [astro-ph.HE]}
  \BibitemShut {NoStop}%
\bibitem [{\citenamefont {Abbott}\ \emph
  {et~al.}(2021{\natexlab{b}})\citenamefont {Abbott} \emph
  {et~al.}}]{LIGOScientific:2020ibl}%
  \BibitemOpen
  \bibfield  {author} {\bibinfo {author} {\bibfnamefont {R.}~\bibnamefont
  {Abbott}} \emph {et~al.} (\bibinfo {collaboration} {LIGO Scientific,
  Virgo}),\ }\bibfield  {title} {\bibinfo {title} {{GWTC-2: Compact Binary
  Coalescences Observed by LIGO and Virgo During the First Half of the Third
  Observing Run}},\ }\href {https://doi.org/10.1103/PhysRevX.11.021053}
  {\bibfield  {journal} {\bibinfo  {journal} {Phys. Rev. X}\ }\textbf {\bibinfo
  {volume} {11}},\ \bibinfo {pages} {021053} (\bibinfo {year}
  {2021}{\natexlab{b}})},\ \Eprint {https://arxiv.org/abs/2010.14527}
  {arXiv:2010.14527 [gr-qc]} \BibitemShut {NoStop}%
\bibitem [{\citenamefont {Abbott}\ \emph {et~al.}(2024)\citenamefont {Abbott}
  \emph {et~al.}}]{LIGOScientific:2021usb}%
  \BibitemOpen
  \bibfield  {author} {\bibinfo {author} {\bibfnamefont {R.}~\bibnamefont
  {Abbott}} \emph {et~al.} (\bibinfo {collaboration} {LIGO Scientific,
  VIRGO}),\ }\bibfield  {title} {\bibinfo {title} {{GWTC-2.1: Deep extended
  catalog of compact binary coalescences observed by LIGO and Virgo during the
  first half of the third observing run}},\ }\href
  {https://doi.org/10.1103/PhysRevD.109.022001} {\bibfield  {journal} {\bibinfo
   {journal} {Phys. Rev. D}\ }\textbf {\bibinfo {volume} {109}},\ \bibinfo
  {pages} {022001} (\bibinfo {year} {2024})},\ \Eprint
  {https://arxiv.org/abs/2108.01045} {arXiv:2108.01045 [gr-qc]} \BibitemShut
  {NoStop}%
\bibitem [{\citenamefont {Abbott}\ \emph
  {et~al.}(2023{\natexlab{a}})\citenamefont {Abbott} \emph
  {et~al.}}]{KAGRA:2021vkt}%
  \BibitemOpen
  \bibfield  {author} {\bibinfo {author} {\bibfnamefont {R.}~\bibnamefont
  {Abbott}} \emph {et~al.} (\bibinfo {collaboration} {KAGRA, VIRGO, LIGO
  Scientific}),\ }\bibfield  {title} {\bibinfo {title} {{GWTC-3: Compact Binary
  Coalescences Observed by LIGO and Virgo during the Second Part of the Third
  Observing Run}},\ }\href {https://doi.org/10.1103/PhysRevX.13.041039}
  {\bibfield  {journal} {\bibinfo  {journal} {Phys. Rev. X}\ }\textbf {\bibinfo
  {volume} {13}},\ \bibinfo {pages} {041039} (\bibinfo {year}
  {2023}{\natexlab{a}})},\ \Eprint {https://arxiv.org/abs/2111.03606}
  {arXiv:2111.03606 [gr-qc]} \BibitemShut {NoStop}%
\bibitem [{\citenamefont {Abbott}\ \emph
  {et~al.}(2021{\natexlab{c}})\citenamefont {Abbott} \emph
  {et~al.}}]{LIGOScientific:2019lzm}%
  \BibitemOpen
  \bibfield  {author} {\bibinfo {author} {\bibfnamefont {R.}~\bibnamefont
  {Abbott}} \emph {et~al.} (\bibinfo {collaboration} {LIGO Scientific,
  Virgo}),\ }\bibfield  {title} {\bibinfo {title} {{Open data from the first
  and second observing runs of Advanced LIGO and Advanced Virgo}},\ }\href
  {https://doi.org/10.1016/j.softx.2021.100658} {\bibfield  {journal} {\bibinfo
   {journal} {SoftwareX}\ }\textbf {\bibinfo {volume} {13}},\ \bibinfo {pages}
  {100658} (\bibinfo {year} {2021}{\natexlab{c}})},\ \Eprint
  {https://arxiv.org/abs/1912.11716} {arXiv:1912.11716 [gr-qc]} \BibitemShut
  {NoStop}%
\bibitem [{\citenamefont {{LIGO Scientific Collaboration and Virgo
  Collaboration}}(2022)}]{ligo_scientific_collaboration_and_virgo_2022_6513631}%
  \BibitemOpen
  \bibfield  {author} {\bibinfo {author} {\bibnamefont {{LIGO Scientific
  Collaboration and Virgo Collaboration}}},\ }\bibfield  {title} {\bibinfo
  {title} {{GWTC-2.1: Deep Extended Catalog of Compact Binary Coalescences
  Observed by LIGO and Virgo During the First Half of the Third Observing Run -
  Parameter Estimation Data Release }},\ }\href
  {https://doi.org/10.5281/zenodo.6513631} {10.5281/zenodo.6513631} (\bibinfo
  {year} {2022})\BibitemShut {NoStop}%
\bibitem [{\citenamefont {Abbott}\ \emph
  {et~al.}(2023{\natexlab{b}})\citenamefont {Abbott} \emph
  {et~al.}}]{KAGRA:2023pio}%
  \BibitemOpen
  \bibfield  {author} {\bibinfo {author} {\bibfnamefont {R.}~\bibnamefont
  {Abbott}} \emph {et~al.} (\bibinfo {collaboration} {KAGRA, VIRGO, LIGO
  Scientific}),\ }\bibfield  {title} {\bibinfo {title} {{Open Data from the
  Third Observing Run of LIGO, Virgo, KAGRA, and GEO}},\ }\href
  {https://doi.org/10.3847/1538-4365/acdc9f} {\bibfield  {journal} {\bibinfo
  {journal} {Astrophys. J. Suppl.}\ }\textbf {\bibinfo {volume} {267}},\
  \bibinfo {pages} {29} (\bibinfo {year} {2023}{\natexlab{b}})},\ \Eprint
  {https://arxiv.org/abs/2302.03676} {arXiv:2302.03676 [gr-qc]} \BibitemShut
  {NoStop}%
\bibitem [{\citenamefont {{LIGO Scientific Collaboration and Virgo
  Collaboration and KAGRA
  Collaboration}}(2021)}]{ligo_scientific_collaboration_and_virgo_2021_5546663}%
  \BibitemOpen
  \bibfield  {author} {\bibinfo {author} {\bibnamefont {{LIGO Scientific
  Collaboration and Virgo Collaboration and KAGRA Collaboration}}},\ }\bibfield
   {title} {\bibinfo {title} {{GWTC-3: Compact Binary Coalescences Observed by
  LIGO and Virgo During the Second Part of the Third Observing Run —
  Parameter estimation data release }},\ }\href
  {https://doi.org/10.5281/zenodo.5546663} {10.5281/zenodo.5546663} (\bibinfo
  {year} {2021})\BibitemShut {NoStop}%
\bibitem [{\citenamefont {Hannam}\ \emph {et~al.}(2014)\citenamefont {Hannam},
  \citenamefont {Schmidt}, \citenamefont {Boh{\'e}}, \citenamefont {Haegel},
  \citenamefont {Husa}, \citenamefont {Ohme}, \citenamefont {Pratten},\ and\
  \citenamefont {P{\"u}rrer}}]{Hannam:2013oca}%
  \BibitemOpen
  \bibfield  {author} {\bibinfo {author} {\bibfnamefont {M.}~\bibnamefont
  {Hannam}}, \bibinfo {author} {\bibfnamefont {P.}~\bibnamefont {Schmidt}},
  \bibinfo {author} {\bibfnamefont {A.}~\bibnamefont {Boh{\'e}}}, \bibinfo
  {author} {\bibfnamefont {L.}~\bibnamefont {Haegel}}, \bibinfo {author}
  {\bibfnamefont {S.}~\bibnamefont {Husa}}, \bibinfo {author} {\bibfnamefont
  {F.}~\bibnamefont {Ohme}}, \bibinfo {author} {\bibfnamefont {G.}~\bibnamefont
  {Pratten}},\ and\ \bibinfo {author} {\bibfnamefont {M.}~\bibnamefont
  {P{\"u}rrer}},\ }\bibfield  {title} {\bibinfo {title} {{Simple Model of
  Complete Precessing Black-Hole-Binary Gravitational Waveforms}},\ }\href
  {https://doi.org/10.1103/PhysRevLett.113.151101} {\bibfield  {journal}
  {\bibinfo  {journal} {Phys. Rev. Lett.}\ }\textbf {\bibinfo {volume} {113}},\
  \bibinfo {pages} {151101} (\bibinfo {year} {2014})},\ \Eprint
  {https://arxiv.org/abs/1308.3271} {arXiv:1308.3271 [gr-qc]} \BibitemShut
  {NoStop}%
\bibitem [{\citenamefont {Boh{\'e}}\ \emph {et~al.}(2016)\citenamefont
  {Boh{\'e}}, \citenamefont {Hannam}, \citenamefont {Husa}, \citenamefont
  {Ohme}, \citenamefont {Puerrer},\ and\ \citenamefont {Schmidt}}]{Bohe:PPv2}%
  \BibitemOpen
  \bibfield  {author} {\bibinfo {author} {\bibfnamefont {A.}~\bibnamefont
  {Boh{\'e}}}, \bibinfo {author} {\bibfnamefont {M.}~\bibnamefont {Hannam}},
  \bibinfo {author} {\bibfnamefont {S.}~\bibnamefont {Husa}}, \bibinfo {author}
  {\bibfnamefont {F.}~\bibnamefont {Ohme}}, \bibinfo {author} {\bibfnamefont
  {M.}~\bibnamefont {Puerrer}},\ and\ \bibinfo {author} {\bibfnamefont
  {P.}~\bibnamefont {Schmidt}},\ }\href
  {https://dcc.ligo.org/LIGO-T1500602/public} {\emph {\bibinfo {title}
  {PhenomPv2 - Technical Notes for LAL Implementation}}},\ \bibinfo {type}
  {Tech. Rep.}\ \bibinfo {number} {{LIGO}-T1500602}\ (\bibinfo  {institution}
  {{LIGO} Project},\ \bibinfo {year} {2016})\BibitemShut {NoStop}%
\bibitem [{\citenamefont {Abac}\ \emph
  {et~al.}(2026{\natexlab{f}})\citenamefont {Abac} \emph
  {et~al.}}]{LIGOScientific:2025slb}%
  \BibitemOpen
  \bibfield  {author} {\bibinfo {author} {\bibfnamefont {A.~G.}\ \bibnamefont
  {Abac}} \emph {et~al.} (\bibinfo {collaboration} {LIGO Scientific, VIRGO,
  KAGRA}),\ }\bibfield  {title} {\bibinfo {title} {{GWTC-4.0: Updating the
  Gravitational-Wave Transient Catalog with Observations from the First Part of
  the Fourth LIGO-Virgo-KAGRA Observing Run}},\ }\href
  {https://doi.org/10.3847/2041-8213/ae2c74} {\bibfield  {journal} {\bibinfo
  {journal} {Astrophys. J. Lett.}\ }\textbf {\bibinfo {volume} {1004}},\
  \bibinfo {pages} {L22} (\bibinfo {year} {2026}{\natexlab{f}})},\ \Eprint
  {https://arxiv.org/abs/2508.18082} {arXiv:2508.18082 [gr-qc]} \BibitemShut
  {NoStop}%
\bibitem [{\citenamefont {{LIGO Scientific Collaboration and Virgo
  Collaboration and KAGRA
  Collaboration}}(2026)}]{ligo_scientific_collaboration_2026_21403342}%
  \BibitemOpen
  \bibfield  {author} {\bibinfo {author} {\bibnamefont {{LIGO Scientific
  Collaboration and Virgo Collaboration and KAGRA Collaboration}}},\ }\bibfield
   {title} {\bibinfo {title} {{Data release for GWTC-4.0: Tests of General
  Relativity. II. Parameterized Tests }},\ }\href
  {https://doi.org/10.5281/zenodo.21403342} {10.5281/zenodo.21403342} (\bibinfo
  {year} {2026})\BibitemShut {NoStop}%
\bibitem [{\citenamefont {Isi}\ \emph {et~al.}(2019)\citenamefont {Isi},
  \citenamefont {Chatziioannou},\ and\ \citenamefont {Farr}}]{Isi:2019asy}%
  \BibitemOpen
  \bibfield  {author} {\bibinfo {author} {\bibfnamefont {M.}~\bibnamefont
  {Isi}}, \bibinfo {author} {\bibfnamefont {K.}~\bibnamefont {Chatziioannou}},\
  and\ \bibinfo {author} {\bibfnamefont {W.~M.}\ \bibnamefont {Farr}},\
  }\bibfield  {title} {\bibinfo {title} {{Hierarchical test of general
  relativity with gravitational waves}},\ }\href
  {https://doi.org/10.1103/PhysRevLett.123.121101} {\bibfield  {journal}
  {\bibinfo  {journal} {Phys. Rev. Lett.}\ }\textbf {\bibinfo {volume} {123}},\
  \bibinfo {pages} {121101} (\bibinfo {year} {2019})},\ \Eprint
  {https://arxiv.org/abs/1904.08011} {arXiv:1904.08011 [gr-qc]} \BibitemShut
  {NoStop}%
\bibitem [{\citenamefont {Abac}\ \emph
  {et~al.}(2026{\natexlab{g}})\citenamefont {Abac} \emph
  {et~al.}}]{LIGOScientific:2026qni}%
  \BibitemOpen
  \bibfield  {author} {\bibinfo {author} {\bibfnamefont {A.~G.}\ \bibnamefont
  {Abac}} \emph {et~al.} (\bibinfo {collaboration} {LIGO Scientific, VIRGO,
  KAGRA}),\ }\bibfield  {title} {\bibinfo {title} {{GWTC-4.0: Tests of General
  Relativity. I. Overview and General Tests}},\ }\href@noop {} {\  (\bibinfo
  {year} {2026}{\natexlab{g}})},\ \Eprint {https://arxiv.org/abs/2603.19019}
  {arXiv:2603.19019 [gr-qc]} \BibitemShut {NoStop}%
\bibitem [{\citenamefont {Hild}\ \emph {et~al.}(2011)\citenamefont {Hild} \emph
  {et~al.}}]{Hild:2010id}%
  \BibitemOpen
  \bibfield  {author} {\bibinfo {author} {\bibfnamefont {S.}~\bibnamefont
  {Hild}} \emph {et~al.},\ }\bibfield  {title} {\bibinfo {title} {{Sensitivity
  Studies for Third-Generation Gravitational Wave Observatories}},\ }\href
  {https://doi.org/10.1088/0264-9381/28/9/094013} {\bibfield  {journal}
  {\bibinfo  {journal} {Class. Quant. Grav.}\ }\textbf {\bibinfo {volume}
  {28}},\ \bibinfo {pages} {094013} (\bibinfo {year} {2011})},\ \Eprint
  {https://arxiv.org/abs/1012.0908} {arXiv:1012.0908 [gr-qc]} \BibitemShut
  {NoStop}%
\bibitem [{\citenamefont {Srivastava}\ \emph {et~al.}(2022)\citenamefont
  {Srivastava}, \citenamefont {Davis}, \citenamefont {Kuns}, \citenamefont
  {Landry}, \citenamefont {Ballmer}, \citenamefont {Evans}, \citenamefont
  {Hall}, \citenamefont {Read},\ and\ \citenamefont
  {Sathyaprakash}}]{Srivastava:2022slt}%
  \BibitemOpen
  \bibfield  {author} {\bibinfo {author} {\bibfnamefont {V.}~\bibnamefont
  {Srivastava}}, \bibinfo {author} {\bibfnamefont {D.}~\bibnamefont {Davis}},
  \bibinfo {author} {\bibfnamefont {K.}~\bibnamefont {Kuns}}, \bibinfo {author}
  {\bibfnamefont {P.}~\bibnamefont {Landry}}, \bibinfo {author} {\bibfnamefont
  {S.}~\bibnamefont {Ballmer}}, \bibinfo {author} {\bibfnamefont
  {M.}~\bibnamefont {Evans}}, \bibinfo {author} {\bibfnamefont {E.~D.}\
  \bibnamefont {Hall}}, \bibinfo {author} {\bibfnamefont {J.}~\bibnamefont
  {Read}},\ and\ \bibinfo {author} {\bibfnamefont {B.~S.}\ \bibnamefont
  {Sathyaprakash}},\ }\bibfield  {title} {\bibinfo {title} {{Science-driven
  Tunable Design of Cosmic Explorer Detectors}},\ }\href
  {https://doi.org/10.3847/1538-4357/ac5f04} {\bibfield  {journal} {\bibinfo
  {journal} {Astrophys. J.}\ }\textbf {\bibinfo {volume} {931}},\ \bibinfo
  {pages} {22} (\bibinfo {year} {2022})},\ \Eprint
  {https://arxiv.org/abs/2201.10668} {arXiv:2201.10668 [gr-qc]} \BibitemShut
  {NoStop}%
\bibitem [{\citenamefont {Poisson}\ and\ \citenamefont
  {Will}(1995)}]{Poisson:1995ef}%
  \BibitemOpen
  \bibfield  {author} {\bibinfo {author} {\bibfnamefont {E.}~\bibnamefont
  {Poisson}}\ and\ \bibinfo {author} {\bibfnamefont {C.~M.}\ \bibnamefont
  {Will}},\ }\bibfield  {title} {\bibinfo {title} {{Gravitational waves from
  inspiraling compact binaries: Parameter estimation using second postNewtonian
  wave forms}},\ }\href {https://doi.org/10.1103/PhysRevD.52.848} {\bibfield
  {journal} {\bibinfo  {journal} {Phys. Rev. D}\ }\textbf {\bibinfo {volume}
  {52}},\ \bibinfo {pages} {848} (\bibinfo {year} {1995})},\ \Eprint
  {https://arxiv.org/abs/gr-qc/9502040} {arXiv:gr-qc/9502040} \BibitemShut
  {NoStop}%
\bibitem [{\citenamefont {Borhanian}(2021)}]{Borhanian:2020ypi}%
  \BibitemOpen
  \bibfield  {author} {\bibinfo {author} {\bibfnamefont {S.}~\bibnamefont
  {Borhanian}},\ }\bibfield  {title} {\bibinfo {title} {{GWBENCH: a novel
  Fisher information package for gravitational-wave benchmarking}},\ }\href
  {https://doi.org/10.1088/1361-6382/ac1618} {\bibfield  {journal} {\bibinfo
  {journal} {Class. Quant. Grav.}\ }\textbf {\bibinfo {volume} {38}},\ \bibinfo
  {pages} {175014} (\bibinfo {year} {2021})},\ \Eprint
  {https://arxiv.org/abs/2010.15202} {arXiv:2010.15202 [gr-qc]} \BibitemShut
  {NoStop}%
\bibitem [{\citenamefont {Morisaki}(2021)}]{Morisaki:2021ngj}%
  \BibitemOpen
  \bibfield  {author} {\bibinfo {author} {\bibfnamefont {S.}~\bibnamefont
  {Morisaki}},\ }\bibfield  {title} {\bibinfo {title} {{Accelerating parameter
  estimation of gravitational waves from compact binary coalescence using
  adaptive frequency resolutions}},\ }\href
  {https://doi.org/10.1103/PhysRevD.104.044062} {\bibfield  {journal} {\bibinfo
   {journal} {Phys. Rev. D}\ }\textbf {\bibinfo {volume} {104}},\ \bibinfo
  {pages} {044062} (\bibinfo {year} {2021})},\ \Eprint
  {https://arxiv.org/abs/2104.07813} {arXiv:2104.07813 [gr-qc]} \BibitemShut
  {NoStop}%
\bibitem [{\citenamefont {Adhikari}\ and\ \citenamefont
  {Morisaki}(2022)}]{Adhikari:2022mbj}%
  \BibitemOpen
  \bibfield  {author} {\bibinfo {author} {\bibfnamefont {N.}~\bibnamefont
  {Adhikari}}\ and\ \bibinfo {author} {\bibfnamefont {S.}~\bibnamefont
  {Morisaki}},\ }\bibfield  {title} {\bibinfo {title} {{Accelerating
  gravitational-wave parametrized tests of general relativity using a multiband
  decomposition of likelihood}},\ }\href
  {https://doi.org/10.1103/PhysRevD.106.104053} {\bibfield  {journal} {\bibinfo
   {journal} {Phys. Rev. D}\ }\textbf {\bibinfo {volume} {106}},\ \bibinfo
  {pages} {104053} (\bibinfo {year} {2022})},\ \Eprint
  {https://arxiv.org/abs/2208.03731} {arXiv:2208.03731 [gr-qc]} \BibitemShut
  {NoStop}%
\bibitem [{\citenamefont {Chandra}(2025)}]{Chandra:2024dhf}%
  \BibitemOpen
  \bibfield  {author} {\bibinfo {author} {\bibfnamefont {K.}~\bibnamefont
  {Chandra}},\ }\bibfield  {title} {\bibinfo {title} {{gwforge: a user-friendly
  package to generate gravitational-wave mock data}},\ }\href
  {https://doi.org/10.1088/1361-6382/ad9b68} {\bibfield  {journal} {\bibinfo
  {journal} {Class. Quant. Grav.}\ }\textbf {\bibinfo {volume} {42}},\ \bibinfo
  {pages} {025003} (\bibinfo {year} {2025})},\ \Eprint
  {https://arxiv.org/abs/2407.21109} {arXiv:2407.21109 [gr-qc]} \BibitemShut
  {NoStop}%
\bibitem [{\citenamefont {Madau}\ and\ \citenamefont
  {Dickinson}(2014)}]{Madau:2014bja}%
  \BibitemOpen
  \bibfield  {author} {\bibinfo {author} {\bibfnamefont {P.}~\bibnamefont
  {Madau}}\ and\ \bibinfo {author} {\bibfnamefont {M.}~\bibnamefont
  {Dickinson}},\ }\bibfield  {title} {\bibinfo {title} {{Cosmic Star-Formation
  History}},\ }\href {https://doi.org/10.1146/annurev-astro-081811-125615}
  {\bibfield  {journal} {\bibinfo  {journal} {Ann. Rev. Astron. Astrophys.}\
  }\textbf {\bibinfo {volume} {52}},\ \bibinfo {pages} {415} (\bibinfo {year}
  {2014})},\ \Eprint {https://arxiv.org/abs/1403.0007} {arXiv:1403.0007
  [astro-ph.CO]} \BibitemShut {NoStop}%
\bibitem [{\citenamefont {Abbott}\ \emph
  {et~al.}(2023{\natexlab{c}})\citenamefont {Abbott} \emph
  {et~al.}}]{KAGRA:2021duu}%
  \BibitemOpen
  \bibfield  {author} {\bibinfo {author} {\bibfnamefont {R.}~\bibnamefont
  {Abbott}} \emph {et~al.} (\bibinfo {collaboration} {KAGRA, VIRGO, LIGO
  Scientific}),\ }\bibfield  {title} {\bibinfo {title} {{Population of Merging
  Compact Binaries Inferred Using Gravitational Waves through GWTC-3}},\ }\href
  {https://doi.org/10.1103/PhysRevX.13.011048} {\bibfield  {journal} {\bibinfo
  {journal} {Phys. Rev. X}\ }\textbf {\bibinfo {volume} {13}},\ \bibinfo
  {pages} {011048} (\bibinfo {year} {2023}{\natexlab{c}})},\ \Eprint
  {https://arxiv.org/abs/2111.03634} {arXiv:2111.03634 [astro-ph.HE]}
  \BibitemShut {NoStop}%
\end{thebibliography}%

\end{document}